\documentclass[twocolumn]{aastex701}

\usepackage{xspace,xcolor,ulem}
\usepackage{multirow}
\usepackage{amsmath}
\usepackage{txfonts}

\graphicspath{{./}{Figs/}}
\newcommand{\wT}{0.32\hsize}
\newcommand{\wH}{0.45\hsize}
\newcommand{\wO}{0.9\hsize}

\newcommand{\ObsID}[1]{OBS-#1\xspace}

\newcommand{\xshooter}{X-Shooter\xspace}

\newcommand{\Ms}{M_\mathrm{obj}}
\newcommand{\Rs}{R_\mathrm{obj}}
\newcommand{\Bs}{B_\mathrm{dipole}}
\newcommand{\Msun}{M_\odot}
\newcommand{\Rsun}{R_\odot}
\newcommand{\Lsun}{L_\odot}

\newcommand{\MJ}{M_\mathrm{J}}

\newcommand{\RS}{R_\odot}

\newcommand{\Mdot}{\dot{M}}
\newcommand{\yr}{\mathrm{yr}}
\newcommand{\AV}{A_\mathrm{V}}
\newcommand{\AVtot}{A_\mathrm{V,\,tot}}
\newcommand{\AVCA}{A_\mathrm{V,\,CASPAR}}
\newcommand{\RV}{R_\mathrm{V}}
\newcommand{\vff}{{\ifmmode v_\mathrm{ff} \else $v_\mathrm{ff}$ \xspace\fi}}
\newcommand{\ff}{f_\mathrm{f}}
\newcommand{\kms}{~\mathrm{km~s^{-1}}}
\newcommand{\cc}{~\mathrm{cm^{-3}}}
\newcommand{\Rt}{R_\mathrm{t}}
\newcommand{\Lacc}{L_\mathrm{acc}}

\newcommand{\LaccCA}{L_\mathrm{acc,\,CASPAR}}

\newcommand{\LHa}{L_{\Ha}}

\newcommand{\feff}{{\ifmmode f_\mathrm{eff} \else $f_\mathrm{eff}$ \fi}\xspace}
\newcommand{\lami}{\lambda_i}
\newcommand{\Semit}{{\ifmmode S_\mathrm{emit} \else $S_\mathrm{emit}$\fi}\xspace}
\newcommand{\Lya}{{\ifmmode \text{Ly-}\alpha \else Ly-$\alpha$\fi}\xspace}
\newcommand{\Pa}{{\ifmmode \text{Pa-}\alpha \else Pa-$\alpha$\fi}\xspace}
\newcommand{\Pb}{{\ifmmode \text{Pa-}\beta \else Pa-$\beta$\fi}\xspace}
\newcommand{\Pg}{{\ifmmode \text{Pa-}\gamma \else Pa-$\gamma$\fi}\xspace}
\newcommand{\Pd}{{\ifmmode \text{Pa-}\delta \else Pa-$\delta$\fi}\xspace}

\newcommand{\Bra}{{\ifmmode \text{Br-}\alpha \else Br-$\alpha$\fi}\xspace}
\newcommand{\Brb}{{\ifmmode \text{Br-}\beta \else Br-$\beta$\fi}\xspace}
\newcommand{\Brg}{{\ifmmode \text{Br-}\gamma \else Br-$\gamma$\fi}\xspace}
\newcommand{\Brd}{{\ifmmode \text{Br-}\delta \else Br-$\delta$\fi}\xspace}

\newcommand{\Ha}{{\ifmmode \text{H}\alpha \else H$\alpha$\fi}\xspace}
\newcommand{\Hb}{{\ifmmode \text{H}\beta \else H$\beta$\fi}\xspace}
\newcommand{\Hg}{{\ifmmode \text{H}\gamma \else H$\gamma$\fi}\xspace}
\newcommand{\Hd}{{\ifmmode \text{H}\delta \else H$\delta$\fi}\xspace}

\newcommand{\Fobs}{F_\mathrm{obs}}
\newcommand{\Fmod}{F_\mathrm{mod}}
\newcommand{\Fsyn}{F_\mathrm{syn}}

\newcommand{\FC}{F_\mathrm{C}}
\newcommand{\FN}{F_\mathrm{N}}

\newcommand{\FCN}{F_\mathrm{CN}}

\newcommand{\LBC}{L_\mathrm{BC}}
\newcommand{\Lshock}{L_\mathrm{shock}}

\newcommand{\Tf}{ {\ifmmode T_\mathrm{f} \else $T_\mathrm{f}$\fi}\xspace}
\newcommand{\nf}{ {\ifmmode n_\mathrm{f} \else $n_\mathrm{f}$\fi}\xspace}

\newcommand{\ConcShock}{``S''\xspace}
\newcommand{\ConcNot}{``N''\xspace}
\newcommand{\ConcBC}{``S+BC''\xspace}
\newcommand{\ConcBCd}{``S+BC-like''\xspace}
\newcommand{\flgU}{U\xspace}
\newcommand{\flgQ}{Q\xspace}
\newcommand{\flgDBC}{DBC\xspace}
\newcommand{\redchi}{{\ifmmode \tilde{\chi}^2 \else $\tilde{\chi}^2$\fi}\xspace}
\newcommand{\Pvalue}{{\ifmmode p \else $p$ \fi}\xspace}

\newcommand{\revise}[1]{#1}

\begin{document}

\title{Hydrogen Line Emission in Accreting Low-Mass Objects I: Spectral Analysis of Shock-Origin Narrow Component}

\shorttitle{Hydrogen Lines in Accreting Low-mass Objects I: Spectral Analysis}
\shortauthors{Aoyama et al.}

\author[0000-0003-0568-9225]{Yuhiko Aoyama}
\affiliation{
School of Physics and Astronomy, Sun Yat-sen University, Guangdong 519082, People's Republic of China
}
\affiliation{
Department of Earth and Planetary Science, Graduate School of Science, The University of Tokyo, 7-3-1 Hongo, Bunkyo-ku, Tokyo 113-0033, Japan}
\email[show]{aoyama@sysu.edu.cn}
\correspondingauthor{Yuhiko Aoyama}

\author[0000-0002-3053-3575]{Jun Hashimoto}
\affil{Academia Sinica Institute of Astronomy \& Astrophysics (ASIAA), AS/NTU, No.1, Sec. 4, Roosevelt Rd., Taipei 106319, Taiwan}
\affil{Astrobiology Center, National Institutes of Natural Sciences, 2-21-1 Osawa, Mitaka, Tokyo 181-8588, Japan}
\email{jhashimoto@asiaa.sinica.edu.tw}

\author[0000-0002-2919-7500]{Gabriel-Dominique Marleau}
\affiliation{Fakult\"at f\"ur Physik, Universit\"at Duisburg-Essen, Lotharstraße 1, 47057 Duisburg, Germany}
\affiliation{Division of Space Research and Planetary Sciences, Physics Institute, University of Bern, Gesellschaftsstr.~6, 3012 Bern, Switzerland}
\affiliation{Max-Planck-Institut f\"ur Astronomie, K\"onigstuhl 17, 69117 Heidelberg, Germany}
\email{gabriel.marleau@uni-due.de}

\author[0000-0003-3562-262X]{Carlo F. Manara}
\affiliation{European Southern Observatory, Garching bei München, Germany}
\email{}

\author[0000-0001-8657-095X]{Juan Manuel Alcal\'a}
\affiliation{INAF - Osservatorio Astronomico di Capodimonte, Via Moiariello 16, I-80131, Napoli, Italy}
\email{}

\author[0000-0002-7154-6065]{Gregory Herczeg}
\affiliation{Kavli Institute for Astronomy and Astrophysics, Peking University, Beijing 100084, China}
\email{}

\author[0000-0002-1787-7883]{Camille Bergez-Casalou}
\affiliation{Université Paris-Saclay, CNRS, Institut d’Astrophysique Spatiale, 91405 Orsay, France}
\email{}

\author[0000-0002-5658-5971]{Masahiro Ikoma}
\affiliation{Astrobiology Center, National Institutes of Natural Sciences, 2-21-1 Osawa, Mitaka, Tokyo 181-8588, Japan}
\affiliation{Division of Science, National Astronomical Observatory of Japan, 2-21-1 Osawa, Mitaka, Tokyo 181-8588, Japan}
\affiliation{Department of Earth and Planetary Science, Graduate School of Science, The University of Tokyo, 7-3-1 Hongo, Bunkyo-ku, Tokyo 113-0033, Japan}
\email{masahiro.ikoma@nao.ac.jp}

\begin{abstract}
Hydrogen lines are widely used as tracers of stellar and planetary accretion.
In classical T~Tauri stars, hydrogen lines are usually interpreted as arising from magnetospheric accretion columns, whereas in lower-mass counterparts (subsolar-mass objects including brown dwarfs and gas giant planets), the post-accretion-shock region can directly emit a substantial fraction of the hydrogen-line luminosity. However, the boundary between non-shock-dominated and shock-dominated cases has remained unclear.
In this study, we compare hydrogen-line profiles predicted by the shock emission model with 254 observations of 164 low-mass accreting objects ($\lesssim 0.5\,M_\odot$) in the VLT/X-Shooter archive. We simultaneously fit seven hydrogen lines (H$\beta$, H$\gamma$, H6, H8, H9, Pa$\beta$, and Br$\gamma$), testing both line profiles and flux ratios within a single framework, and introduce a phenomenological broad-component-subtracted fit for cases with mixed non-shock and shock contributions.
We find that shock emission dominates the hydrogen-line emission at object masses $M\lesssim0.05\,M_\odot$ or free-fall velocities $v_\mathrm{ff}<175\,\mathrm{km\,s^{-1}}$, but becomes minor at $M \gtrsim 0.2\,M_\odot$.
The inferred flow velocities at the shock front are often significantly smaller than the free-fall velocity from infinity, implying smaller truncation radii and surface dipole magnetic field strengths of sub-kG.
The accretion luminosities inferred from the shock-emission fitting are systematically larger than literature values, often by orders of magnitude, likely because conventional estimates neglect line emission that is non-negligible in low-mass objects.
We also confirm that H$\alpha$ is more susceptible than the other hydrogen lines to additional non-shock components.
\end{abstract}

\keywords{
\uat{Accretion}{14},
\uat{Brown dwarfs}{185},
\uat{H I line emission}{690},
\uat{Low mass stars}{2050},
\uat{Spectral energy distribution}{2129},
\uat{Young stellar objects}{1834}
}

\section{Introduction}

Young stars and gas-giant planets undergo gas accretion from their surrounding disks during the final stages of their formation. This phase is important not only because it controls the mass growth of the central object, but also because it shapes the structure and evolution of the surrounding disk \citep{Bouvier2014}.
Additionally, because these surrounding disks are the sites of planet/satellite formation around stars/planets, the accretion process also sets the inner boundary conditions for those formation environments \citep[e.g.,][]{manara2023,canup2002}.

Magnetospheric accretion is the most widely accepted framework of accretion in classical T~Tauri stars (CTTS) \citep[for reviews, see][]{Hartmann+2016}. In this scenario, the large-scale magnetic field of the central object truncates the inner disk and channels disk material onto the surface along magnetic field lines \citep[e.g.,][]{Koenigl1991}.
A similar accretion geometry has also been suggested for some accreting gas-giant planets on the basis of hydrogen-line observations and modeling \citep[e.g.,][]{thanathibodee2019,Aoyama+2019,Aoyama+2021,Aoyama+2024}.

\subsection{Hydrogen Line Emission Mechanisms}
\label{sec:I_HLMechanism}
\begin{figure}
    \centering
    \includegraphics[width=\linewidth]{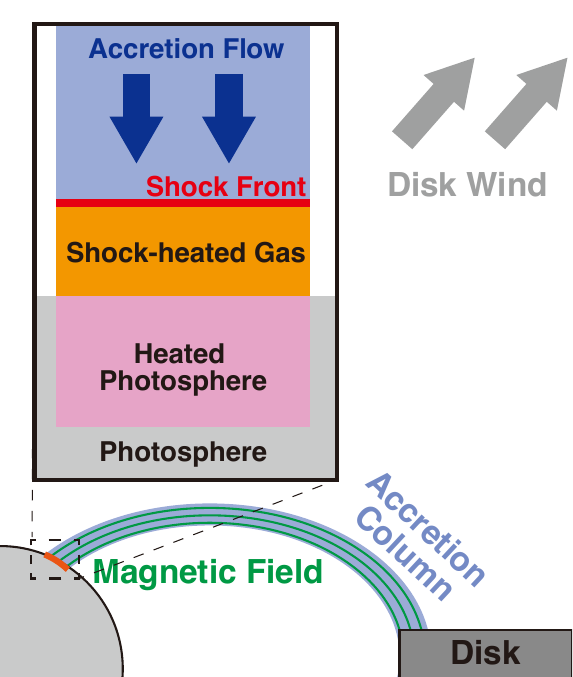}
    \caption{Schematic illustration of magnetospheric accretion and accretion shock. In this paper, ``shock emission'' refers to emission from the shock-heated gas (orange). Emission from the accretion column and/or disk wind is referred to as non-shock emission.}
    \label{fig:Schematic}
\end{figure}

Hydrogen lines are widely used as tracers of accretion \citep[e.g.,][]{bertout1988}. The accretion energy originates from the gravitational potential energy of the inflowing gas and is released at the stellar or planetary surface through a strong accretion shock. The accretion luminosity, which measures this energy-release rate, correlates well with both the fluxes \citep{Gullbring1998,Herczeg+2008,rigliaco2012,alcala2014,Alcala+2017,Fiorellino2025,Shridharan2026} and the spectral widths \citep{natta2004,fang2009,White2003} of hydrogen lines.

Nevertheless, in CTTSs, hydrogen lines are usually thought not to directly originate from the accretion-shock heated gas itself.
Figure~\ref{fig:Schematic} schematically illustrates an accretion shock in magnetospheric accretion. The ``shock-heated gas'' shown in orange refers to the rapidly cooling post-shock flow immediately downstream of the shock front (often referred to as the Zeldovich spike in the context of radiative shocks).
In this paper, we refer to the direct emission from this region, where accretion energy is converted into radiation, as ``shock emission''. This shock-heated gas
is generally too hot to emit hydrogen lines efficiently at CTTSs \citep[e.g.,][]{Lamzin1998,calvet1998}.
Instead, the shock-heated gas primarily emits higher-energy photons, such as X-rays, which
are subsequently absorbed and reprocessed in the pre-shock gas \citep{Lamzin1998,calvet1998} and in the photosphere beneath the shock, \citep[i.e., the heated photosphere;][]{Drake1980,Cranmer2009,dodin2015}, producing optical and infrared continuum emission.
In particular, hydrogen lines in CTTSs are often interpreted as originating in the magnetospheric accretion columns \citep{Hartmann+1994,Muzerolle+1998,Muzerolle+2001,Alencar2012}, because their broad profiles require broadening by the velocity field of the accretion flow, whereas thermal Doppler broadening alone would imply temperatures far above the hydrogen ionization threshold.
Moreover, VLTI/GRAVITY observations constrain the size of the Br$\gamma$-emitting region to within several stellar radii, supporting an origin in the magnetospheric accretion columns \citep{gravitycollaboration_9,GravityCollaboration_10}.

In accreting planets, by contrast, a large fraction of the hydrogen-line emission can arise as shock emission \citep{Aoyama+2018}. Because the free-fall velocity is lower than in CTTSs, the shock is weaker and the shock-heated gas temperature is correspondingly lower.
In this regime, hydrogen lines become efficient coolants of the shock-heated gas.
Indeed, the flux ratios of multiple hydrogen lines (i.e., line-color diagnostics) suggest that some planetary-mass objects favor shock emission rather than non-shock emission as the dominant origin of their hydrogen lines \citep{Betti2022}.
This suggests that the overall accretion geometry may remain broadly magnetospheric in both cases, while the dominant line-emitting region within that framework differs.

This distinction in emission mechanism matters directly for accretion diagnostics, especially for the measurement of accretion luminosity. In many studies, $\Lacc$ is estimated from the UV continuum using slab models \citep{valenti1993,Herczeg+2008,manara2013,manara2014,alcala2014,Alcala+2017}, which effectively represent emission from the heated photosphere. This is a good approximation when the shock-heated gas radiates mainly in X-rays and the accretion energy is reprocessed into optically thick continuum emission.
However, if the shock-heated gas cools efficiently through hydrogen-line emission, a non-negligible fraction of the accretion energy can escape directly in lines rather than being reprocessed by the heated photosphere. Observationally, at low-mass and/or weakly-accreting objects, cases have been reported in which the energy carried by hydrogen-line emission is comparable to that in the continuum \citep[e.g.,][]{Zhou2014},
implying that $\Lacc$ may be underestimated by a factor of a few. Moreover, shock-emission models predict that the unobservable hydrogen line Ly$\alpha$ can be tens of times brighter than \Ha, the brightest observable hydrogen line \citep{Aoyama+2018}, suggesting that $\Lacc$ may in some cases be underestimated by orders of magnitude \citep{Aoyama+2021}.

The transition between these regimes is expected to occur around the shock velocity at which a nearly fully ionized layer appears and hydrogen-line cooling becomes inefficient, namely around $200\,\kms$ \citep{Aoyama+2018,Aoyama+2020}.
For free-fall velocity from infinity at the surface of a $1\,\Rsun$ object, this threshold corresponds to a mass of roughly $0.1\,\Msun$. Therefore, the regime where shock emission dominates hydrogen lines may extend beyond planets into brown dwarfs and even the low-mass end of M dwarfs. Motivated by this, \citet{Hashimoto+Aoyama2025} applied hydrogen-line flux-ratio diagnostics to brown dwarfs and very-low-mass stars in order to identify the boundary between shock-dominated and non-shock-dominated regimes. Although they found some objects whose hydrogen lines appear to be dominated by shock emission, the results showed large scatter against object properties such as mass and free-fall velocity, and the boundary between the two emission regimes remained unclear.
This may indicate that the dominant emission mechanism changes only gradually and that mixed contributions are common over a broad parameter space. To investigate such intermediate cases further, it may be useful to consider whether different contributions can be separated, at least phenomenologically, in the observed line profiles.

\subsection{Broad and Narrow Components of Excess Lines}
\label{sec:I_BCNC}

One possible phenomenological way to describe such mixed profiles is in terms of a narrow component (NC) and a broad component (BC).
Metal lines, such as \ion{He}{1}, sometimes show a mixture of a broad component (BC) and a narrow component (NC), while in other cases only one of these components is present \citep[e.g.,][]{Alencar2000,Beristain2001}. The narrow component often shows strong circular polarization, likely caused by the Zeeman effect under a strong magnetic field, whereas the broader component shows weak or no polarization \citep[e.g.,][]{Symington2005,Johns-Krull2013}. This is usually interpreted as indicating stronger (weaker) magnetic fields in the NC (BC) emitting region, and thus as evidence that the NC and BC originate closer to and farther from the object surface, respectively \citep[e.g.,][]{Yang2006,Johns-Krull2013}.
In this picture, the larger width of the BC is naturally understood if it samples a spatially extended velocity field \citep[e.g.,][]{Hartmann+1994,Muzerolle+1998,Muzerolle+2001}, whereas the NC, arising from a more localized region, is governed mainly by local line broadening. Accordingly, the NC of metal lines is often thought to arise in the post-shock region \citep[e.g.,][]{Kwan+Fischer2011}.
Although ``post-shock region'' often refers mainly to the heated photosphere in the CTTS literature, we use it more broadly to include the shock-heated gas immediately downstream of the shock front.

Hydrogen lines, on the other hand, often show more complex profiles than metal lines \citep[e.g.,][]{Reipurth1996,Folha+Emerson2001,Wilson2022a}. The complex spectral profiles of hydrogen lines are well reproduced by modeling the accretion-column emission \citep[e.g.,][]{Hartmann+1994,Muzerolle+1998,Muzerolle+2001} (see also \S~\ref{sec:I_HLMechanism}).
In the context of NC and BC decomposition, the accretion-column emission corresponds to part of the BC. In addition, stellar winds can contribute to the BC in optically thicker cases \citep[e.g.,][]{Kurosawa2006a,Kurosawa2011}. Therefore, NC from post-shock gas is often minor in hydrogen lines.

Nevertheless, shock-origin hydrogen-line emission can become dominant in planetary-mass objects (see \S~\ref{sec:I_HLMechanism}).
Even beyond the planetary-mass regime, \citet{antoniucci2017} found that, among their low-mass ($<2\,\Msun$) and low mass-accretion rate ($<10^{-9}\,\Msun\,\mathrm{yr}^{-1}$) samples, about one-third shows symmetric narrow hydrogen-line profiles. These widths are broadly consistent with the shock-model predictions \citep{Aoyama+2019,Aoyama+2020}, suggesting that they may correspond to shock-origin NC as in metal lines.
If such a shock-origin component is dominant, additional non-shock emission from accretion flows and/or winds may appear as a broader component, by analogy with metal-line profiles.

\subsection{Summary and Objective of This Study}

Hydrogen lines are likely to convert accretion energy directly from the post-accretion-shock region in low-mass and/or low-accretion-rate objects, but indirectly from the accretion column, heated photosphere, and/or stellar wind in high-mass and/or high-accretion-rate objects.
The ambiguous boundary between shock-origin and non-shock-origin cases inferred by \citet{Hashimoto+Aoyama2025} suggests that these components can coexist over a broad parameter range. However, because hydrogen lines have large opacity, the non-shock component often shows a complex spectral profile, making it difficult to decompose the individual contributions. In lower-opacity cases, on the other hand, which are likely to occur in the intermediate co-existing regime, the non-shock component may instead show a broader but still relatively symmetric profile, similar to what is often seen in lower-opacity metal lines.

If this is the case, subtracting such a broad excess may enable a cleaner comparison between the residual narrow component and the shock emission model, and may therefore help clarify how the dominant hydrogen-line emission mechanism changes across parameter space. In addition, direct spectral fitting may allow the estimation of model parameters and thus a better characterization of accretion properties.

In this study, we compare the spectral profiles predicted by the shock-origin hydrogen-line emission model of \citet{Aoyama+2018} with VLT/\xshooter archival spectra of low-mass objects. We simultaneously fit seven hydrogen lines (\Hb, \Hg, H6, H8, H9, \Pb, and \Brg), excluding \Ha\ because of its likely more complex origin.
This simultaneous spectral fitting allows us to test whether a single shock-emission model can reproduce both the line profiles and the hydrogen-line flux ratios; the latter have already been shown to distinguish shock-origin hydrogen lines \citep{Betti2022,Hashimoto+Aoyama2025}, making this a more stringent test. Moreover, this spectral analysis may allow us to separate a possible BC overlapping the shock emission and thereby assess intermediate cases in which shock emission is only partially responsible for the observed hydrogen lines.

This paper is structured as follows. Section~\ref{sec:Method} describes the shock-emission model, target selection, and spectral-fitting procedures. Section~\ref{sec:FitResult} presents the fitting results and the resulting classification of hydrogen-line origins. Section~\ref{sec:SED_Estimate} summarizes the parameter estimates inferred from the spectral fitting. Section~\ref{sec:D} discusses the implications of the results. Finally, Section~\ref{sec:Conclusion} summarizes this work.

\section{Methods}
\label{sec:Method}

\subsection{Overview of the Shock-Emission Model}
\label{sec:Emodel}

We model the hydrogen-line emission from accretion-shock-heated gas following \citet{Aoyama+2018}.
In this model, the shock heating is assumed to be instantaneous, and the maximum temperature immediately downstream of the shock front is computed from the Rankine--Hugoniot relations. The post-shock cooling flow (shock-heated gas in Figure~\ref{fig:Schematic}) is then modeled with non-equilibrium chemistry, including the time-dependent hydrogen level populations and ionization state. Hydrogen level populations are calculated up to principal quantum number $n=10$, together with the ionized state, accounting for the relevant collisional and radiative processes of excitation, de-excitation, ionization, and recombination.
Radiative transfer is computed for the hydrogen lines and recombination continua on wavelength grids defined for each line and continuum. In the shock-heated gas, the dominant cooling channel is hydrogen collisional excitation, followed by hydrogen-line emission.
Because this excitation itself drives the cooling, the gas evolves thermally before the hydrogen level populations can approach local thermodynamic equilibrium (LTE).
Accordingly, this model solves the coupled time evolution of the hydrodynamics, chemistry, and radiative transfer.

This shock-emission model is characterized primarily by two input parameters: the pre-shock flow velocity and the hydrogen-nuclei number density, $(v_0,n_0)$, which later serve as the fitting parameters. Here, $n_0$ denotes the number density of hydrogen nuclei, including those bound in molecules, rather than the total particle number density. In the strong-shock limit, the pre-shock gas temperature has only a minor effect.
The elemental abundances follow the cosmic abundance of \citet{allen1973}, and hydrogen is assumed to be in molecular form when reaching the shock surface, that is, entering the shock.
Although this assumption is unlikely to be appropriate for accreting stars, the initial form of hydrogen is important mainly for weak shocks with $v_0\sim30$\,km\,s$^{-1}$ \citep{Aoyama+2020} and has only a minor effect under the stronger shock conditions relevant to stellar accretion.
He and metals are assumed to be in atomic form.

An important limitation of the model is that it becomes inappropriate, and numerically unstable, for excessively strong shocks. When the ionization fraction approaches unity, hydrogen-line cooling becomes inefficient, while the present model does not include an equally effective coolant for such nearly fully ionized regions. 
In reality, metal-line cooling can become important in these regions but is not included in this model\footnote{%
The region where metal-line cooling becomes important is precisely the region that contributes little to the hydrogen-line emission. Therefore, the omission of metal-line cooling is not expected to affect significantly the modeled hydrogen lines, unless such nearly fully ionized regions occupy a substantial fraction of the post-shock cooling zone.%
}.
However, \citet{Aoyama+2018,Aoyama+2020} showed that this regime is reached only for sufficiently strong shocks, roughly $v_0\gtrsim200\,\kms$. 
At lower velocities, even if the immediate post-shock temperature is high enough to initiate hydrogen ionization, the ionization process takes finite time, and the gas cools back to a low temperature ($\sim10^4$\,K) before becoming fully ionized. This is why hydrogen-line emission can remain significant up to substantially stronger shocks than previously expected.

\subsection{Target Selection}
\label{sec:M_Target}
The observation samples are selected using the CASPAR (Comprehensive Archive of Substellar and Planetary Accretion Rates; \citealp{Betti+2023}) database. We select objects observed with VLT/\xshooter and with free-fall velocity from infinity $\vff<350~\kms$.
For spectral analysis, we excluded the observations with a low spectral resolution ($R<4000$) as well as the low signal-to-noise ratio (peak S/N$<5$ at \Hb).
The selected samples include M-type stars with masses up to $\lesssim0.6~\Msun$ ($600~\MJ$), reaching masses likely beyond the shock-dominated regime for hydrogen-line emission. Also in the theoretical-modeling studies, accretion shock with pre-shock velocities $v_0 \gtrsim 200~\kms$ becomes less efficient at converting accretion energy into hydrogen-line emission \citep{Aoyama+2018,Aoyama+2020}. However, the $v_0$ can be significantly lower than \vff, when the starting point of free-fall accretion (i.e., the disk truncation radius) is close to the object surface.
Therefore, to avoid missing the transition from shock-dominant to flow-dominant cases, we adopt a relatively large threshold of $\vff<350~\kms$.
This distinction between $v_0$ and $\vff$ is revisited in \S~\ref{sec:v0_related}.

In total, we analyze the 254 observations of 164 objects. The targets and their basic properties are listed in Table~\ref{tab:Targets} in Appendix~\ref{sec:A_Targets}.

\subsection{Data processing}
\label{sec:M_Data}

We retrieved the VLT/\xshooter archival spectra from the ESO Science Portal.\footnote{\url{https://archive.eso.org/scienceportal/home}}
When spectra had already been flux-calibrated as described by \citet{Manara2021}, we adopted those spectra. Whether this calibration was applied to each arm is indicated in Column~6 of Table~\ref{tab:Summary}.
We then estimated the local continuum level, subtracted it from the spectra, and added the standard deviation of the continuum in quadrature to the flux uncertainty in each spectral bin (see also Appendix~\ref{sec:A_FluxMeasure}).
We also convert air wavelengths to vacuum wavelengths using \texttt{specutils}.

We consider two types of extinction separately: large-scale interstellar extinction and small-scale local extinction in the vicinity of each object, because they may have different extinction curves. For the local extinction, both extinction amount ($\AV$) and curve ($\RV$) are included as free parameters in the fitting (see \S~\ref{sec:SpectralFit_method}). For the interstellar extinction, we corrected the observed fluxes before fitting using the three-dimensional Milky Way dust map of \citet{Edenhofer+2024} via the Python module \texttt{dustmaps} \citep{dustmaps}. The extinction curve was taken from \citet{ZGR2023}. The Galactic coordinates and distances were adopted from Gaia DR3 \citep{Gaia2023}, while objects without Gaia DR3 values were supplemented using CASPAR. For nearby members of the TW Hydra Association (TWA), which lie outside the coverage of the 3D dust-extinction map, we assumed zero interstellar extinction.

\subsection{Fitting and Evaluation Procedure}
\label{sec:SpectralFit_method}

We fit the shock-model spectra to the observations within a Bayesian framework.
The fit includes six free parameters: two model parameters ($v_0$ and $n_0$), a normalization parameter $\Semit$ (emitting area; scale factor), a wavelength-scale parameter $f_\lambda$, and two extinction parameters $\AV$ and $\RV$.
As discussed in the previous section, these fitted extinction parameters describe only the local extinction, after the interstellar extinction has been corrected for.
The synthetic spectrum is written as
\begin{equation}
\label{eq:fsyn}
\Fsyn(\lami;\theta)=\left(\frac{\Semit}{4\pi d^2}\right)\,
\Fmod\left(\frac{\lami}{f_\lambda};v_0,n_0\right)\,
10^{-0.4\,A_\lambda(\lami;\AV,\RV)},
\end{equation}
where $d$ is the distance to the object, $\Fmod$ is the model flux density (\S~\ref{sec:Emodel}), and $A_\lambda(\lami;\AV,\RV)$ is the extinction curve function. For the local extinction, we adopt the extinction law of \citet{Cardelli1989}.
We assume a single $f_\lambda$ for all fitted lines,
which captures a global wavelength shift due to, for example, wavelength-calibration offsets or the systemic velocity.
Relative wavelength shifts between different lines are of the order of a few $\kms$ or less \citep{Vernet+2011,Gonneau2020}, which is substantially smaller than the instrumental spectral resolution. We therefore neglect their effects in the fitting.
Further details of the fitting procedure are given in Appendix~\ref{sec:AFit_FitDetail}.

The fit is performed for seven hydrogen lines (\Hb, \Hg, \Hd, H8, H9, \Pb, and \Brg). We exclude H7 because it overlaps with the \ion{Ca}{2} H line; it is shown in the spectral figures for reference. We also exclude \Ha\ from the fit because it is more susceptible than the other lines to additional emission components \citep[e.g.,][]{Folha+Emerson2001,White2003,Kurosawa2006a}. Further discussion of \Ha\ is given in \S~\ref{sec:D_Ha}.

\subsection{Broad-component-subtracted fit}
\label{sec:M_BCSFit}

Motivated by the possibility that some observed hydrogen-line profiles contain both a relatively narrow shock-related component and a broader excess component, we also perform a broad-component-subtracted (BCS) fit. The purpose of this test is to examine whether removing the broad excess makes the remaining narrow component more consistent with the shock-emission model.

In this procedure, we first estimate the shock-related narrow component by fitting the shock model to the line-core region. We then fit the residual broad excess with a single Gaussian profile and subtract it from the observed spectrum. Finally, we refit the shock-emission model to the broad-component-subtracted spectrum using the same fitting framework as in \S~\ref{sec:SpectralFit_method}.
Because the broad component is represented by a simple Gaussian, this procedure should be regarded as a phenomenological test rather than a physical decomposition of the line profile. The technical details are described in Appendix~\ref{sec:AFit_BCS}.

\section{Identifying Hydrogen Lines of Shock Origin}
\label{sec:FitResult}

This study assesses whether the shock-emission model can reproduce the observed spectra, thereby evaluating whether the hydrogen lines originate from the shock-heated gas or not.
Previous studies have shown that flux ratios among hydrogen lines can help distinguish shock-origin from non-shock-origin emission \citep{Betti2022,Betti2022Erratum,Aoyama+2024,Hashimoto+Aoyama2025}. In our analysis, the flux ratios are naturally taken into account because we test whether all seven fitted hydrogen lines can be reproduced self-consistently by the shock-emission model with a single parameter set. They therefore remain an important constraint on the emission origin. However, previous analyses based on the integrated fluxes of only three hydrogen lines leave some degeneracy between shock-origin and non-shock-origin emission.
By fitting the spectral profiles of seven lines, our approach provides further constraints and reduces this degeneracy, as discussed in \S~\ref{sec:CompFluxRatio}. On the other hand, unlike the previous flux-ratio studies, we do not attempt to identify the specific mechanism of emission inconsistent with the shock model; we simply classify them as ``non-shock emission.’’

\subsection{Representative Fitting Examples}

\begin{figure*}[p]
    \centering
    \includegraphics[width=\wO]{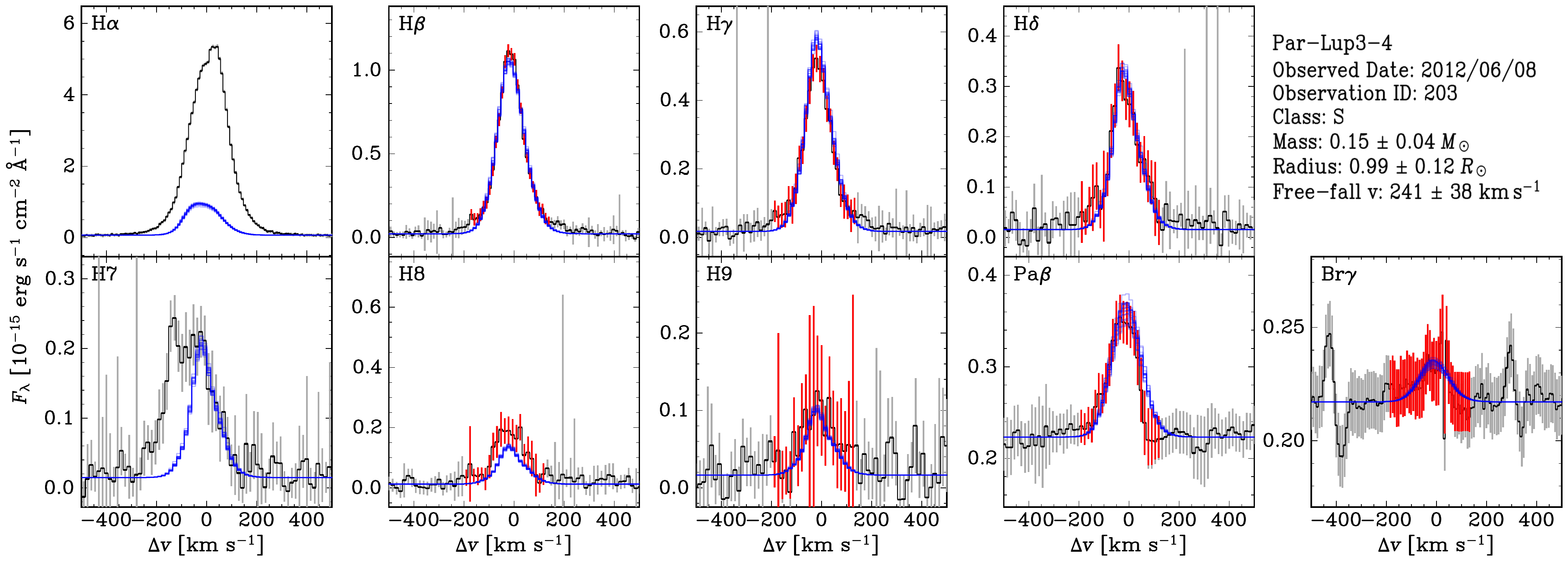}
    \includegraphics[width=\wO]{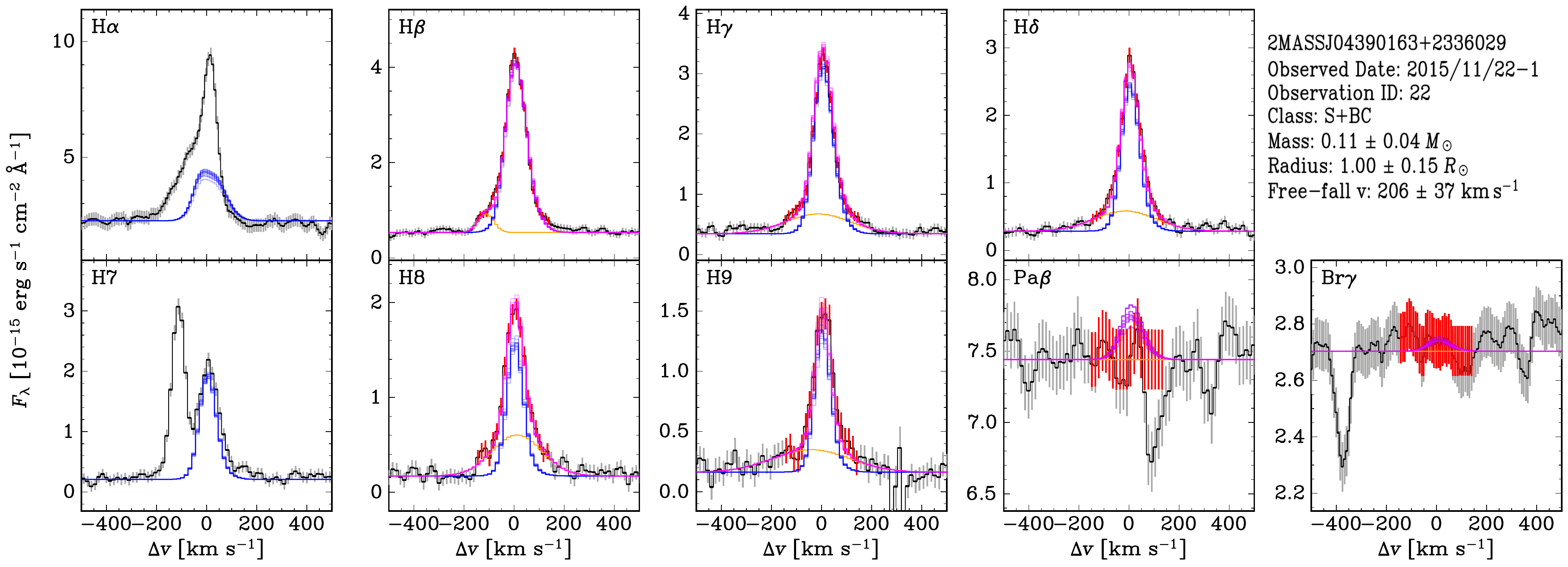}\\
    \includegraphics[width=\wO]{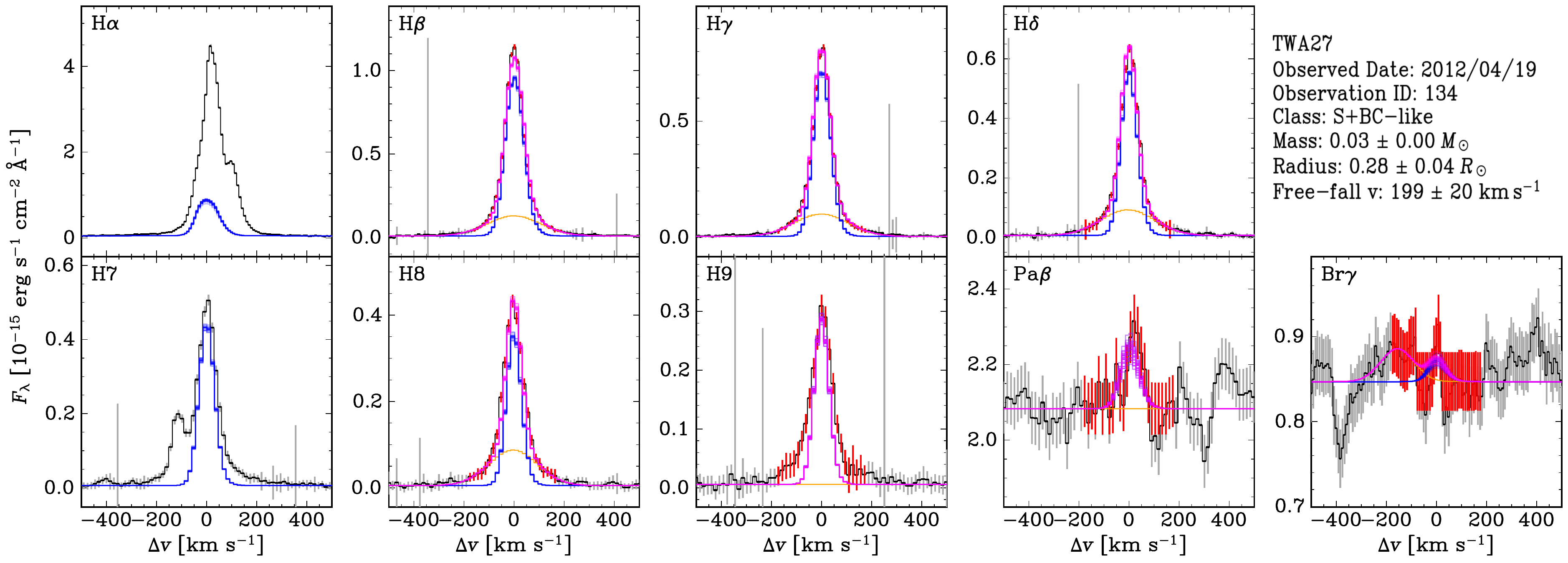}
    \caption{
    Spectral profiles of hydrogen lines for \ObsID{203} (Par-Lup3-4), and \ObsID{22} (2MASS J04390163+2336029), and \ObsID{134} (TWA27) from top to bottom. With $\Delta v$ and $F_\lambda$, we denote the velocity offset from line center and the spectral flux density, respectively.
    For \ObsID{203} (Par-Lup3-4), the top panel shows a good agreement between the model (blue line), which is randomly drawn from the MultiNest posterior, and observation (red error bars).
    For \ObsID{22} (2MASS J04390163+2336029), the middle panel shows the shock-emission model alone fails to reproduce the observation especially in the wing part, consistent with the previous flux-ratio analysis \citep{Hashimoto+Aoyama2025}. However, with considering a Gaussian Broad-Component (BC, orange line), the shock-model + BC (pink) well reproduces the observation.
    \ObsID{134} (TWA27) in the bottom panel is also largely reproduced by shock-model + BC, but the fit is statistically rejected.
    In either case, only \Ha is much fainter in the model than the observation, indicating a strong additional emission component.
    }
    \label{fig:SEDs_good}
\end{figure*}

\begin{figure*}[htb]
    \centering
    \includegraphics[width=\wO]{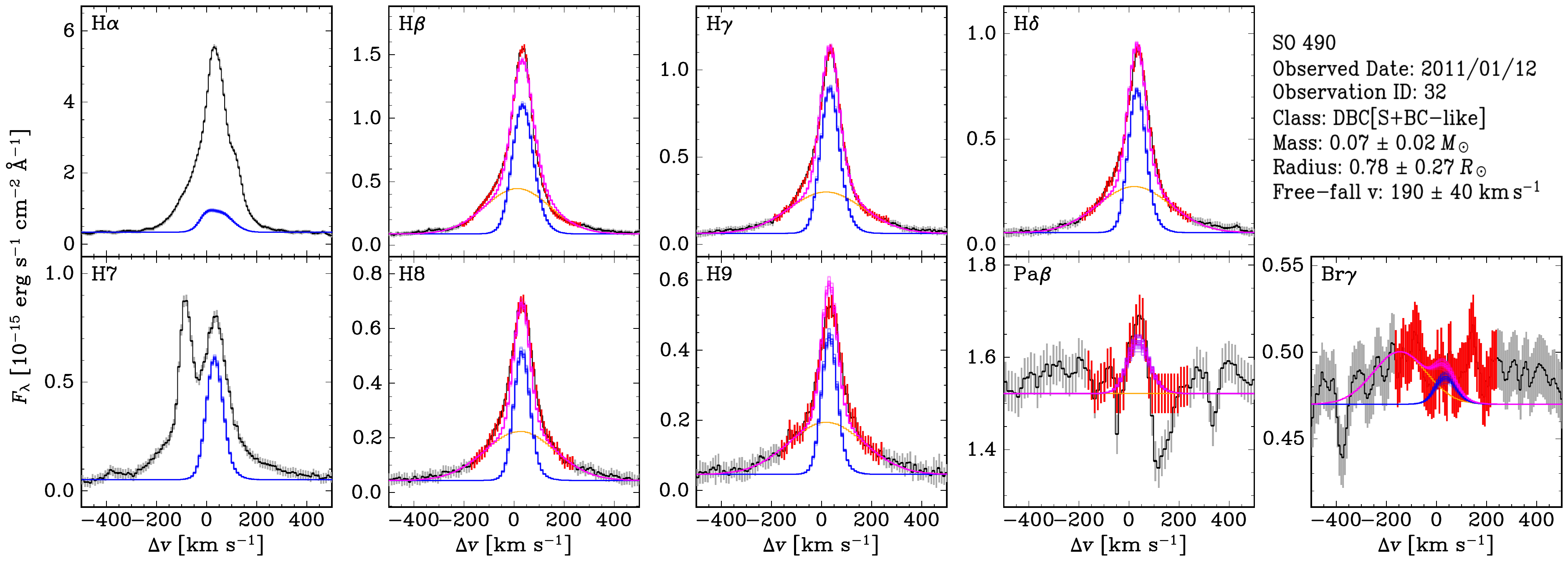}
    \includegraphics[width=\wO]{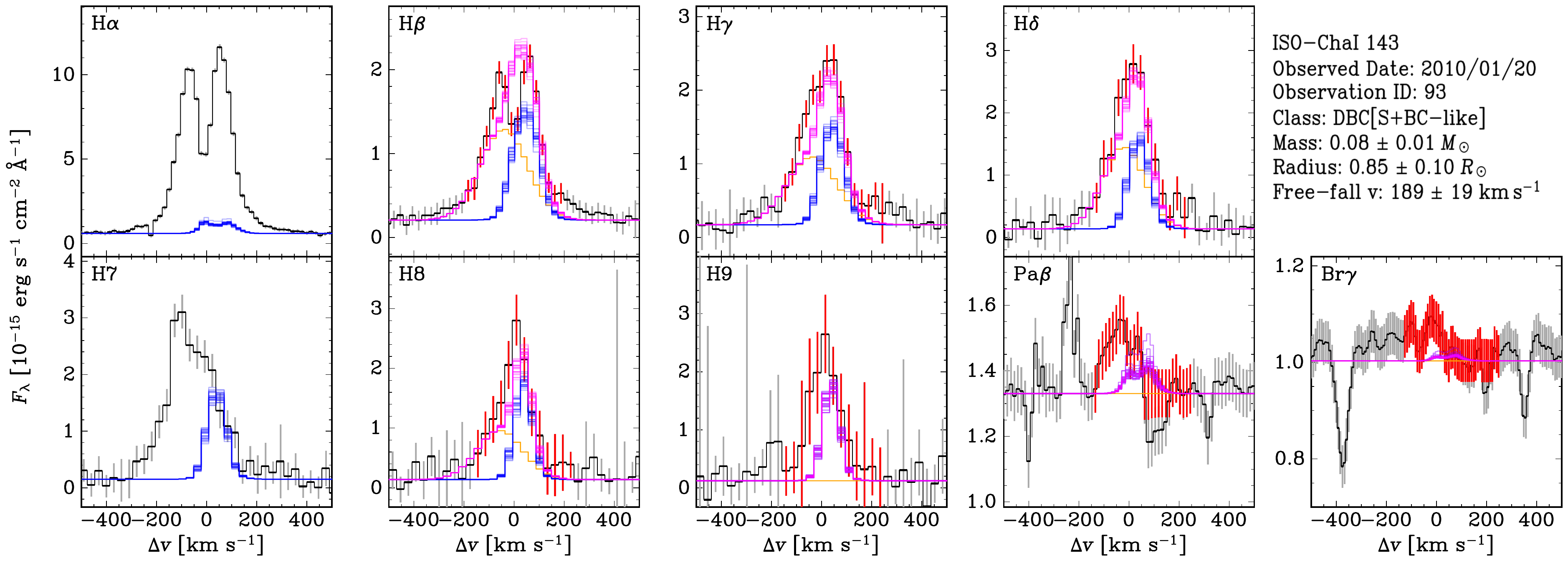}
    \caption{
    Same as Figure~\ref{fig:SEDs_good} but for \ObsID{32} (SO 490, upper row) and \ObsID{93} (ISO ChaI-143, lower row).
    Both are classified as \flgDBC[\ConcBCd], indicating that the broad component dominates while part of the spectrum may arise from shock-origin emission.
    Although the upper-row example (\ObsID{32}; SO490) appears consistent with this interpretation, this interpretation is unlikely for the lower-row example (\ObsID{93}; ISOChaI-143), because the fitting fails to reproduce the double-peaked \Hb\ profile.
    }
    \label{fig:SEDs_DBC}
\end{figure*}

The top row of Figure~\ref{fig:SEDs_good} shows a representative example of shock-origin hydrogen lines. The shock-emission model (blue) simultaneously reproduces the shapes of the individual line profiles and their relative amplitudes across the different panels, which reflect the line flux ratios.
Although nine hydrogen lines (\Ha, \Hb, \Hg, H6, H7, H8, H9, \Pb, and \Brg) are shown in the figure, only seven of them (\Hb, \Hg, H6, H8, H9, \Pb, and \Brg) are used in the fitting.
H7 is excluded from the fitting because of its overlap with the \ion{Ca}{2} H line.
Nevertheless, the shock-model prediction based on the fitted parameters naturally reproduces the observed H7 profile in these examples, providing an additional consistency check on the fitted shock solution.
In contrast, the predicted \Ha\ line is significantly fainter than the observed one in all three examples, and this tendency is also seen in most otherwise well-fitted spectra. We therefore treat \Ha\ separately from the other lines and discuss its origin in \S~\ref{sec:D_Ha}.

We also attempt to identify a shock-origin component in spectra containing an additional emission component. Assuming that this additional emission appears as a smooth broad component, we fit the broad-component-subtracted (BCS) spectra described in \S~\ref{sec:M_BCSFit}. As shown in the middle and bottom rows of Figure~\ref{fig:SEDs_good}, the shock-emission model simultaneously reproduces all the fitted hydrogen lines in the BCS spectra, in the same manner as in the pure-shock case. The upper row of Figure~\ref{fig:SEDs_DBC} further demonstrates that fitting the BCS spectra can reveal an underlying shock-origin component even when the observed profiles are dominated by the non-shock broad component.

These examples demonstrate that the combined spectral fitting of multiple hydrogen lines can identify spectra consistent with pure shock emission and can also isolate a shock-origin component when it is superposed on broader non-shock emission.
All spectral profiles other than the examples shown in Figures~\ref{fig:SEDs_good} and \ref{fig:SEDs_DBC} are available in the online material at \doi{10.5281/zenodo.19398530}.

\subsection{Assessment of Fit Quality}
\label{sec:FitGoodness}
Quantitatively, we assess the goodness of fit using the $\Pvalue$-value, i.e., the probability of obtaining a (reduced-)$\chi^2$ value at least as large as the observed one if the model is correct.
Adopting a 2-$\sigma$ criterion, we regard fits with $\Pvalue>0.05$ as acceptable. The $\Pvalue$-values for all fits are listed in Table~\ref{tab:Summary}, together with the reduced $\chi^2$ values ($\redchi$).

The three examples in Figure~\ref{fig:SEDs_good} illustrate the range of fit outcomes.
In the top-panel example (\ObsID{203}, Par-Lup3-4), the fiducial fit yields ($\redchi$, $\Pvalue$) = (0.72, 0.99), satisfying the criterion.
The middle-panel example (\ObsID{22}, 2MASS J04390163+2336029) fails the fiducial fit with ($\redchi, \Pvalue$) = (2.9, $<$0.05) but passes the BCS fit with ($\redchi, \Pvalue$) = (0.74, 0.99).
On the other hand, the bottom-panel example (\ObsID{134}, TWA27) fails both the fiducial and the BCS fits, with ($\redchi, \Pvalue$) = (8.0, $<$0.05) and (2.1, $<$0.05), respectively, even though the latter appears visually reasonable.

The discrepancy between the modeled and observed spectra in the last example (\ObsID{134}, TWA27) is at least partly due to the simplicity of our broad-component model. In BCS fit, different from NC fitted by the model prediction, BC is assumed to be a single Gaussian profile. In reality, however, the BC is likely shaped by non-thermal broadening processes and is therefore not necessarily Gaussian \citep[e.g.,][]{Hartmann+1994}.
Therefore, a statistical discrepancy from the shock-NC + Gaussian-BC model does not necessarily rule out the possibility that the observed line profile consists of a shock-origin narrow component together with a broad component that is not well described by a Gaussian. This motivates our decision to distinguish such formally rejected but visually plausible cases from more clearly discrepant spectra in the classification scheme below.

\begin{startlongtable}
\begin{deluxetable*}{rrllllcc|cc|cc}
\tablewidth{0pt} 
\tablecaption{Summary of Classification of Observed Hydrogen Lines} \label{tab:Summary}
\tablehead{ 
 \multicolumn{2}{c}{IDs}& \multicolumn{4}{c}{Observation Information} & \multicolumn{2}{c}{Fiducial}&
 \multicolumn{2}{c}{BC subtracted} %& 
 &
\\
\colhead{Obs} & \colhead{Obj} & \colhead{2MASS Name} & \colhead{Ref. Name} & 
\colhead{Date} & \colhead{Calib.} &
\colhead{\redchi} &\colhead{\Pvalue} &
\colhead{\redchi} &\colhead{\Pvalue} &
\colhead{HA25} & 
\colhead{Conclusion}
}
\decimalcolnumbers
\startdata  
  1 &  1 & SR12C & SR12C & 2016/05/03 & --- & $0.31$ & $0.99$ & $0.29$ & $0.99$ & --- & Q[S] \\
  2 &  2 & J04141458+2827580 & FN Tau & 2010/01/18 & UV$^{1}$ & $12$ & $<0.05$ & $6.4$ & $<0.05$ & --- & N \\
  3 &  3 & J04141760+2806096 & [BCG93] 1 & 2020/10/31 & UVN$^{2}$ & $180$ & $<0.05$ & $15$ & $<0.05$ & --- & N \\
  4 &  4 & J04215563+2755060 & DE Tau & 2021/11/26 & UVN$^{3}$ & $390$ & $<0.05$ & $110$ & $<0.05$ & --- & N \\
  5 &  5 & J04233919+2456141 & FT Tau & 2023/09/23 & --- & $31$ & $<0.05$ & $3.3$ & $<0.05$ & --- & DBC[S+BC-like] \\
  6 &  6 & J04245708+2711565 & IP Tau & 2024/03/01 & --- & $16$ & $<0.05$ & $1.3$ & $<0.05$ & --- & DBC[S+BC-like] \\
  7 &  7 & J04262939+2624137 & KPNO-Tau 3 & 2015/10/25-1 & --- & $0.85$ & $0.86$ & $0.77$ & $0.96$ & F & Q[S] \\
  8 &  7 & J04262939+2624137 & KPNO-Tau 3 & 2015/10/25-2 & --- & $0.8$ & $0.94$ & $0.72$ & $0.98$ & F & Q[S] \\
  9 &  7 & J04262939+2624137 & KPNO-Tau 3 & 2015/10/25-3 & --- & $0.95$ & $0.64$ & $0.75$ & $0.98$ & F & Q[S] \\
 10 &  8 & J04322210+1827426 & MHO 6 & 2015/10/25-1 & --- & $0.56$ & $0.99$ & $0.5$ & $0.99$ & --- & S \\
 11 &  8 & J04322210+1827426 & MHO 6 & 2015/10/25-2 & --- & $0.71$ & $0.97$ & $0.87$ & $0.8$ & --- & S \\
 12 &  8 & J04322210+1827426 & MHO 6 & 2015/10/25-3 & --- & $0.49$ & $0.99$ & $0.38$ & $0.99$ & --- & S \\
 13 &  9 & J04334871+1810099 & DM Tau & 2012/11/13 & UVN$^{4}$ & $19$ & $<0.05$ & $11$ & $<0.05$ & --- & N \\
 14 &  9 & J04334871+1810099 & DM Tau & 2012/11/24 & --- & $5.3$ & $<0.05$ & $2.5$ & $<0.05$ & --- & S+BC-like \\
 15 &  9 & J04334871+1810099 & DM Tau & 2021/11/28 & UVN$^{3}$ & $110$ & $<0.05$ & $66$ & $<0.05$ & --- & N \\
 16 & 10 & J04381486+2611399 & ITG 3 & 2015/10/26-1 & --- & $0.38$ & $0.99$ & $0.39$ & $0.99$ & --- & U[S] \\
 17 & 10 & J04381486+2611399 & ITG 3 & 2015/10/26-2 & --- & $0.44$ & $0.99$ & $0.44$ & $0.99$ & --- & S \\
 18 & 10 & J04381486+2611399 & ITG 3 & 2015/10/26-3 & --- & $0.41$ & $0.99$ & $0.41$ & $0.99$ & --- & Q[S] \\
 19 & 10 & J04381486+2611399 & ITG 3 & 2015/12/17-1 & --- & $0.61$ & $0.99$ & $0.53$ & $0.99$ & --- & S \\
 20 & 10 & J04381486+2611399 & ITG 3 & 2015/12/17-2 & --- & $0.32$ & $0.99$ & $0.36$ & $0.99$ & --- & S \\
 21 & 10 & J04381486+2611399 & ITG 3 & 2015/12/17-3 & --- & $0.3$ & $0.99$ & $0.28$ & $0.99$ & --- & S \\
 22 & 11 & J04390163+2336029 & --- & 2015/11/22-1 & --- & $2.9$ & $<0.05$ & $0.74$ & $0.99$ & --- & S+BC \\
 23 & 11 & J04390163+2336029 & --- & 2015/11/22-2 & --- & $3.6$ & $<0.05$ & $0.87$ & $0.85$ & --- & S+BC \\
 24 & 11 & J04390163+2336029 & --- & 2015/11/22-3 & --- & $3.2$ & $<0.05$ & $0.73$ & $0.99$ & --- & S+BC \\
 25 & 12 & J04390396+2544264 & --- & 2016/01/11-1 & --- & $1.0$ & $0.25$ & $0.49$ & $0.99$ & F & S \\
 26 & 12 & J04390396+2544264 & --- & 2016/01/11-2 & --- & $1.1$ & $0.1$ & $0.65$ & $0.99$ & F & S \\
 27 & 12 & J04390396+2544264 & --- & 2016/01/11-3 & --- & $1.3$ & $<0.05$ & $0.66$ & $0.99$ & F & S+BC \\
 28 & 13 & J05320638-0111000 & CVSO 104 & 2020/11/27 & UVN$^{5}$ & $35$ & $<0.05$ & $6.8$ & $<0.05$ & --- & N \\
 29 & 14 & J05380097-0226079 & SO 1362 & 2019/11/18 & UVN$^{6}$ & $24$ & $<0.05$ & $6.7$ & $<0.05$ & --- & N \\
 30 & 15 & J05380674-0230227 & SO 341 & 2020/02/07 & UVN$^{6}$ & $8.1$ & $<0.05$ & $3.8$ & $<0.05$ & --- & DBC[S+BC-like] \\
 31 & 16 & J05381319-0226088 & SO 397 & 2011/01/12 & UVN$^{7}$ & $12$ & $<0.05$ & $1.3$ & $<0.05$ & --- & S+BC-like \\
 32 & 17 & J05382358-0220475 & SO 490 & 2011/01/12 & UVN$^{7}$ & $17$ & $<0.05$ & $2.3$ & $<0.05$ & F & DBC[S+BC-like] \\
 33 & 18 & J05382543-0242412 & SO 500 & 2009/12/24 & UVN$^{7}$ & $20$ & $<0.05$ & $2.3$ & $<0.05$ & F & S+BC-like \\
 34 & 19 & J05382751-0235042 & SO 520 & 2020/02/13 & UVN$^{6}$ & $78$ & $<0.05$ & $48$ & $<0.05$ & --- & N \\
 35 & 20 & J05383141-0236338 & SO 562 & 2021/12/29 & UVN$^{6}$ & $190$ & $<0.05$ & $12$ & $<0.05$ & --- & N \\
 36 & 21 & J05383405-0236375 & SO 587 & 2009/12/22 & UVN$^{7}$ & $3.5$ & $<0.05$ & $1.4$ & $<0.05$ & --- & S+BC-like \\
 37 & 22 & J05383902-0245321 & SO 646 & 2009/12/24 & UVN$^{7}$ & $43$ & $<0.05$ & $14$ & $<0.05$ & --- & N \\
 38 & 23 & J05384386-0237068 & SO 694 & 2020/02/14 & UVN$^{6}$ & $31$ & $<0.05$ & $19$ & $<0.05$ & --- & N \\
 39 & 24 & J05384746-0235252 & SO 726 & 2021/12/15 & --- & $5.2$ & $<0.05$ & $0.74$ & $0.99$ & --- & DBC[S+BC] \\
 40 & 24 & J05384746-0235252 & SO 726 & 2022/01/10 & --- & $3.3$ & $<0.05$ & $1.5$ & $<0.05$ & --- & DBC[S+BC-like] \\
 41 & 24 & J05384746-0235252 & SO 726 & 2022/01/12 & UVN$^{6}$ & $14$ & $<0.05$ & $3.6$ & $<0.05$ & --- & S+BC-like \\
 42 & 25 & J05384818-0244007 & SO 739 & 2020/02/14 & UVN$^{6}$ & $3.3$ & $<0.05$ & $2.6$ & $<0.05$ & B & S+BC-like \\
 43 & 26 & J05385831-0216101 & SO 818 & 2020/02/16 & UVN$^{6}$ & $7.5$ & $<0.05$ & $1.8$ & $<0.05$ & --- & DBC[S+BC-like] \\
 44 & 27 & J05390193-0235029 & SO 848 & 2009/12/22 & UVN$^{7}$ & $30$ & $<0.05$ & $6.6$ & $<0.05$ & --- & N \\
 45 & 28 & J05390297-0241272 & SO 859 & 2021/12/22 & --- & $29$ & $<0.05$ & $9.0$ & $<0.05$ & --- & N \\
 46 & 28 & J05390297-0241272 & SO 859 & 2022/01/10 & --- & $33$ & $<0.05$ & $14$ & $<0.05$ & --- & N \\
 47 & 28 & J05390297-0241272 & SO 859 & 2022/01/12 & UVN$^{6}$ & $86$ & $<0.05$ & $240$ & $<0.05$ & --- & N \\
 48 & 29 & J05393938-0217045 & SO 1152 & 2019/10/12 & UVN$^{6}$ & $4.8$ & $<0.05$ & $2.5$ & $<0.05$ & --- & DBC[S+BC-like] \\
 49 & 30 & J05395173-0222472 & SO 1248 & 2021/12/30 & UVN$^{6}$ & $140$ & $<0.05$ & $73$ & $<0.05$ & F & N \\
 50 & 31 & J05395362-0233426 & SO 1260 & 2011/01/13 & UVN$^{7}$ & $40$ & $<0.05$ & $1.9$ & $<0.05$ & --- & DBC[S+BC-like] \\
 51 & 32 & J05400195-0221325 & SO 1327 & 2021/11/29 & --- & $16$ & $<0.05$ & $2.6$ & $<0.05$ & --- & DBC[S+BC-like] \\
 52 & 32 & J05400195-0221325 & SO 1327 & 2021/12/30 & UVN$^{6}$ & $22$ & $<0.05$ & $7.1$ & $<0.05$ & --- & N \\
 53 & 33 & J05402414-0031213 & CVSO 176 & 2020/12/02 & UVN$^{5}$ & $16$ & $<0.05$ & $6.3$ & $<0.05$ & --- & N \\
 54 & 34 & J08361073-7908184 & RECX 18 & 2010/01/20 & UV$^{1}$ & $0.87$ & $0.71$ & $0.54$ & $0.99$ & F & S \\
 55 & 35 & J08385150-7916137 & RECX 17 & 2010/01/18 & UV$^{1}$ & $0.31$ & $0.99$ & $0.31$ & $0.99$ & F & S \\
 56 & 36 & J08413030-7853064 & RECX 14 & 2010/01/19 & UV$^{1}$ & $0.59$ & $0.98$ & $0.67$ & $0.95$ & F & S \\
 57 & 37 & J08422710-7857479 & RECX 5 & 2010/01/19 & UV$^{1}$ & $2.7$ & $<0.05$ & $1.2$ & $<0.05$ & F & S+BC-like \\
 58 & 37 & J08422710-7857479 & RECX 5 & 2018/06/18 & --- & $3.2$ & $<0.05$ & $1.9$ & $<0.05$ & F & S+BC-like \\
 59 & 37 & J08422710-7857479 & RECX 5 & 2022/01/28 & UVN$^{8}$ & $2.8$ & $<0.05$ & $2.8$ & $<0.05$ & F & S+BC-like \\
 60 & 38 & J08431857-7905181 & RECX 15 & 2010/01/18 & UV$^{1}$ & $30$ & $<0.05$ & $21$ & $<0.05$ & --- & N \\
 61 & 38 & J08431857-7905181 & RECX 15 & 2018/06/17 & --- & $64$ & $<0.05$ & $5.4$ & $<0.05$ & --- & N \\
 62 & 38 & J08431857-7905181 & RECX 15 & 2022/04/09 & UVN$^{8}$ & $56$ & $<0.05$ & $3.0$ & $<0.05$ & --- & DBC[S+BC-like] \\
 63 & 39 & J08440914-7833457 & RECX 16 & 2010/01/18 & UV$^{1}$ & $14$ & $<0.05$ & $5.8$ & $<0.05$ & F & N \\
 64 & 39 & J08440914-7833457 & RECX 16 & 2021/04/27 & UVN$^{8}$ & $8.4$ & $<0.05$ & $2.7$ & $<0.05$ & S & S+BC-like \\
 65 & 39 & J08440914-7833457 & RECX 16 & 2021/05/01 & UVN$^{8}$ & $8.4$ & $<0.05$ & $2.7$ & $<0.05$ & S & S+BC-like \\
 66 & 40 & J08441637-7859080 & RECX 9 & 2010/01/19 & UV$^{1}$ & $1.5$ & $<0.05$ & $0.96$ & $0.58$ & F & S+BC \\
 67 & 40 & J08441637-7859080 & RECX 9 & 2022/01/26 & UVN$^{8}$ & $10$ & $<0.05$ & $2.0$ & $<0.05$ & F & S+BC-like \\
 68 & 41 & J10172689-5354265 & TWA22 & 2011/01/13 & UVN$^{9}$ & $13$ & $<0.05$ & $4.6$ & $<0.05$ & --- & S+BC-like \\
 69 & 42 & J10561638-7630530 & ESO-HA 553 & 2015/04/05 & UV$^{10}$ & $0.38$ & $0.99$ & $0.38$ & $0.99$ & C & S \\
 70 & 43 & J10574219-7659356 & Sz 4 & 2015/04/03 & UV$^{10}$ & $1.5$ & $<0.05$ & $1.4$ & $<0.05$ & --- & S+BC-like \\
 71 & 44 & J11004022-7619280 & Sz 8 & 2015/05/01 & UV$^{10}$ & $16$ & $<0.05$ & $3.0$ & $<0.05$ & --- & DBC[S+BC-like] \\
 72 & 45 & J11020983-3430355 & TWA28 & 2010/03/23 & UVN$^{9}$ & $1.3$ & $<0.05$ & $1.1$ & $0.13$ & --- & S+BC \\
 73 & 46 & J11023265-7729129 & CHXR 71 & 2015/04/07 & UV$^{10}$ & $0.27$ & $0.99$ & $0.27$ & $0.99$ & --- & S \\
 74 & 47 & J11025504-7721508 & Sz 10 & 2010/01/18 & UV$^{11}$ & $14$ & $<0.05$ & $5.3$ & $<0.05$ & --- & N \\
 75 & 47 & J11025504-7721508 & Sz 10 & 2021/04/29 & UVN$^{8}$ & $44$ & $<0.05$ & $4.0$ & $<0.05$ & --- & DBC[S+BC-like] \\
 76 & 48 & J11040425-7639328 & CHSM 1715 & 2016/02/16 & UV$^{10}$ & $1.3$ & $<0.05$ & $0.95$ & $0.63$ & --- & DBC[S+BC] \\
 77 & 49 & J11044258-7741571 & ISO-ChaI 52 & 2010/01/18 & UV$^{11}$ & $12$ & $<0.05$ & $4.1$ & $<0.05$ & --- & DBC[S+BC-like] \\
 78 & 50 & J11045701-7715569 & Sz 13 & 2016/01/29 & UV$^{10}$ & $1.3$ & $<0.05$ & $0.98$ & $0.51$ & --- & DBC[S+BC] \\
 79 & 51 & J11063276-7625210 & CHSM 7869 & 2015/04/03 & UV$^{10}$ & $0.22$ & $0.99$ & $0.22$ & $0.99$ & --- & S \\
 80 & 52 & J11064180-7635489 & GX Cha & 2010/01/18 & UV$^{11}$ & $8.6$ & $<0.05$ & $3.1$ & $<0.05$ & F & DBC[S+BC-like] \\
 81 & 52 & J11064180-7635489 & GX Cha & 2021/06/08 & UVN$^{8}$ & $20$ & $<0.05$ & $6.2$ & $<0.05$ & N & N \\
 82 & 53 & J11065939-7530559 & --- & 2015/04/20 & UV$^{10}$ & $2.5$ & $<0.05$ & $0.65$ & $0.99$ & B & DBC[S+BC] \\
 83 & 54 & J11071181-7625501 & CHSM 9484 & 2015/04/04 & UV$^{10}$ & $0.093$ & $0.99$ & $0.093$ & $0.99$ & --- & S \\
 84 & 55 & J11071668-7735532 & Cha HA 1 & 2010/01/18 & UV$^{11}$ & $0.59$ & $0.98$ & $0.48$ & $0.99$ & --- & S \\
 85 & 56 & J11072825-7652118 & Sz 20 & 2015/04/03 & UV$^{10}$ & $19$ & $<0.05$ & $2.1$ & $<0.05$ & --- & DBC[S+BC-like] \\
 86 & 57 & J11074245-7733593 & Cha HA 2 & 2016/02/16 & UV$^{10}$ & $1.0$ & $0.27$ & $0.79$ & $0.96$ & --- & Q[S] \\
 87 & 58 & J11074366-7739411 & Sz 21 & 2015/04/07 & --- & $4.9$ & $<0.05$ & $1.3$ & $<0.05$ & --- & DBC[S+BC-like] \\
 88 & 58 & J11074366-7739411 & Sz 21 & 2015/04/14 & UV$^{10}$ & $10$ & $<0.05$ & $2.0$ & $<0.05$ & --- & DBC[S+BC-like] \\
 89 & 59 & J11075809-7742413 & Sz 23 & 2016/01/29 & UV$^{10}$ & $32$ & $<0.05$ & $34$ & $<0.05$ & --- & N \\
 90 & 60 & J11080234-7640343 & --- & 2016/01/28 & UVN$^{12}$ & $0.56$ & $0.99$ & $0.56$ & $0.99$ & --- & S \\
 91 & 61 & J11080297-7738425 & HO Cha & 2010/01/19 & UV$^{11}$ & $16$ & $<0.05$ & $4.7$ & $<0.05$ & --- & S+BC-like \\
 92 & 62 & J11081850-7730408 & ISO-ChaI 138 & 2015/04/23 & UV$^{10}$ & $0.34$ & $0.99$ & $0.34$ & $0.99$ & C & S \\
 93 & 63 & J11082238-7730277 & ISO-ChaI 143 & 2010/01/20 & UV$^{11}$ & $4.5$ & $<0.05$ & $1.7$ & $<0.05$ & S & DBC[S+BC-like] \\
 94 & 64 & J11083952-7734166 & HQ Cha & 2010/01/18 & UV$^{11}$ & $1.4$ & $<0.05$ & $1.0$ & $0.46$ & F & Q[S+BC] \\
 95 & 65 & J11085090-7625135 & Sz 28 & 2015/05/03 & UV$^{10}$ & $1.3$ & $<0.05$ & $0.9$ & $0.78$ & S & S+BC \\
 96 & 66 & J11085367-7521359 & PU Car & 2015/04/14 & UV$^{10}$ & $55$ & $<0.05$ & $3.0$ & $<0.05$ & --- & DBC[S+BC-like] \\
 97 & 67 & J11085464-7702129 & Sz 29 & 2010/01/20 & UV$^{11}$ & $5.8$ & $<0.05$ & $3.5$ & $<0.05$ & --- & S+BC-like \\
 98 & 68 & J11085497-7632410 & HS Cha & 2015/04/06 & UV$^{10}$ & $0.54$ & $0.99$ & $0.54$ & $0.99$ & --- & Q[S] \\
 99 & 69 & J11094621-7634463 & Hn 10E & 2013/03/15 & UVN$^{10}$ & $4.7$ & $<0.05$ & $3.4$ & $<0.05$ & --- & DBC[S+BC-like] \\
100 & 70 & J11094742-7726290 & ESO-HA 567 & 2013/03/14 & UVN$^{10}$ & $8.8$ & $<0.05$ & $4.8$ & $<0.05$ & --- & U[S+BC-like] \\
101 & 71 & J11095215-7639128 & ISO-ChaI 217 & 2012/04/17 & UVN$^{7}$ & $0.28$ & $0.99$ & $0.22$ & $1.0$ & --- & Q[S] \\
102 & 71 & J11095215-7639128 & ISO-ChaI 217 & 2012/04/18 & UVN$^{7}$ & $0.52$ & $0.99$ & $0.47$ & $0.99$ & --- & Q[S] \\
103 & 72 & J11095407-7629253 & Sz 33 & 2013/03/14 & UVN$^{10}$ & $6.1$ & $<0.05$ & $1.7$ & $<0.05$ & --- & DBC[S+BC-like] \\
104 & 73 & J11095873-7737088 & Sz 35 & 2015/04/03 & UV$^{10}$ & $62$ & $<0.05$ & $9.9$ & $<0.05$ & --- & N \\
105 & 74 & J11102226-7625138 & CHSM 17173 & 2016/01/28 & UVN$^{12}$ & $0.54$ & $0.99$ & $0.54$ & $0.99$ & --- & S \\
106 & 75 & J11104141-7720480 & ISO-ChaI 252 & 2015/05/03 & UV$^{10}$ & $0.21$ & $0.99$ & $0.21$ & $0.99$ & --- & U[S] \\
107 & 76 & J11104959-7717517 & Sz 37 & 2013/03/14 & UVN$^{10}$ & $81$ & $<0.05$ & $10$ & $<0.05$ & --- & N \\
108 & 77 & J11105333-7634319 & Sz 38 & 2015/04/14 & UV$^{10}$ & $190$ & $<0.05$ & $9.6$ & $<0.05$ & --- & N \\
109 & 77 & J11105333-7634319 & Sz 38 & 2022/05/04 & UVN$^{8}$ & $130$ & $<0.05$ & $11$ & $<0.05$ & --- & N \\
110 & 78 & J11105597-7645325 & Hn 13 & 2015/04/05 & UV$^{10}$ & $0.24$ & $0.99$ & $0.19$ & $0.99$ & --- & S \\
111 & 79 & J11113965-7620152 & Sz 39 & 2010/01/19 & UVN$^{11}$ & $25$ & $<0.05$ & $6.9$ & $<0.05$ & --- & N \\
112 & 79 & J11113965-7620152 & Sz 39 & 2021/06/05 & UVN$^{8}$ & $41$ & $<0.05$ & $6.8$ & $<0.05$ & --- & N \\
113 & 80 & J11120351-7726009 & IM Cha & 2015/05/28 & UV$^{10}$ & $0.19$ & $0.99$ & $0.19$ & $0.99$ & --- & S \\
114 & 81 & J11120984-7634366 & Sz 40 & 2015/04/14 & UV$^{10}$ & $10$ & $<0.05$ & $4.6$ & $<0.05$ & F & S+BC-like \\
115 & 81 & J11120984-7634366 & Sz 40 & 2021/06/08 & UVN$^{8}$ & $9.7$ & $<0.05$ & $3.9$ & $<0.05$ & F & S+BC-like \\
116 & 82 & J11123092-7644241 & Sz 43 & 2013/03/14 & UVN$^{10}$ & $40$ & $<0.05$ & $6.1$ & $<0.05$ & --- & N \\
117 & 83 & J11132446-7629227 & Hn 18 & 2010/01/20 & UV$^{11}$ & $2.5$ & $<0.05$ & $1.6$ & $<0.05$ & --- & S+BC-like \\
118 & 84 & J11175211-7629392 & --- & 2015/04/18 & UVN$^{10}$ & $0.8$ & $0.87$ & $0.8$ & $0.87$ & --- & S \\
119 & 85 & J11183572-7935548 & --- & 2015/01/04 & --- & $2.4$ & $<0.05$ & $2.0$ & $<0.05$ & --- & S+BC-like \\
120 & 85 & J11183572-7935548 & --- & 2015/04/03 & UV$^{10}$ & $6.8$ & $<0.05$ & $3.5$ & $<0.05$ & --- & S+BC-like \\
121 & 85 & J11183572-7935548 & --- & 2018/07/14 & --- & $14$ & $<0.05$ & $11$ & $<0.05$ & --- & N \\
122 & 86 & J11241186-7630425 & --- & 2015/04/06 & UV$^{10}$ & $0.18$ & $0.99$ & $0.18$ & $0.99$ & --- & S \\
123 & 87 & J11321831-3019518 & TWA30 & 2011/04/23-1 & UVN$^{9}$ & $0.29$ & $1.0$ & $0.29$ & $1.0$ & --- & S \\
124 & 87 & J11321831-3019518 & TWA30 & 2011/04/23-2 & UVN$^{9}$ & $0.17$ & $1.0$ & $0.13$ & $1.0$ & --- & S \\
125 & 87 & J11321831-3019518 & TWA30 & 2011/07/15 & UVN$^{9}$ & $0.85$ & $0.86$ & $0.73$ & $0.98$ & --- & S \\
126 & 87 & J11321831-3019518 & TWA30 & 2012/04/15 & --- & $0.35$ & $0.99$ & $0.35$ & $0.99$ & --- & S \\
127 & 87 & J11321831-3019518 & TWA30 & 2012/04/16 & --- & $0.46$ & $0.99$ & $0.42$ & $0.99$ & --- & S \\
128 & 88 & J11324116-2652090 & TWA08B & 2011/01/13 & UVN$^{9}$ & $2.4$ & $<0.05$ & $1.7$ & $<0.05$ & --- & S+BC-like \\
129 & 88 & J11324116-2652090 & TWA08B & 2018/05/20 & --- & $9.5$ & $<0.05$ & $9.1$ & $<0.05$ & --- & N \\
130 & 89 & J11432669-7804454 & --- & 2015/01/04 & --- & $5.3$ & $<0.05$ & $2.2$ & $<0.05$ & F & DBC[S+BC-like] \\
131 & 89 & J11432669-7804454 & --- & 2015/04/20 & UV$^{10}$ & $25$ & $<0.05$ & $5.8$ & $<0.05$ & F & N \\
132 & 90 & J12071089-3230537 & TWA31 & 2014/04/18 & UVN$^{9}$ & $98$ & $<0.05$ & $8.7$ & $<0.05$ & --- & N \\
133 & 91 & J12073346-3932539 & TWA27 & 2010/03/23 & UVN$^{9}$ & $2.7$ & $<0.05$ & $2.5$ & $<0.05$ & S & S+BC-like \\
134 & 91 & J12073346-3932539 & TWA27 & 2012/04/19 & UVN$^{9}$ & $7.3$ & $<0.05$ & $2.5$ & $<0.05$ & F & S+BC-like \\
135 & 92 & J12265135-3316124 & TWA32 & 2014/04/18 & UVN$^{9}$ & $55$ & $<0.05$ & $13$ & $<0.05$ & --- & N \\
136 & 93 & J13005532-7710222 & Sz 50 & 2010/01/20 & UV$^{1}$ & $0.97$ & $0.54$ & $0.62$ & $0.99$ & --- & S \\
137 & 94 & J15354856-2958551 & --- & 2018/05/19 & --- & $27$ & $<0.05$ & $3.5$ & $<0.05$ & --- & S+BC-like \\
138 & 95 & J15392828-3446180 & Sz 66 & 2012/04/18 & UVN$^{13}$ & $8.9$ & $<0.05$ & $5.5$ & $<0.05$ & --- & N \\
139 & 95 & J15392828-3446180 & Sz 66 & 2021/05/16 & UVN$^{14}$ & $18$ & $<0.05$ & $4.2$ & $<0.05$ & --- & DBC[S+BC-like] \\
140 & 96 & J15414081-3345188 & AKC 18 & 2015/04/20 & UVN$^{15}$ & $1.9$ & $<0.05$ & $0.95$ & $0.65$ & F & DBC[S+BC] \\
141 & 97 & J15445789-3423392 & AKC 19 & 2011/04/23 & UVN$^{13}$ & $2.8$ & $<0.05$ & $1.1$ & $0.14$ & F & S+BC \\
142 & 98 & J15450887-3417333 & --- & 2015/06/15 & UVN$^{15}$ & $1.6$ & $<0.05$ & $1.3$ & $<0.05$ & --- & DBC[S+BC-like] \\
143 & 99 & J15451741-3418283 & Sz 69 & 2011/04/23 & UVN$^{13}$ & $100$ & $<0.05$ & $18$ & $<0.05$ & --- & N \\
144 & 99 & J15451741-3418283 & Sz 69 & 2021/05/02 & UVN$^{14}$ & $84$ & $<0.05$ & $26$ & $<0.05$ & --- & N \\
145 & 99 & J15451741-3418283 & Sz 69 & 2021/05/03 & UVN$^{14}$ & $540$ & $<0.05$ & $15$ & $<0.05$ & --- & N \\
146 &100 & J15451851-3421246 & --- & 2015/06/25 & UVN$^{15}$ & $4.7$ & $<0.05$ & $1.5$ & $<0.05$ & F & S+BC-like \\
147 &101 & J15475062-3528353 & Sz 72 & 2012/04/18 & UVN$^{13}$ & $450$ & $<0.05$ & $100$ & $<0.05$ & --- & N \\
148 &101 & J15475062-3528353 & Sz 72 & 2021/05/03 & UVN$^{14}$ & $340$ & $<0.05$ & $71$ & $<0.05$ & --- & N \\
149 &102 & J15493074-3549514 & Sz 76 & 2014/04/28 & --- & $12$ & $<0.05$ & $1.3$ & $<0.05$ & --- & DBC[S+BC-like] \\
150 &102 & J15493074-3549514 & Sz 76 & 2021/05/08 & UVN$^{14}$ & $5.7$ & $<0.05$ & $3.5$ & $<0.05$ & --- & S+BC-like \\
151 &102 & J15493074-3549514 & Sz 76 & 2021/08/08-1 & UVN$^{14}$ & $5.7$ & $<0.05$ & $4.0$ & $<0.05$ & --- & S+BC-like \\
152 &102 & J15493074-3549514 & Sz 76 & 2021/08/08-2 & --- & $8.5$ & $<0.05$ & $2.4$ & $<0.05$ & --- & S+BC-like \\
153 &103 & J15514032-2146103 & --- & 2018/05/21 & --- & $1.1$ & $0.19$ & $1.2$ & $0.075$ & F & S \\
154 &104 & J15530132-2114135 & --- & 2018/05/20 & --- & $12$ & $<0.05$ & $2.3$ & $<0.05$ & B & S+BC-like \\
155 &105 & J15534211-2049282 & --- & 2016/07/24 & --- & $24$ & $<0.05$ & $5.4$ & $<0.05$ & --- & N \\
156 &106 & J15580252-3736026 & Sz 84 & 2012/04/18 & UVN$^{13}$ & $35$ & $<0.05$ & $11$ & $<0.05$ & N & N \\
157 &106 & J15580252-3736026 & Sz 84 & 2022/05/11 & UVN$^{14}$ & $19$ & $<0.05$ & $4.6$ & $<0.05$ & B & DBC[S+BC-like] \\
158 &107 & J15582981-2310077 & UScoCTIO 33 & 2018/05/20 & --- & $40$ & $<0.05$ & $7.6$ & $<0.05$ & B & N \\
159 &108 & J15591135-2338002 & UScoCTIO 128 & 2014/04/25-1 & --- & $2.8$ & $<0.05$ & $1.9$ & $<0.05$ & S & S+BC-like \\
160 &108 & J15591135-2338002 & UScoCTIO 128 & 2014/04/25-2 & --- & $0.8$ & $0.92$ & $0.99$ & $0.5$ & S & S \\
161 &108 & J15591135-2338002 & UScoCTIO 128 & 2014/04/25-3 & --- & $0.53$ & $0.99$ & $0.5$ & $0.99$ & S & S \\
162 &109 & J15592523-4235066 & --- & 2015/06/27 & UVN$^{15}$ & $0.64$ & $0.98$ & $0.64$ & $0.98$ & B & S \\
163 &110 & J16000060-4221567 & --- & 2015/04/03 & UVN$^{15}$ & $5.2$ & $<0.05$ & $0.92$ & $0.74$ & --- & DBC[S+BC] \\
164 &110 & J16000060-4221567 & --- & 2021/07/21 & UVN$^{14}$ & $1.7$ & $<0.05$ & $1.0$ & $0.32$ & --- & S+BC \\
165 &111 & J16000236-4222145 & --- & 2015/07/01 & UVN$^{15}$ & $1.6$ & $<0.05$ & $1.4$ & $<0.05$ & --- & DBC[S+BC-like] \\
166 &112 & J16001844-2230114 & --- & 2016/08/09 & --- & $170$ & $<0.05$ & $28$ & $<0.05$ & --- & N \\
167 &113 & J16002612-4153553 & --- & 2015/06/28 & UVN$^{15}$ & $0.57$ & $0.99$ & $0.57$ & $0.99$ & B & S \\
168 &114 & J16024152-2138245 & --- & 2018/05/21 & --- & $21$ & $<0.05$ & $4.9$ & $<0.05$ & F & DBC[S+BC-like] \\
169 &115 & J16030548-4018254 & EX Lup & 2010/05/04 & UVN$^{7}$ & $15$ & $<0.05$ & $1.7$ & $<0.05$ & --- & DBC[S+BC-like] \\
170 &115 & J16030548-4018254 & EX Lup & 2022/03/27 & --- & $54$ & $<0.05$ & $14$ & $<0.05$ & --- & N \\
171 &115 & J16030548-4018254 & EX Lup & 2022/07/29 & --- & $18$ & $<0.05$ & $1.3$ & $<0.05$ & --- & DBC[S+BC-like] \\
172 &115 & J16030548-4018254 & EX Lup & 2023/03/24 & --- & $12$ & $<0.05$ & $1.2$ & $<0.05$ & --- & DBC[S+BC-like] \\
173 &115 & J16030548-4018254 & EX Lup & 2023/03/25 & --- & $7.5$ & $<0.05$ & $1.2$ & $<0.05$ & --- & DBC[S+BC-like] \\
174 &115 & J16030548-4018254 & EX Lup & 2023/03/26 & --- & $8.6$ & $<0.05$ & $1.4$ & $<0.05$ & --- & DBC[S+BC-like] \\
175 &115 & J16030548-4018254 & EX Lup & 2023/03/30 & --- & $11$ & $<0.05$ & $1.2$ & $<0.05$ & --- & DBC[S+BC-like] \\
176 &116 & J16053215-1933159 & --- & 2021/06/28 & --- & $2.9$ & $<0.05$ & $0.78$ & $0.98$ & S & S+BC \\
177 &117 & J16060391-2056443 & --- & 2014/04/08-1 & --- & $2.1$ & $<0.05$ & $1.8$ & $<0.05$ & F & S+BC-like \\
178 &117 & J16060391-2056443 & --- & 2014/04/08-2 & --- & $2.5$ & $<0.05$ & $1.5$ & $<0.05$ & F & S+BC-like \\
179 &117 & J16060391-2056443 & --- & 2014/04/08-3 & --- & $2.1$ & $<0.05$ & $1.6$ & $<0.05$ & F & S+BC-like \\
180 &118 & J16063539-2516510 & --- & 2018/05/20 & UVN$^{16}$ & $4.9$ & $<0.05$ & $4.1$ & $<0.05$ & C & S+BC-like \\
181 &119 & J16070384-3911113 & --- & 2016/06/03 & UVN$^{7}$ & $0.29$ & $0.99$ & $0.31$ & $0.99$ & --- & S \\
182 &120 & J16071159-3903475 & Sz 91 & 2012/04/18 & UVN$^{13}$ & $66$ & $<0.05$ & $12$ & $<0.05$ & --- & N \\
183 &121 & J16073773-3921388 & Lup 713s & 2010/04/06 & UVN$^{13}$ & $54$ & $<0.05$ & $12$ & $<0.05$ & N & N \\
184 &121 & J16073773-3921388 & Lup 713s & 2012/04/18 & UVN$^{7}$ & $22$ & $<0.05$ & $5.8$ & $<0.05$ & F & N \\
185 &121 & J16073773-3921388 & Lup 713s & 2012/04/19 & UVN$^{7}$ & $21$ & $<0.05$ & $4.4$ & $<0.05$ & N & DBC[S+BC-like] \\
186 &122 & J16075230-3858059 & Sz 95 & 2015/07/12 & UVN$^{15}$ & $2.1$ & $<0.05$ & $1.5$ & $<0.05$ & --- & S+BC-like \\
187 &123 & J16080017-3902595 & Lup 604s & 2010/04/06 & UVN$^{13}$ & $0.82$ & $0.9$ & $0.73$ & $0.98$ & F & S \\
188 &124 & J16081263-3908334 & Sz 96 & 2015/07/03 & UVN$^{15}$ & $2.5$ & $<0.05$ & $1.6$ & $<0.05$ & --- & DBC[S+BC-like] \\
189 &125 & J16081497-3857145 & --- & 2015/04/20 & UVN$^{15}$ & $4.2$ & $<0.05$ & $0.94$ & $0.71$ & B & DBC[S+BC] \\
190 &126 & J16082180-3904214 & Sz 97 & 2011/04/23 & UVN$^{13}$ & $23$ & $<0.05$ & $3.0$ & $<0.05$ & --- & DBC[S+BC-like] \\
191 &126 & J16082180-3904214 & Sz 97 & 2022/05/13 & UVN$^{14}$ & $100$ & $<0.05$ & $22$ & $<0.05$ & --- & N \\
192 &127 & J16082576-3906011 & Sz 100 & 2011/04/23 & --- & $11$ & $<0.05$ & $2.7$ & $<0.05$ & F & S+BC-like \\
193 &127 & J16082576-3906011 & Sz 100 & 2022/06/24 & UVN$^{14}$ & $42$ & $<0.05$ & $7.3$ & $<0.05$ & F & N \\
194 &128 & J16082751-1949047 & --- & 2018/05/21 & UVN$^{16}$ & $6.7$ & $<0.05$ & $3.2$ & $<0.05$ & F & S+BC-like \\
195 &129 & J160828.1-391310 & Lup 607 & 2015/05/23 & UVN$^{15}$ & $0.32$ & $0.99$ & $0.32$ & $0.99$ & --- & S \\
196 &130 & J16083026-3906111 & Sz 103 & 2011/04/23 & UVN$^{13}$ & $20$ & $<0.05$ & $8.1$ & $<0.05$ & --- & N \\
197 &130 & J16083026-3906111 & Sz 103 & 2022/05/01 & UVN$^{14}$ & $28$ & $<0.05$ & $5.5$ & $<0.05$ & --- & N \\
198 &131 & J16083081-3905488 & Sz 104 & 2010/04/06 & UVN$^{13}$ & $4.2$ & $<0.05$ & $1.6$ & $<0.05$ & F & S+BC-like \\
199 &131 & J16083081-3905488 & Sz 104 & 2022/06/24 & UVN$^{14}$ & $17$ & $<0.05$ & $3.1$ & $<0.05$ & S & DBC[S+BC-like] \\
200 &132 & J16083733-3923109 & Lup 706 & 2010/04/06 & UVN$^{13}$ & $1.2$ & $<0.05$ & $0.51$ & $0.99$ & F & Q[S+BC] \\
201 &133 & J16084940-3905393 & Par-Lup3-3 & 2010/04/06 & UVN$^{13}$ & $0.33$ & $0.99$ & $0.31$ & $0.99$ & --- & S \\
202 &134 & J16085143-3905304 & Par-Lup3-4 & 2010/04/07 & UVN$^{13}$ & $2.7$ & $<0.05$ & $0.91$ & $0.79$ & --- & DBC[S+BC] \\
203 &134 & J16085143-3905304 & Par-Lup3-4 & 2012/06/08 & UVN$^{2}$ & $0.67$ & $0.99$ & $0.67$ & $0.99$ & --- & S \\
204 &134 & J16085143-3905304 & Par-Lup3-4 & 2013/05/09-1 & UVN$^{2}$ & $0.68$ & $0.99$ & $0.55$ & $0.99$ & --- & S \\
205 &134 & J16085143-3905304 & Par-Lup3-4 & 2013/05/09-2 & UVN$^{2}$ & $0.74$ & $0.99$ & $0.44$ & $0.99$ & --- & S \\
206 &134 & J16085143-3905304 & Par-Lup3-4 & 2013/05/10-1 & UVN$^{2}$ & $1.3$ & $<0.05$ & $1.3$ & $<0.05$ & --- & S+BC-like \\
207 &134 & J16085143-3905304 & Par-Lup3-4 & 2013/05/10-2 & UVN$^{2}$ & $0.79$ & $0.97$ & $0.43$ & $0.99$ & --- & S \\
208 &134 & J16085143-3905304 & Par-Lup3-4 & 2013/05/11-1 & UVN$^{2}$ & $5.7$ & $<0.05$ & $1.1$ & $<0.05$ & --- & DBC[S+BC-like] \\
209 &134 & J16085143-3905304 & Par-Lup3-4 & 2013/05/11-2 & UVN$^{2}$ & $4.0$ & $<0.05$ & $0.79$ & $0.98$ & --- & DBC[S+BC] \\
210 &134 & J16085143-3905304 & Par-Lup3-4 & 2013/05/20-1 & UVN$^{2}$ & $2.4$ & $<0.05$ & $1.2$ & $<0.05$ & --- & DBC[S+BC-like] \\
211 &134 & J16085143-3905304 & Par-Lup3-4 & 2013/05/20-2 & UVN$^{2}$ & $9.6$ & $<0.05$ & $1.7$ & $<0.05$ & --- & DBC[S+BC-like] \\
212 &134 & J16085143-3905304 & Par-Lup3-4 & 2014/06/24 & --- & $0.58$ & $0.99$ & $0.45$ & $0.99$ & --- & S \\
213 &135 & J16085157-3903177 & Sz 110 & 2011/04/23 & UVN$^{13}$ & $27$ & $<0.05$ & $1.2$ & $<0.05$ & --- & DBC[S+BC-like] \\
214 &135 & J16085157-3903177 & Sz 110 & 2022/05/24 & UVN$^{14}$ & $43$ & $<0.05$ & $5.8$ & $<0.05$ & --- & N \\
215 &136 & J16085324-3914401 & --- & 2015/07/12 & UVN$^{15}$ & $0.24$ & $0.99$ & $0.24$ & $0.99$ & --- & S \\
216 &137 & J16085529-3848481 & --- & 2015/07/12 & UVN$^{15}$ & $1.8$ & $<0.05$ & $1.0$ & $0.28$ & F & S+BC \\
217 &138 & J16085553-3902339 & Sz 112 & 2012/04/18 & UVN$^{13}$ & $3.3$ & $<0.05$ & $3.3$ & $<0.05$ & F & S+BC-like \\
218 &138 & J16085553-3902339 & Sz 112 & 2022/07/24 & --- & $10$ & $<0.05$ & $3.7$ & $<0.05$ & F & S+BC-like \\
219 &139 & J16085953-3856275 & --- & 2011/04/23 & UVN$^{13}$ & $1.0$ & $0.44$ & $0.73$ & $0.98$ & F & S \\
220 &140 & J16090002-1908368 & --- & 2018/05/21 & UVN$^{16}$ & $5.0$ & $<0.05$ & $3.4$ & $<0.05$ & F & S+BC-like \\
221 &141 & J16090141-3925119 & --- & 2011/04/23 & UVN$^{13}$ & $1.1$ & $0.2$ & $0.67$ & $0.98$ & --- & Q[S] \\
222 &142 & J16090185-3905124 & Sz 114 & 2011/04/23 & UVN$^{7}$ & $13$ & $<0.05$ & $2.7$ & $<0.05$ & --- & S+BC-like \\
223 &142 & J16090185-3905124 & Sz 114 & 2013/03/15 & UVN$^{10}$ & $4.4$ & $<0.05$ & $2.9$ & $<0.05$ & --- & DBC[S+BC-like] \\
224 &142 & J16090185-3905124 & Sz 114 & 2022/05/26 & UVN$^{14}$ & $25$ & $<0.05$ & $9.7$ & $<0.05$ & --- & N \\
225 &143 & J16090621-3908518 & Sz 115 & 2012/04/18 & UVN$^{13}$ & $3.1$ & $<0.05$ & $0.95$ & $0.66$ & --- & S+BC \\
226 &143 & J16090621-3908518 & Sz 115 & 2022/06/24 & UVN$^{14}$ & $0.95$ & $0.6$ & $0.56$ & $0.99$ & --- & S \\
227 &144 & J16092697-3836269 & --- & 2015/07/13 & UVN$^{15}$ & $54$ & $<0.05$ & $12$ & $<0.05$ & F & N \\
228 &145 & J16094434-3913301 & Sz 117 & 2015/07/13 & UVN$^{15}$ & $12$ & $<0.05$ & $0.98$ & $0.53$ & --- & DBC[S+BC] \\
229 &145 & J16094434-3913301 & Sz 117 & 2022/05/30 & UVN$^{14}$ & $22$ & $<0.05$ & $2.5$ & $<0.05$ & --- & DBC[S+BC-like] \\
230 &146 & J16095361-1754474 & --- & 2018/05/20 & UVN$^{16}$ & $7.8$ & $<0.05$ & $3.4$ & $<0.05$ & F & S+BC-like \\
231 &147 & J16095628-3859518 & Lup 818s & 2011/04/23 & UVN$^{13}$ & $2.1$ & $<0.05$ & $1.0$ & $0.32$ & --- & S+BC \\
232 &148 & J16100133-3906449 & --- & 2015/07/13 & UVN$^{15}$ & $0.29$ & $0.99$ & $0.29$ & $0.99$ & --- & S \\
233 &149 & J16101857-3836125 & --- & 2015/08/18 & UVN$^{15}$ & $1.5$ & $<0.05$ & $0.49$ & $0.99$ & B & S+BC \\
234 &150 & J16101984-3836065 & --- & 2015/08/08 & UVN$^{15}$ & $0.28$ & $0.99$ & $0.22$ & $0.99$ & --- & S \\
235 &151 & J16102955-3922144 & --- & 2014/05/06 & --- & $10$ & $<0.05$ & $1.7$ & $<0.05$ & --- & DBC[S+BC-like] \\
236 &151 & J16102955-3922144 & --- & 2015/08/13 & UVN$^{15}$ & $1.4$ & $<0.05$ & $0.99$ & $0.5$ & --- & S+BC \\
237 &152 & J16104636-1840598 & --- & 2018/05/20 & UVN$^{16}$ & $3.3$ & $<0.05$ & $2.0$ & $<0.05$ & F & S+BC-like \\
238 &153 & J16115979-3823383 & SST-Lup3-1 & 2010/04/06 & UVN$^{13}$ & $8.8$ & $<0.05$ & $1.6$ & $<0.05$ & B & S+BC-like \\
239 &154 & J16134410-3736462 & --- & 2015/06/26 & UVN$^{7}$ & $17$ & $<0.05$ & $1.5$ & $<0.05$ & B & DBC[S+BC-like] \\
240 &154 & J16134410-3736462 & --- & 2022/05/03 & UVN$^{14}$ & $18$ & $<0.05$ & $1.5$ & $<0.05$ & S & DBC[S+BC-like] \\
241 &155 & J16135434-2320342 & --- & 2018/05/20 & UVN$^{16}$ & $56$ & $<0.05$ & $4.6$ & $<0.05$ & F & DBC[S+BC-like] \\
242 &156 & J16181904-2028479 & --- & 2018/05/20 & UVN$^{16}$ & $2.0$ & $<0.05$ & $1.6$ & $<0.05$ & F & S+BC-like \\
243 &157 & J16262152-2426009 & GY 5 & 2010/08/21 & UVN$^{17}$ & $0.58$ & $0.99$ & $0.56$ & $0.99$ & --- & U[S] \\
244 &158 & J16262189-2444397 & GY 3 & 2010/06/02 & UVN$^{17}$ & $1.8$ & $<0.05$ & $1.1$ & $0.094$ & F & DBC[S+BC] \\
245 &159 & J16270659-2441488 & GY 204 & 2010/07/05 & --- & $1.4$ & $<0.05$ & $1.0$ & $0.22$ & --- & Q[S+BC] \\
246 &159 & J16270659-2441488 & GY 204 & 2010/07/27 & --- & $3.1$ & $<0.05$ & $1.2$ & $<0.05$ & --- & S+BC-like \\
247 &159 & J16270659-2441488 & GY 204 & 2010/09/22 & --- & $2.2$ & $<0.05$ & $2.2$ & $<0.05$ & --- & S+BC-like \\
248 &159 & J16270659-2441488 & GY 204 & 2012/06/08 & --- & $0.72$ & $0.99$ & $0.56$ & $0.99$ & --- & Q[S] \\
249 &160 & J16273526-2438334 & GY 295 & 2010/06/08 & UVN$^{17}$ & $0.17$ & $0.99$ & $0.16$ & $0.99$ & --- & U[S] \\
250 &161 & J16273863-2438391 & GY 310 & 2010/06/09 & UVN$^{17}$ & $1.2$ & $0.085$ & $1.1$ & $0.17$ & --- & Q[S] \\
251 &162 & J16281650-2436579 & ISO-Oph 196 & 2010/07/29-1 & --- & $5.6$ & $<0.05$ & $2.8$ & $<0.05$ & --- & DBC[S+BC-like] \\
252 &162 & J16281650-2436579 & ISO-Oph 196 & 2010/07/29-2 & --- & $2.1$ & $<0.05$ & $1.9$ & $<0.05$ & --- & DBC[S+BC-like] \\
253 &163 & J16284527-2428190 & SR 13 & 2010/05/04 & --- & $23$ & $<0.05$ & $1.5$ & $<0.05$ & --- & DBC[S+BC-like] \\
254 &164 & J19013357-3700304 & LS-RCrA 1 & 2024/05/23 & --- & $0.54$ & $0.99$ & $0.55$ & $0.99$ & --- & S 
%%%%%%%%%%%%%%%%%%%%%%%%%%%%%%%%%%%
\enddata
\tablecomments{
Object-name abbreviations \revise{and corresponding SIMBAD identifiers \citep{SIMBAD}}:
SO = [HHM2007]; 
KPNO-Tau = [BLH2002]\,KPNO-Tau; 
%AssCha = Ass\,Cha\,T2-; 
GY = [GY92]; 
AKC = [AKC2006]; 
Lup = [LEM2005]\,Lup.
\\
Publications related to the data set; 
$^{1}$\citet{Manara2016}, \citet{Rugel2018}; %Herczeg
$^{2}$\citet{Whelan2014}, \citet{Pinilla2021}; %Manara others. %% Par-Lup/Whelan? CIDA1/Pinilla2021?
$^{3}$\citet{PENELLOPE_Tau};
$^{4}$\citet{manara2014}; %  TD14
$^{5}$\citet{PENELLOPE_Ori};
$^{6}$\citet{Mauco2023};%Manara sigOri,
$^{7}$\citet{Alcala2011},\citet{rigliaco2012}, \citet{Whelan2014};
%Alcala [085.C-0764, 087.C-0244, 089.C-0143, 095.C-0134, 097.C-0349] ;
$^{8}$\citet{PENELLOPE_Cha};
$^{9}$\citet{Venuti+2019};
$^{10}$\citet{Rugel2018}; %Manara Cha,
$^{11}$ESO programme 094.C-0805; %(2016 observations) Herczeg Cha
$^{12}$\citet{Manara2017};
$^{13}$\citet{alcala2014};
$^{14}$\citet{PENELLOPE_Lups};
$^{15}$\citet{Alcala+2017};
$^{15}$\citet{Whelan2014}; %Manara others. %% Par-Lup/Whelan? CIDA1/Pinilla2021?
$^{16}$\citet{Manara2020}, \citet{Empey2026}; %% USco
$^{17}$\citet{Manara2015}. %% Oph
}
\vspace{-0.2cm} 
\end{deluxetable*}
\end{startlongtable}

\subsection{Classification of the Fitted Spectra}
\label{sec:Class}

Motivated by the above inspection, we classify the fitted observations into four categories:
\begin{itemize}
    \item a pure shock-emission spectrum (\ConcShock),
    \item a shock-origin narrow component plus a broad Gaussian component (\ConcBC),
    \item a spectrum plausibly consisting of a shock-origin narrow component plus a broad component, but not robustly confirmed as such (\ConcBCd),
    \item none of the above, meaning that the shock component is not detectable and is likely negligible (\ConcNot).
\end{itemize}
We do not count the ``marginal'' category, \ConcBCd, as a successful fit. Instead, we retain it as a separate reference category, because these cases may still contain useful information while remaining distinct from the more clearly discrepant fits.
The number of observations in each class is summarized in Table~\ref{tab:ClassCount}.

\begin{deluxetable}{l ccc  cc}
\label{tab:ClassCount}
\tablecaption{Counts of the Hydrogen-line Classifications}
\tablehead{& Normal & DBC & U & Total & Fraction}
\startdata
S          & 62 & --- & 4  & 66  & 26\%\\
S+BC       & 18 & 11  & 0  & 29  & 11\%\\
S+BC-like  & 50 & 51  & 1  & 102 & 40\%\\
N          & 57 & --- & 0  & 57  & 22\%
\enddata
\end{deluxetable}

In addition to these four main categories, we assign three auxiliary flags to cases that require special caution.
First, some objects that pass the BCS fitting are dominated by the broad component.
Even if part of the emission can be attributed to shock emission, such cases should be separated when assessing which emission mechanism is dominant. Moreover, when an arbitrary Gaussian component becomes comparable to or stronger than the shock-emission component, the fit can sometimes misinterpret a complex line profile as a shock+BC decomposition, as illustrated in the lower row of Figure~\ref{fig:SEDs_DBC}. In other cases, however, it can plausibly capture a dominant BC plus a relatively minor shock component, as shown in the upper row.
We therefore identify such cases separately as ``dominant broad-component’’ (\flgDBC).

Second, some observations pass the statistical criterion only because of large observational uncertainties.
Specifically, when the flux S/N of \Hg is below 3, we label the fit as ``uncertain'' (\flgU).
Because \Hg is often the second most constraining line after \Hb, a poorly measured \Hg typically means that the fit is effectively constrained only by \Hb, which is insufficient for robustly identifying the emission mechanism.

Finally, when using the fitted parameters to infer accretion properties discussed later in \S~\ref{sec:SED_Estimate}, some estimates may fall in unlikely regimes. We flag such cases as ``Questionable estimated quantity’’ (\flgQ). The detailed criteria for this flag are described in Appendix~\ref{sec:AFit_suspicious}.
Also, if an observation satisfies multiple flag criteria, Table~\ref{tab:Summary} lists only the highest-priority flag, in the order \flgDBC, \flgU, and \flgQ.

\subsection{Parameter Dependence of Emission Mechanism}
\label{sec:ClassDependency}

Hydrogen lines in classical T Tauri stars are generally thought to arise mainly from gas surrounding the accretion shock, because the strong shock heats the post-shock region beyond the temperature range favorable for hydrogen-line emission \citep[see the review by][]{Hartmann+2016}.
In contrast, for low-mass objects such as planets, the weaker shock produces post-shock gas at $\sim 10^{4\text{--}5}$\,K, where hydrogen-line emission is efficient \citep{Aoyama+2018}. Our classifications indeed include cases consistent with this shock-origin.
These considerations suggest that the key controlling parameter is the shock strength, which can be approximately captured by the mass and free-fall velocity.

\begin{figure}
    \centering
    \includegraphics[width=\linewidth]{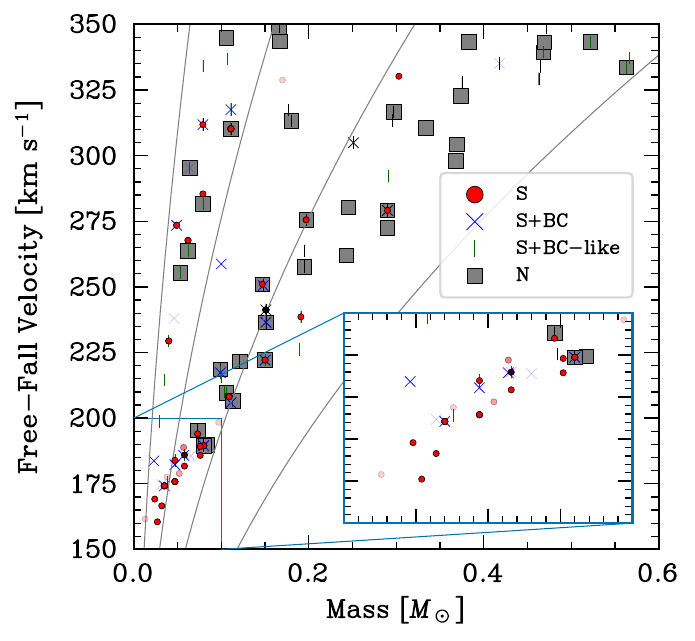}
    \caption{
    Emission-mechanism classifications in the mass--free-fall-velocity plane.
    Red circles, blue crosses, \revise{green} bars, and gray squares denote the four main classes defined in \S~\ref{sec:Class}: \ConcShock, \ConcBC, \ConcBCd, and \ConcNot, respectively.
    Objects with the auxiliary flags for large observational uncertainty (\flgU) or broad-component dominance (\flgDBC) are shown in black, and the unlikely inferred property (\flgQ) are in transparent color,
    while retaining the symbol of their main class.
    Thin gray lines indicate isocontours of object radius: 0.2~$\Rsun$, 0.5~$\Rsun$, 1~$\Rsun$, and 2~$\Rsun$, from left to right.
    Note that points are plotted in the legend order from top to bottom, with the upper entries drawn in the foreground and shown with smaller symbols, so that they are not obscured by the larger symbols plotted behind them. Also, multiple observations of the same object are plotted separately.
    }
    \label{fig:ClassScat}
\end{figure}

Figure~\ref{fig:ClassScat} therefore shows our emission-mechanism classifications in the mass--free-fall-velocity plane. The four main classes are distinguished by symbol and color, while cases with the auxiliary flags \flgU, \flgDBC, or \flgQ are shown in black with their symbols retained.

As expected, pure shock-emission cases (\ConcShock) are concentrated toward the lower-left region of the diagram, whereas non-shock cases (\ConcNot) become more frequent toward higher mass and/or higher free-fall velocity. However, no clear boundary separating these regimes is evident, consistent with \citet{Hashimoto+Aoyama2025}.

\begin{figure*}
    \centering
    \includegraphics[width=0.45\linewidth]{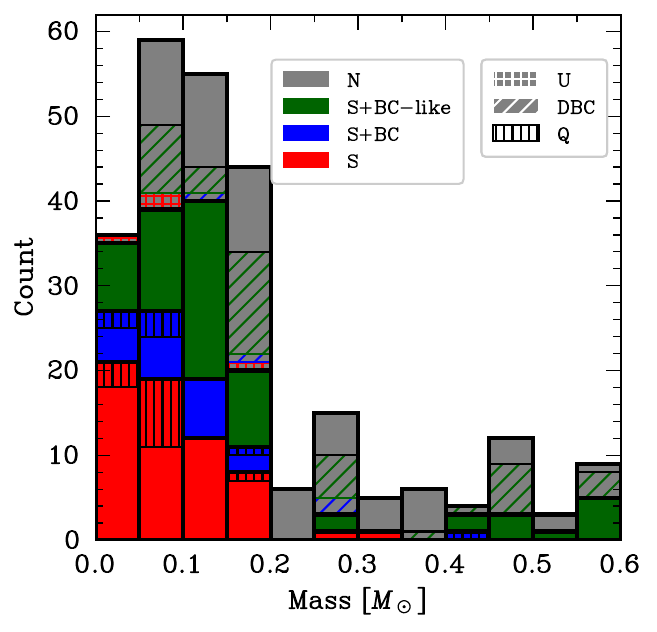}
    \includegraphics[width=0.45\linewidth]{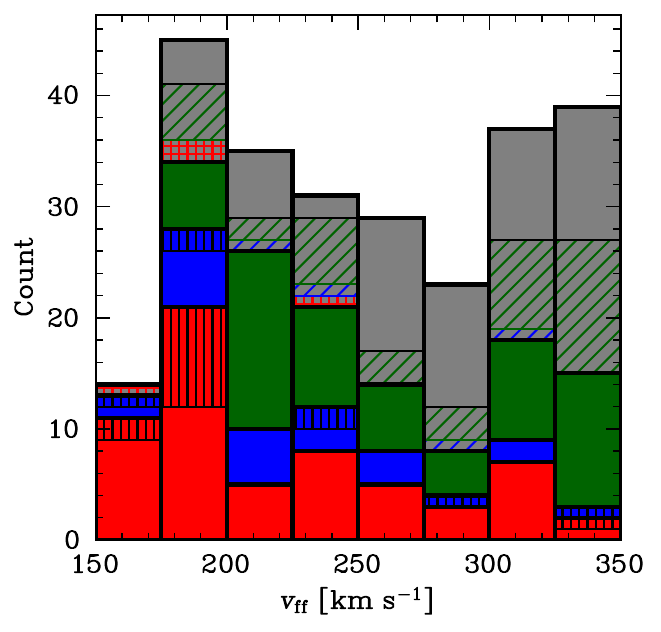}\\
    \includegraphics[width=0.45\linewidth]{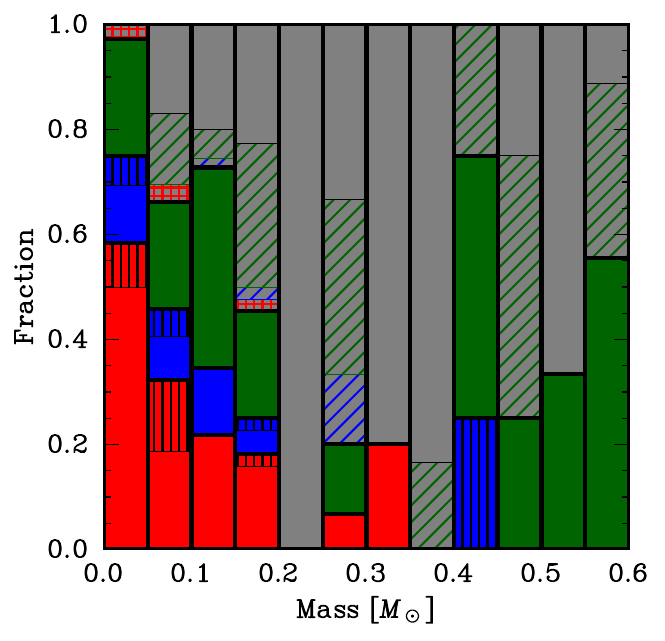}
    \includegraphics[width=0.45\linewidth]{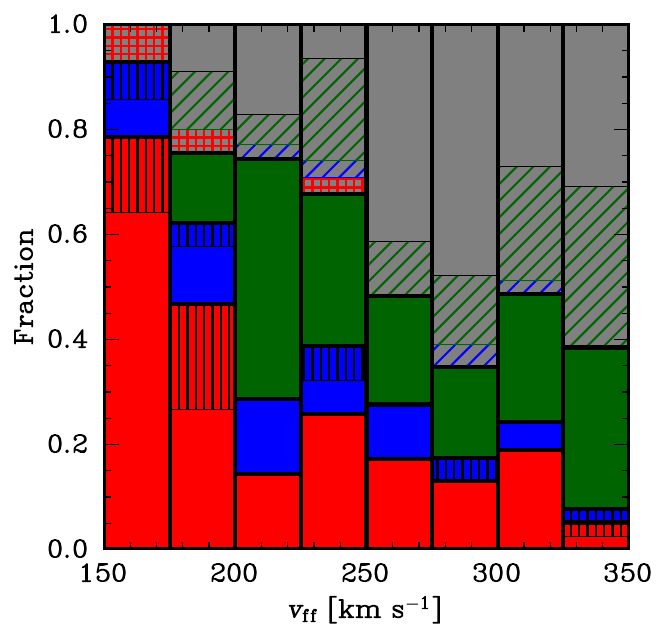}
    \caption{
    Histogram of count (upper) and fraction (lower) of classification of emission mechanism relative to mass (left) and free-fall velocity (right).
    Grid and diagonal hatches denote the \flgU and \flgDBC, and hatch color shows the original category (see \S~\ref{sec:Class}). In addition, a vertical hatching indicates that we use model-parameter extrapolation to get a likely fit.
    }
    \label{fig:ClassHist}
\end{figure*}

For a more quantitative comparison, Figure~\ref{fig:ClassHist} shows the counts and fractions of the classifications as functions of mass and free-fall velocity.
At $M<0.05\,\Msun$, no \ConcNot\ cases appear, and the shock emission is dominant in all cases (i.e., no \flgDBC\ case is present).
The fraction of shock-dominant cases (\ConcShock + \ConcBC) decreases with increasing mass and is nearly zero above $0.2\,\Msun$, with only three exceptions, all of which either have low S/N or show a substantial BC contribution\footnote{
The two \ConcShock cases above $0.2~\Msun$ (\ObsID{73}, CHXR 71, $0.29\,\Msun$; and \ObsID{215}, 2MASS J16085324-3914401, $0.30\,\Msun$) shows low S/N ratio of $<2$ in H8 and H9.
Also, the one \ConcBC case above $0.2~\Msun$ (\ObsID{41}, SO 726, $0.42\,\Msun$) shows relatively large fraction of BC ($35\,\%$).}.
Thus, at least in our samples, the shock-emission dominates below $0.05\Msun$, whereas non-shock emission dominates above $0.2\Msun$.

Even when either the shock or non-shock emission is dominant, the contribution from the other component is often not negligible. At $<0.05\Msun$, around $40\%$ of the spectra includes a BC contribution (\ConcBC), while at $>0.2\Msun$, around half of the spectra may include a shock-origin component even though the non-shock component is the dominant (\flgDBC). This confirms that the shock and non-shock components frequently coexist, which may explain why \citet{Hashimoto+Aoyama2025} did not find a clear boundary between the two regimes.

Previously, \citet{Aoyama+2021} suggested that the shock-emission may become dominant when $v_0 \lesssim 200\kms$, where the shock emission is efficient (see also \S~\ref{sec:Emodel}). Indeed, shock emission is dominant in more than half of the cases with $\vff<200\kms$. However, Figures~\ref{fig:ClassScat} and \ref{fig:ClassHist} show that the classifications (\ConcShock, \ConcBC, \ConcBCd, and \ConcNot) are not simply ordered by $\vff$, and shock-dominant cases remain even at $\vff > 300\kms$. This indicates that the dominant emission mechanism cannot be distinguished by $\vff$ alone. This is partly because the accretion-flow velocity $v_0$ can differ from the free-fall velocity from infinity, $\vff$, as discussed later in \S~\ref{sec:v0_vff}. More fundamentally, however, the transition cannot be understood solely in terms of the shock-emission efficiency. The BC emission efficiency must also play an important role, as indicated by the presence of BC contributions even in low-mass or low-$\vff$ cases. We discuss this point further in \S~\ref{sec:D_BC}.

\section{Parameter Estimate}
\label{sec:SED_Estimate}

In this section, we estimate the physical properties of the accretion flow from the parameters constrained by our spectral fitting. The fit directly constrains six model parameters: the pre-shock accretion-flow velocity ($v_0$) and hydrogen-nuclei number density ($n_0$), the emitting area ($\Semit$), the visual extinction ($\AV$), the extinction-curve slope ($\RV$), and the wavelength-shift parameter ($f_\lambda$).
From these directly fitted parameters, we further derive several physical quantities, including the accretion luminosity ($\Lacc$), the accretion-shock filling factor ($\ff$), disk truncation radius ($\Rt$), and object dipole magnetic flux ($\Bs$). Please see also Appendix~\ref{sec:AFit} for detailed procedures.

The resulting representative parameter values and acceptable ranges for all accepted modes are listed in Tables~\ref{tab:FidFit} and~\ref{tab:BCSFEstimate} for the fiducial and BCS fits, respectively. Also, the combined properties are listed in Table~\ref{tab:combined}.

\subsection{Accretion Luminosity}
\label{sec:Lacc}

\begin{figure}
\def\wtem{\wO}
    \centering
    \includegraphics[width=\wtem]{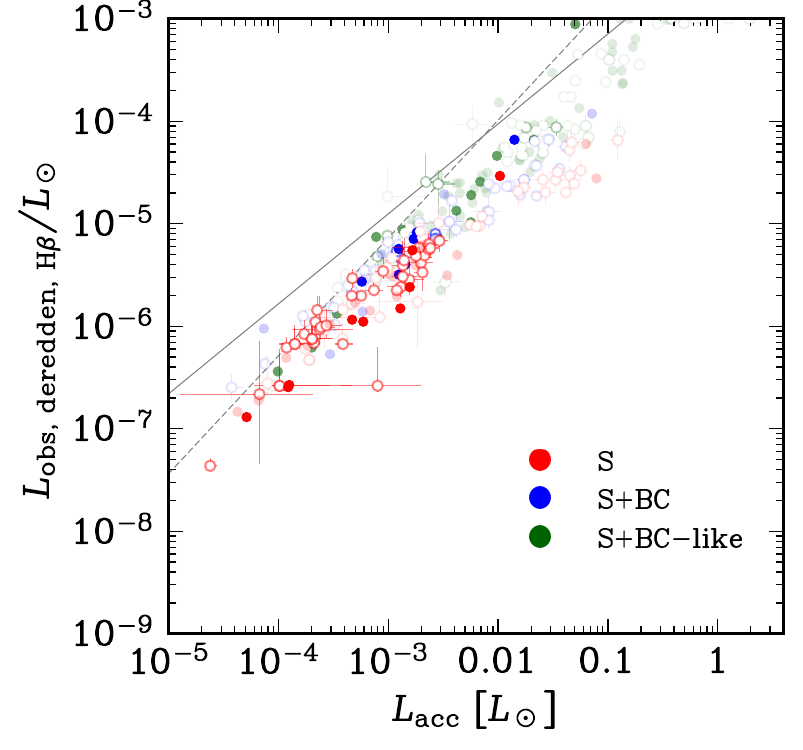}
    \includegraphics[width=\wtem]{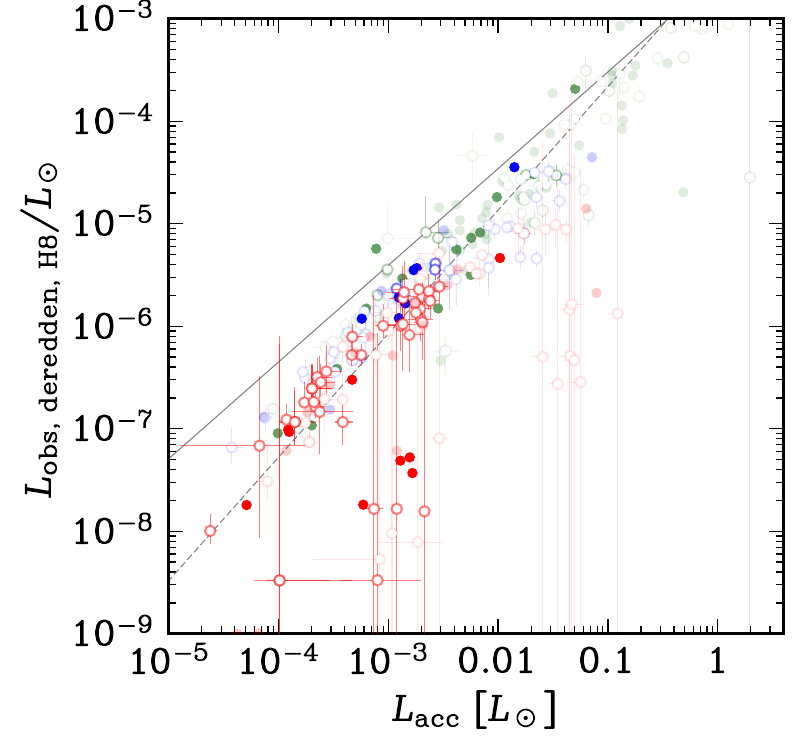}
    \caption{
    Relationship between accretion luminosity ($\Lacc$) and the luminosities of \Hb\ (upper) and H8 (lower), all normalized by the solar luminosity $\Lsun$.
    The subscript ``dereddened'' denotes line luminosities additionally dereddened using the estimated local extinction.
    Red, blue, and \revise{green} symbols correspond to the estimates in \revise{\ConcShock (pure shock emission), \ConcBC (shock emission + broad component), and \ConcBCd (potentially explained by \ConcBC under a relaxed statistical criterion)}, respectively.
    Transparent symbols denote less-confident estimates (see Appendix~\ref{sec:AFit_suspicious}). 
    Although they are individually less reliable, they may still provide useful information on the population trend, particularly near the fitting-grid boundary at large $\Lacc$ in these plots. 
    When the mean value falls outside the plotting range, the corresponding symbol is placed at the edge of the plot.
    Open symbols indicate that another accepted mode exists for the same observation.
    The solid and dashed lines show the empirical $\Lacc$--line-luminosity relations derived from observation \citep[][solid line]{Alcala+2017} and from shock-emission model predictions \citep[][dashed line]{Aoyama+2021}.
    }
    \label{fig:Lacc}
\end{figure}

Accretion luminosity is one of the fundamental properties of accretion shocks. We derive it from the fitted parameters as:
\begin{align}
\label{eq:Lacc1}
    \Lacc &= \frac{1}{2} \rho_0 v_0^3 \Semit,\\
          &= \frac{1}{2} \Mdot v_0^2,
          \label{eq:LaccMdot}
\end{align}
where
\begin{align}
    \rho_0 &= \mu_\mathrm{H} n_0,\\
    \Mdot  &= \rho_0 v_0 \Semit, \label{eq:Mdot}
\end{align}
where \revise{$\mu_\mathrm{H}=m_\mathrm{H}/X = 2.27 \times 10^{-24}$\,g is the mean mass per hydrogen nucleus (not per gas particle), $X=0.74$ is the hydrogen mass fraction, $m_\mathrm{H}$ is the mass of a hydrogen nucleus,} and $\Semit$ is the emitting area.
The derived $\Lacc$ is shown in Figure~\ref{fig:Lacc}, relative to the observed luminosities of \Hb (upper) and H8 (lower).
These line luminosities are dereddened using the extinction inferred from the fitting and include both the shock-origin and non-shock-origin components.
As expected, the samples broadly follow the fitted relationship of \citet{Aoyama+2021} (dashed line), since both are based on the same shock-emission model. The slight upward offset in $L_{\mathrm{H8}}$ (lower panel) largely corresponds to the broad-component excess (BC) contribution. 
On the other hand, $L_{\Hb}$ often shows a downward offset, particularly at large $\Lacc$ when the transparent points are also considered.
This trend is likely caused by saturation of \Hb\ in stronger and denser shocks \citep[see Appendix~C in][]{Aoyama+2021}.

\begin{figure}
    \centering
    \includegraphics[width=\wO]{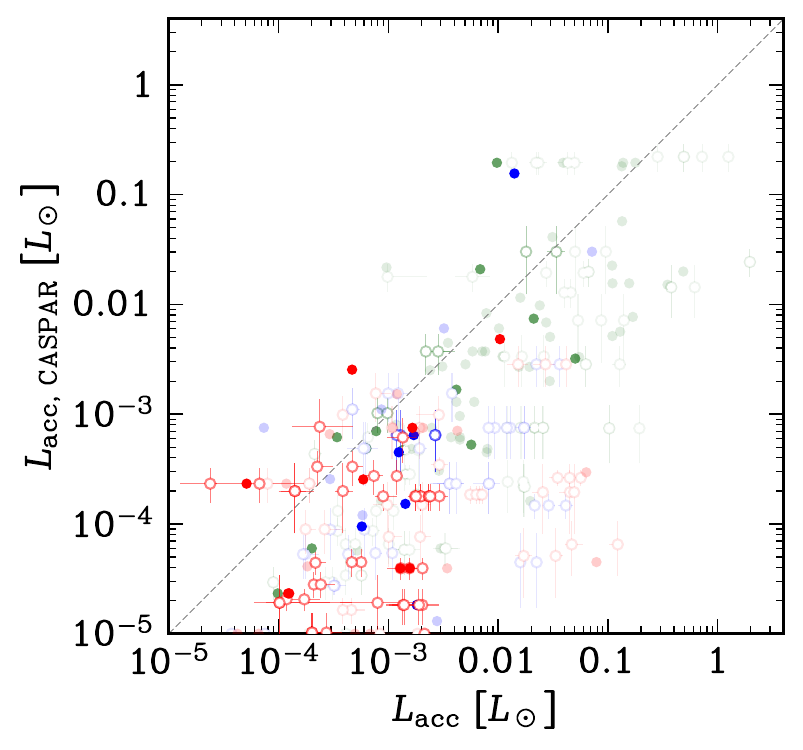}
    \includegraphics[width=\wO]{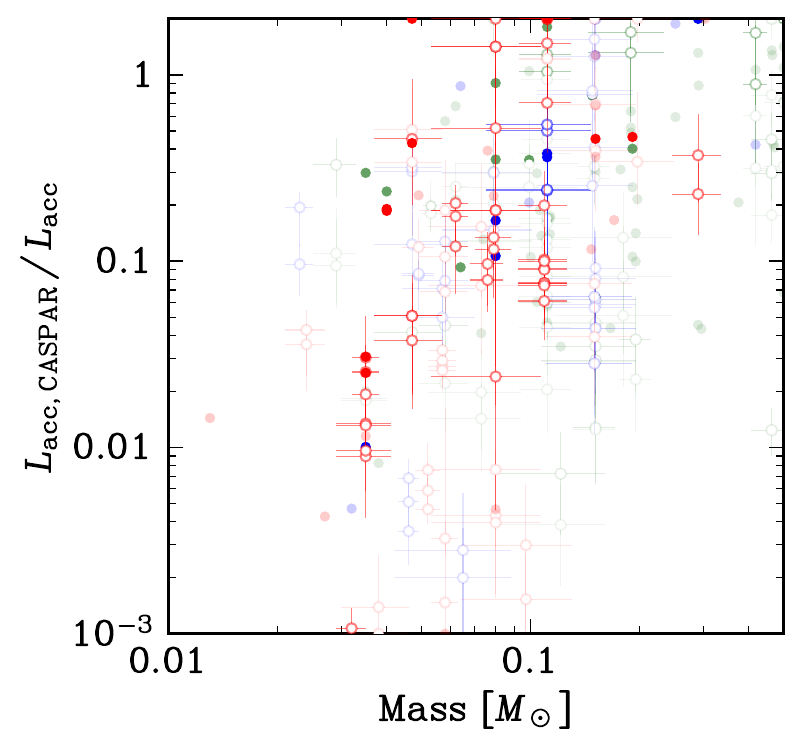}
    \caption{
Comparison between the accretion luminosities derived in this work ($\Lacc$) and those adopted from the literature ($\LaccCA$).
\textit{Upper panel}: Direct comparison of the two estimates, with the shock-model-derived values on the horizontal axis and the literature-based values on the vertical axis.
\textit{Lower panel}: Ratio of $\LaccCA$ to $\Lacc$ as a function of object mass.
The literature-based values are systematically smaller, reflecting the fact that the conventional definition of $\Lacc$ is tied primarily to the UV continuum excess and does not fully account for other emission channels.
}
    \label{fig:Lacc_comp}
\end{figure}

Next, we compare in Figure~\ref{fig:Lacc_comp} the derived $\Lacc$ with literature values of accretion luminosity taken from the CASPAR database \citep{Betti+2023}\footnote{Computed as $\LaccCA = GM\Mdot R^{-1}$, because the CASPAR tables are more complete for $\Mdot$ than for $\Lacc$}.
Hereafter, the literature accretion luminosity is denoted as $\LaccCA$, and $\Lacc$ denotes the value derived in this work.
The upper panel of Figure~\ref{fig:Lacc_comp} shows that $\LaccCA$ is systematically lower than $\Lacc$.
The lower panel shows that the discrepancy is small \revise{(typically factor of a few)} for relatively massive objects but becomes larger toward lower masses. For some very-low-mass objects, the difference reaches orders of magnitude.

The discrepancy between $\LaccCA$ and $\Lacc$ is likely caused by the definition of $\LaccCA$.
In the conventional framework, \revise{$\LaccCA$} is effectively defined by the UV continuum excess \citep{Alcala+2017}, assuming that this excess is the dominant radiative channel of accretion energy \citep{Hartmann+2016}. When other radiative channels become non-negligible, the conventional estimate $\LaccCA$ can therefore underestimate the total accretion luminosity. As shown in \S~\ref{sec:Class}, hydrogen lines in low-mass objects ($<0.2\Msun$) can arise directly from accretion-shock-heated gas, a component not included in the conventional UV-excess framework. This is a likely cause of the small $\LaccCA/\Lacc$ ratios.
Indeed, the shock-emission model predicts that, under some conditions of moderately strong shocks, most of the accretion energy can be radiated through hydrogen lines \citep{Aoyama+2018,Aoyama+2021}, accounting for extremely small $\LaccCA/\Lacc$ ratios, such as $\sim10^{-3}$ in Figure~\ref{fig:Lacc_comp}. This interpretation is also consistent with the large, nearly unity $\LaccCA/\Lacc$ ratios at $M>0.2\Msun$, where shock-origin hydrogen lines become minor (see \S~\ref{sec:ClassDependency}).

The limitation of the conventional definition of accretion luminosity $\LaccCA$ is also supported by observational evidence. Some observational studies of \revise{accreting low-mass objects} have found that the UV continuum and hydrogen-line luminosities can be comparable \citep{Herczeg+2008,Herczeg+2009,Zhou2014}.
This indicates that hydrogen-line emission can carry a non-negligible fraction of the accretion energy, and already points to a limitation of conventional $\LaccCA$ estimates based solely on the UV continuum excess.

The remaining question is whether the missing luminosity can be large enough to explain the most extreme cases in Figure~\ref{fig:Lacc_comp}, where $\LaccCA/\Lacc$ reaches $\sim10^{-3}$.
Such small ratios require a missing luminosity orders of magnitude larger than the UV continuum excess. The shock-emission model predicts hydrogen Lyman-$\alpha$ as a major hidden component. If so, this component can easily be missed observationally, because hydrogen Lyman-$\alpha$ is difficult to observe directly owing to strong interstellar absorption.
Nevertheless, Lyman-$\alpha$ luminosities can be indirectly reconstructed from Lyman-$\alpha$-pumped H$_2$ fluorescence. Such reconstructions suggest that Lyman-$\alpha$ can indeed dominate the far-UV excess \citep{Herczeg+2004,Schindhelm+2012,France+2014}.
Thus, conventional $\LaccCA$ estimates may indeed underestimate the accretion luminosity by orders of magnitude in very-low-mass objects.

\subsection{Mass Accretion Rate}
\label{sec:Mdot}

\begin{figure}
\def\wtem{\wO}
    \centering
    \includegraphics[width=\wtem]{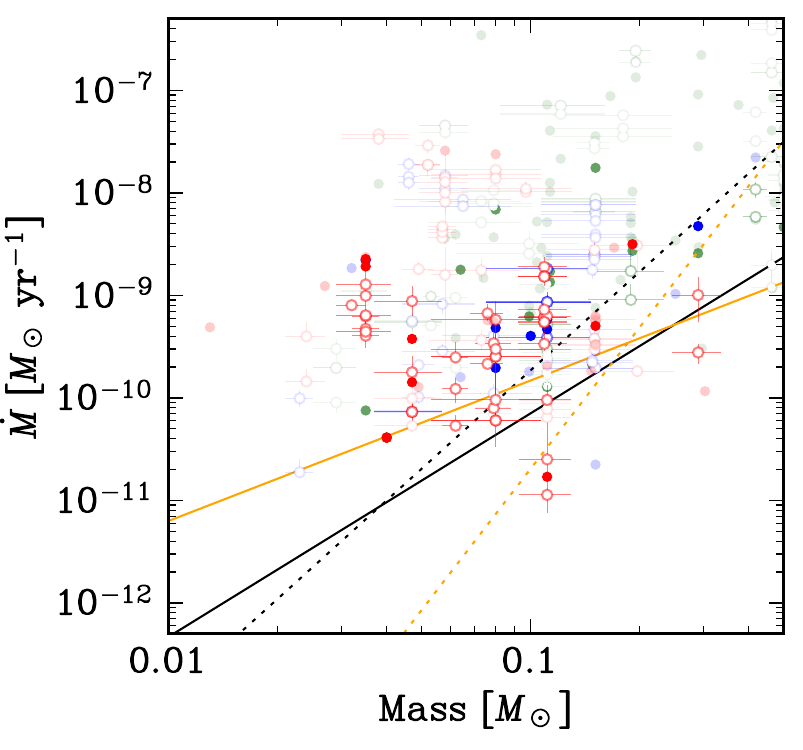}
    \includegraphics[width=\wtem]{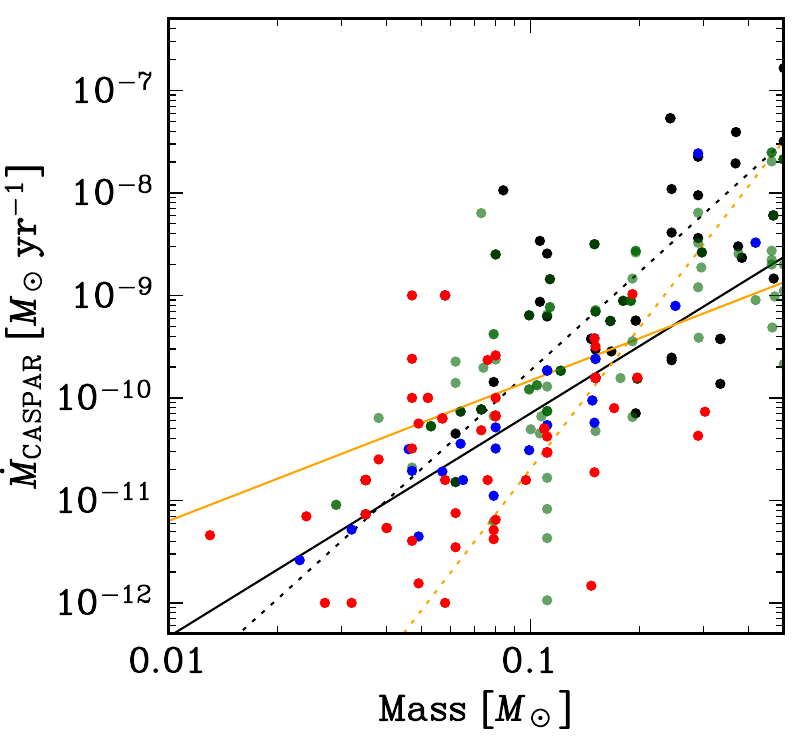}
    \caption{
    Mass accretion rate derived in this work (upper) and in the CASPAR dataset \citep{Betti+2023} (lower), shown as functions of object mass. 
    The black points in the lower panel correspond to the \ConcNot case, which could not be fitted with the shock model and is therefore not shown in the upper panel. Otherwise, the colors and symbols are the same as in Figure~\ref{fig:Lacc}.
    The black lines show the broken power-law fits reported by \citet{Betti+2023}, written as $\log_{10}\Mdot = a\log_{10}M_\ast + b$:
    $(a,b)=(2.18,-7.97)$ for $M_\ast>0.075,\Msun$ (solid) and
    $(a,b)=(3.19,-6.54)$ for $M_\ast<0.075,\Msun$ (dotted).
    The orange lines show the corresponding fits from \citet{Alcala+2017}:
    $(a,b)=(1.37,-8.46)$ for $M_\ast>0.2,\Msun$ (solid) and
    $(a,b)=(4.58,-6.11)$ for $M_\ast<0.2,\Msun$ (dotted).
    }
    \label{fig:Mdot}
\end{figure}

Since the accretion luminosity is proportional to the mass accretion rate (see Eq.\ref{eq:LaccMdot}), the systematic difference in $\Lacc$ discussed in the previous section directly propagates into $\Mdot$. The upper panel of Figure~\ref{fig:Mdot} shows the $\Mdot$ values derived in this work using Eq.~\ref{eq:Mdot}, while the lower panel shows the values taken from the CASPAR database \citep{Betti+2023}. As expected from the difference in $\Lacc$, our derived mass accretion rates are systematically higher than the CASPAR values.
Because the difference in $\Mdot$ is driven only by the difference in $\Lacc$, the ratio between the two $\Mdot$ estimates is identical to the $\LaccCA/\Lacc$ ratio shown in Figure~\ref{fig:Lacc_comp}, and is therefore not shown separately.

Mass accretion rates are known to increase with stellar mass, with an approximately quadratic dependence \citep{manara2023}. Several studies have suggested that this relation may become steeper at the low-mass end, with a break around $\Ms \sim0.2$–$0.3,\Msun$ \citep{Manara2017,Alcala+2017}.
This mass range is intriguingly similar\footnote{\revise{A lower trend-break mass, near the hydrogen burning limit ($0.075\,\Msun$), has also been proposed \citep{Betti+2023}.}} to the regime where shock-origin hydrogen-line emission becomes dominant in our classification (see \S~\ref{sec:ClassDependency}).
The lower panel of Figure~\ref{fig:Mdot} reproduces this behavior: below $M_\sim0.2,\Msun$, many objects lie significantly below the high-mass branch of the literature relation (solid lines). In contrast, this low-mass offset largely disappears when our revised $\Mdot$ values are used, as shown in the upper panel. This is because our derived $\Mdot$ values are larger than the literature values, and the difference increases toward lower masses, as shown in Figure~\ref{fig:Lacc_comp}.
We therefore suggest that the previously reported steepening or break of the $\Mdot$–$\Ms$ relation at low masses may be at least partly caused by a systematic underestimation of $\Mdot$ in objects where the accretion energy is directly radiated as shock-origin hydrogen lines (see the discussion in \S~\ref{sec:Lacc}).

This interpretation may also affect the discussion of the apparent age dependence of the $\Mdot$–$\Ms$ relation. Previous studies reported that the low-mass steepening is more prominent in older star-forming regions \citep{Alcala+2017,Manara2017}, and \citet{Betti+2023} found that older samples tend to show steeper relations. This trend has been interpreted as evidence that disks around low-mass stars evolve faster, causing their accretion rates to decline more rapidly \citep{manara2023}. If, however, the low-mass steepening is partly caused by a systematic difference in the hydrogen-line emission mechanism, the apparent age dependence may instead reflect an evolution in the accretion shock and the emission mechanism of the objects.
Further studies are needed to distinguish between these possibilities.

\subsection{Accretion-Shock Filling Factor}
\label{sec:ff}

\begin{figure}
    \centering
    \def\wtem{\wO}
    \includegraphics[width=\wtem]{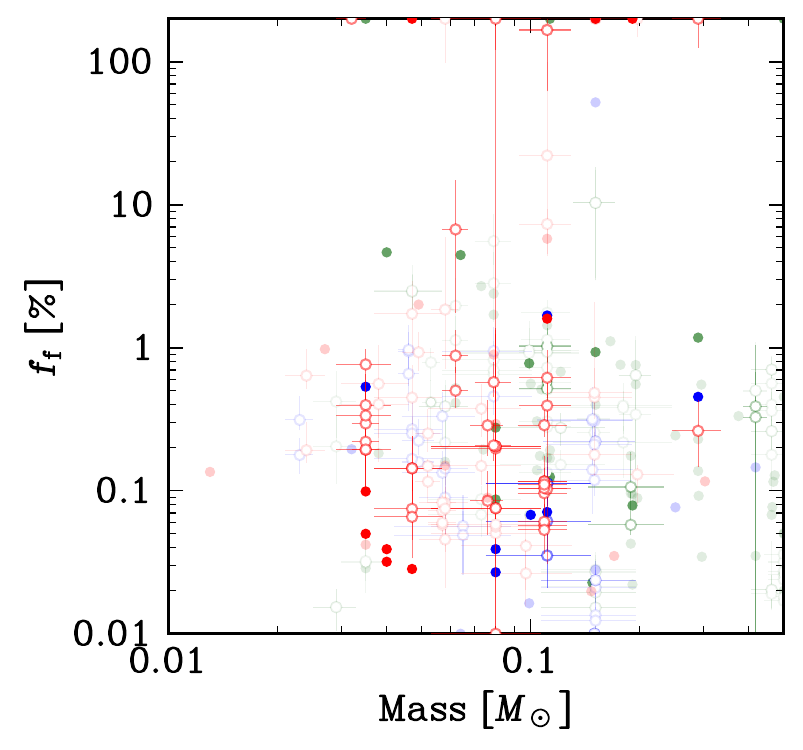}
    \includegraphics[width=\wtem]{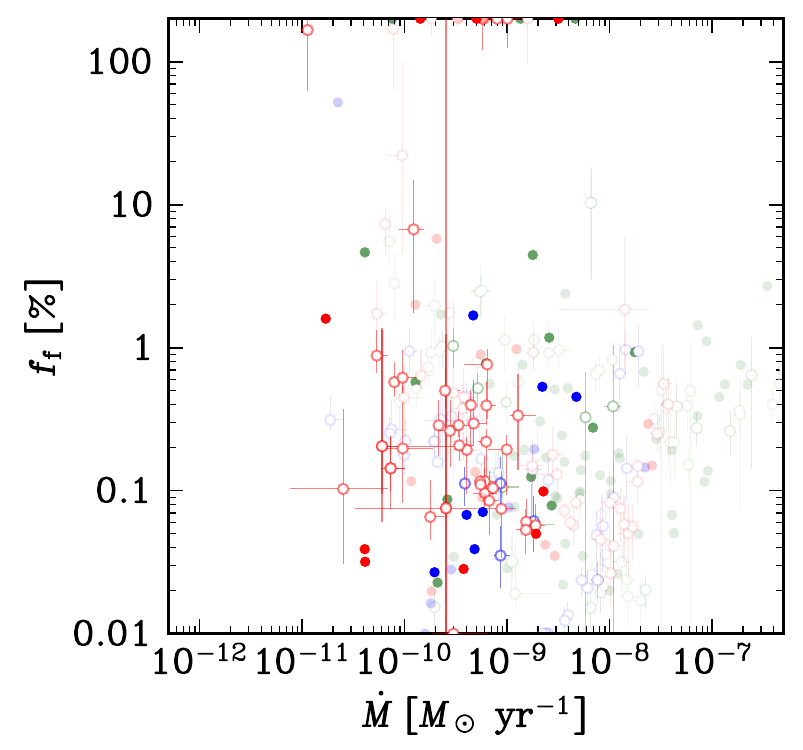}
    \caption{
    Accretion-shock filling factor on the object surface against object mass (upper) and mass accretion rate (lower).
    Colors and symbols are described in the caption of Figure~\ref{fig:Lacc}.
    }
    \label{fig:ff}
\end{figure}

The filling factor is defined as the emitting area normalized by the object surface area:
\begin{equation}
    \ff \equiv \Semit/(4\pi\Rs^{2}).
\end{equation}
Figure~\ref{fig:ff} shows that the inferred $\ff$ values typically span $\sim 0.01$--$1\%$,
with no clear dependence on mass nor mass accretion rate.
For comparison, estimates based on post-shock warm-slab models, which phenomenologically represent the heated photosphere, often give filling factors of order 5--10\% \citep{valenti1993,ingleby2013,Venuti+2015,Pittman+2025_MagSph}. These values are larger than our inferred $\ff$, but this difference is not necessarily a contradiction because the two estimates trace different components of the accretion shock. In the framework of recent multi-component accretion-column models, the large filling factors inferred from warm-slab models are associated with low-density accretion columns with low energy fluxes, whereas high-density accretion columns with high energy fluxes have much smaller filling factors, typically $\lesssim 1\%$ \citep{ingleby2013,Espaillat+2021,Pittman+2025_MagSph}. Indeed, other diagnostics sensitive to the immediate post-shock region (corresponding to the “shock-heated gas” in Figure~\ref{fig:Schematic}) also imply small filling factors of typically $\lesssim 1\%$ \citep{Fischer+2008,Espaillat+2021}. Therefore, our hydrogen-line-based filling factors should be interpreted as tracing the high-density, high-energy-flux component of the accretion columns, and are consistent with previous estimates in this regime.

\subsection{Local Extinction around Objects}
\label{sec:extinction}
\begin{figure*}
    \centering
    \def\wtem{\wT}
    \includegraphics[width=\wtem]{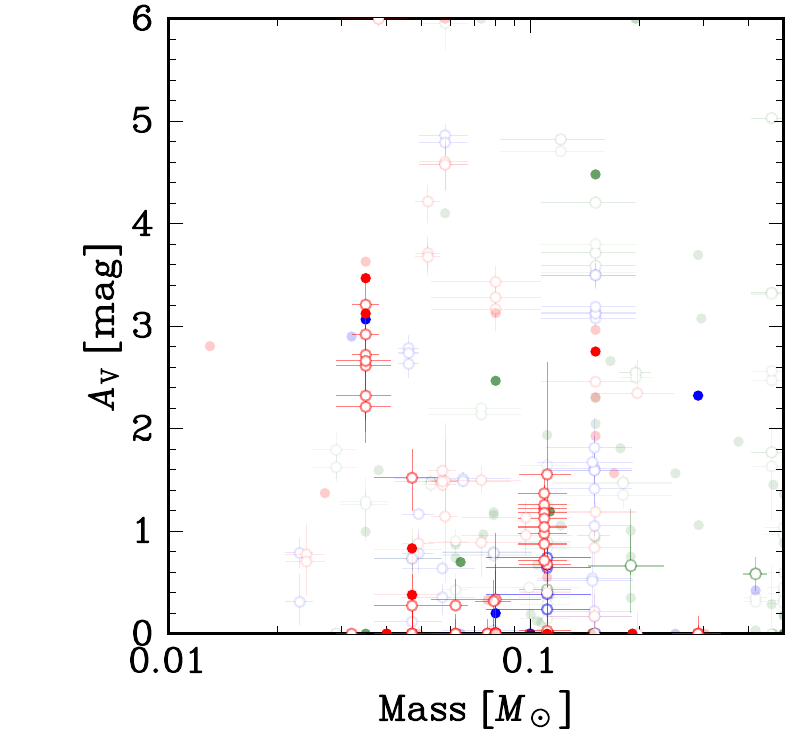}
    \includegraphics[width=\wtem]{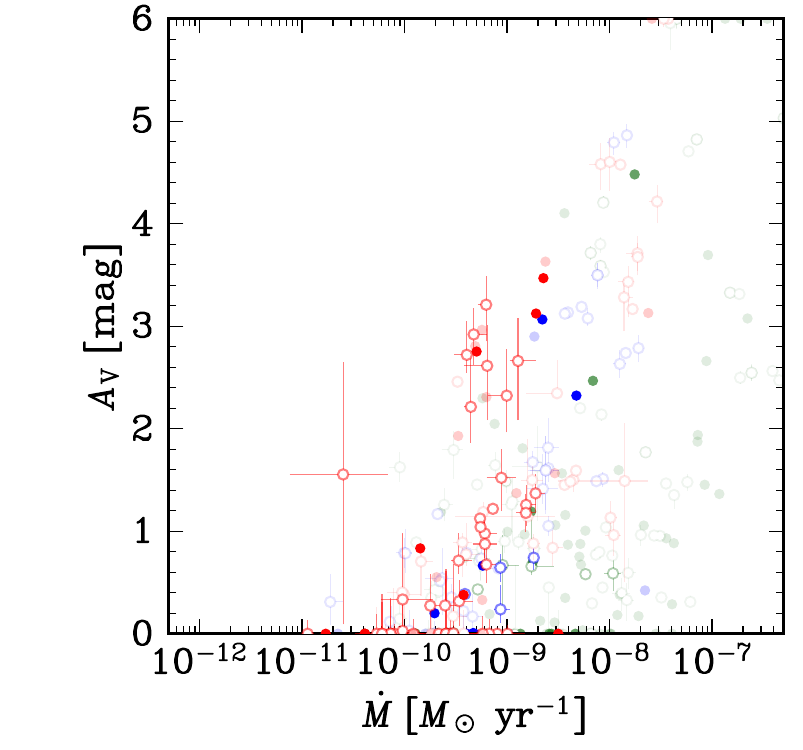}
    \includegraphics[width=\wtem]{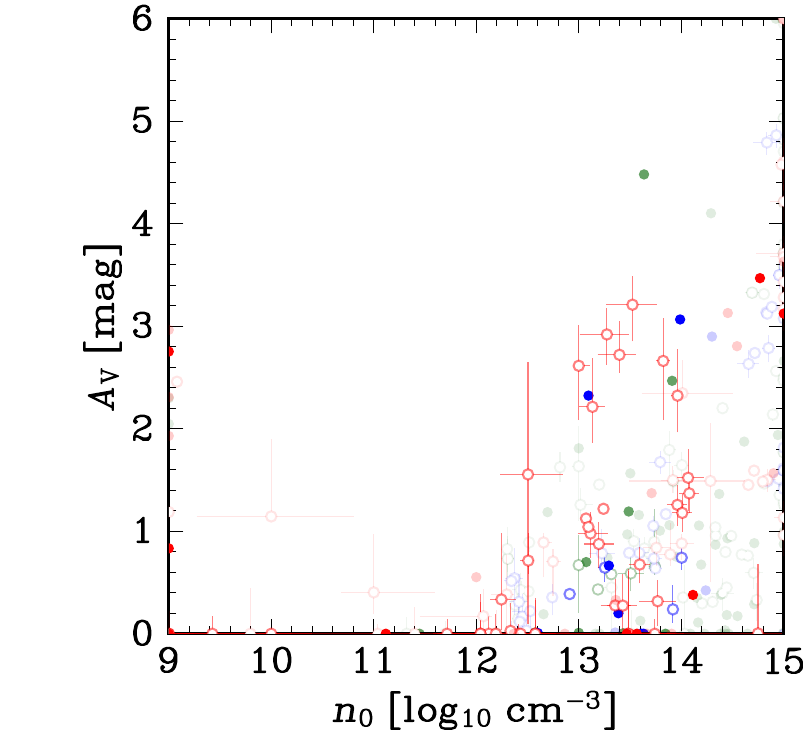}
    \caption{
    Visual extinction $\AV$ as functions of object mass (left), mass accretion rate (middle), and pre-shock number density $n_0$ (right).
    Colors and symbols are described in the caption of Figure~\ref{fig:Lacc}.
    }
    \label{fig:Extinction}
\end{figure*}

The local extinction near each accreting object is treated as a free parameter in our fitting, in addition to the interstellar extinction that is treated separately (see \S~\ref{sec:M_Data}).
Figure~\ref{fig:Extinction} shows that the fitting-derived local visual extinction $\AV$ ranges up to $\sim 6$ mag.
Whereas no strong correlation is evident with object mass (left), the upper envelope of $\AV$ increases with increasing mass accretion rate $\Mdot$ (middle), largely reflecting the corresponding trend with pre-shock density $n_0$ (right). In other words, large extinctions become more common in denser accretion flows.
On the other hand, some dense accretion flows show only weak extinction, suggesting that the extinction source may be spatially localized and may not always overlap with the line of sight. Strong time variability is also consistent with this picture (see also \S~\ref{sec:D_Variability}).

\begin{figure}
    \def\wtem{\wO}
    \centering
    \includegraphics[width=\wtem]{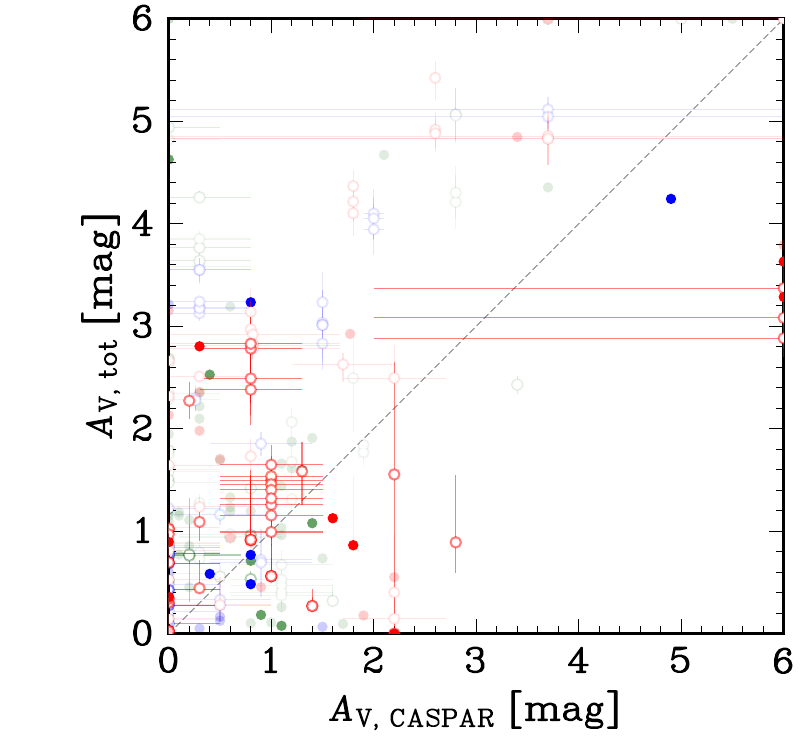}
    \includegraphics[width=\wtem]{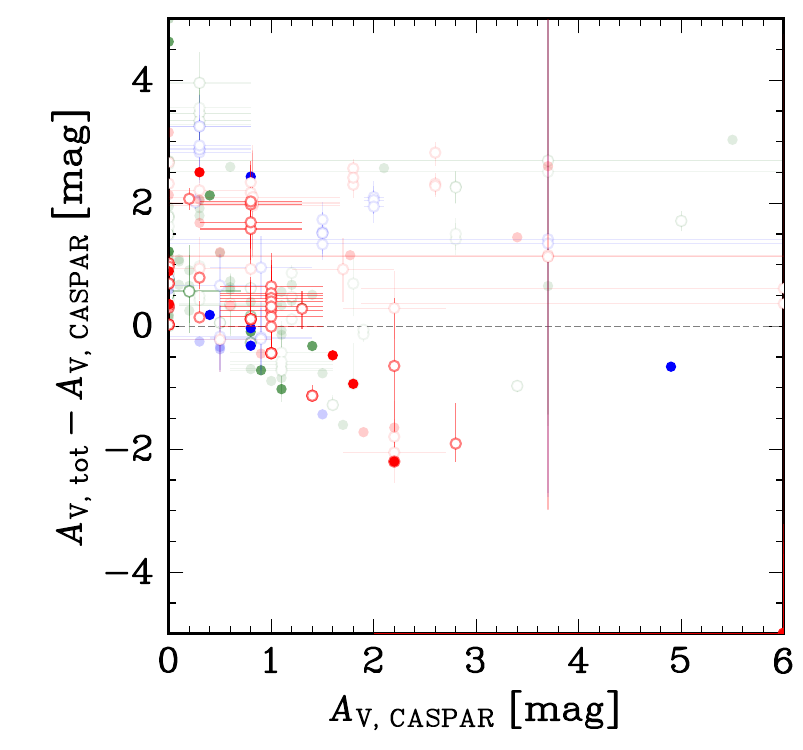}
    \caption{
    Upper panel: Comparison between the literature visual extinction $\AVCA$ \citep[CASPAR;][]{Betti+2023}, and the total extinction $\AVtot$, defined as the sum of the fitted local extinction and the interstellar extinction inferred from the 3D dust-map model of \citet{Edenhofer+2024}.
    Lower panel: $\AVtot - \AVCA$ as a function of $\AVCA$.
    $\AVCA$ values are literature estimates obtained using a variety of methods, which may contribute to the scatter.
    }
    \label{fig:Extinction_comp}
\end{figure}

We compare the derived $\AV$ with the literature values compiled in CASPAR \citep{CASPAR}, $\AVCA$. These values were derived using a variety of methods but generally represent the total extinction along the line of sight. For comparison with $\AVCA$, we therefore compute the total visual extinction, $\AVtot$, as the sum of the fitting-derived local extinction and the interstellar extinction estimated from the Galactic dust map, although these two components are not strictly equivalent. Figure~\ref{fig:Extinction_comp} compares $\AVtot$ with $\AVCA$.
Overall, $\AVtot$ broadly agrees with $\AVCA$, although offsets of up to $\approx 2$\,mag are not uncommon. At small $\AVCA$ of $\lesssim 3$\,mag, the offsets are seen in both directions; namely either $\AVtot$ or $\AVCA$ can be larger depending on the observations.
Part of this scatter may arise from the heterogeneous nature of the CASPAR values, which were compiled from different studies and estimated using different methods, including spectral comparison with templates, broadband colors, and dust-map-based estimates. Consequently, their uncertainties and systematic assumptions are not uniform. In particular, some nearby systems, such as TWA, are assumed to have $\AVCA=0$, causing them to cluster near the left boundary of the figure.
In addition, extinction can vary with time by up to $\approx 2$\,mag, as discussed in \S~\ref{sec:D_Variability}, which may also contribute to the observed scatter between $\AVtot$ and $\AVCA$.

On the other hand, when $\AVCA\gtrsim 3$\,mag, $\AVtot$ more often appears to be larger than $\AVCA$. These points are shown with transparent symbols mostly because of their high inferred densities, which require extrapolation beyond the density grid (see \S~\ref{sec:AFit_suspicious}). Such extrapolation may systematically bias the inferred $\AV$ toward larger values. Alternatively, if the trend is not an artifact of the extrapolation, the high density in the accretion column may produce strongly localized extinction that is less reflected in $\AVCA$, which is mostly derived from photospheric emission. Further investigation of these high-density cases requires extending the model calculations to higher densities.

We also allow $\RV$ to vary using the extinction law of \citet{Cardelli1989}.
However, meaningful constraints on $\RV$ are rare, and in most cases the acceptable range spans nearly the full allowed interval, $2.6$--$5.6$.

\subsection{Disk Truncation Radius and Magnetic Flux Inferred from Accretion Velocity}
\label{sec:v0_related}

One of the fitted model parameters, $v_0$, represents the accretion-flow velocity immediately upstream of the shock.
In magnetospheric accretion, the flow is expected to be close to free fall, so $v_0$ should approach the free-fall velocity from infinity, $\vff$, as the starting point of the flow moves farther outward, i.e., as the disk truncation radius $\Rt$ increases.
For this reason, the approximation $v_0 \simeq \vff$ is often adopted when $\Rt$ is assumed to be sufficiently large.
In this section, we compare the fitted $v_0$ with $\vff$ (\S~\ref{sec:v0_vff}), interpret the inferred $v_0/\vff$ ratios in terms of $\Rt$ (\S~\ref{sec:Rt}), and discuss the magnetic field strengths implied by this interpretation (\S~\ref{sec:BFlux}).

\subsubsection{Accretion Velocity Relative to Free-fall Velocity }
\label{sec:v0_vff}

\begin{figure*}
    \centering
    \includegraphics[width=\wT]{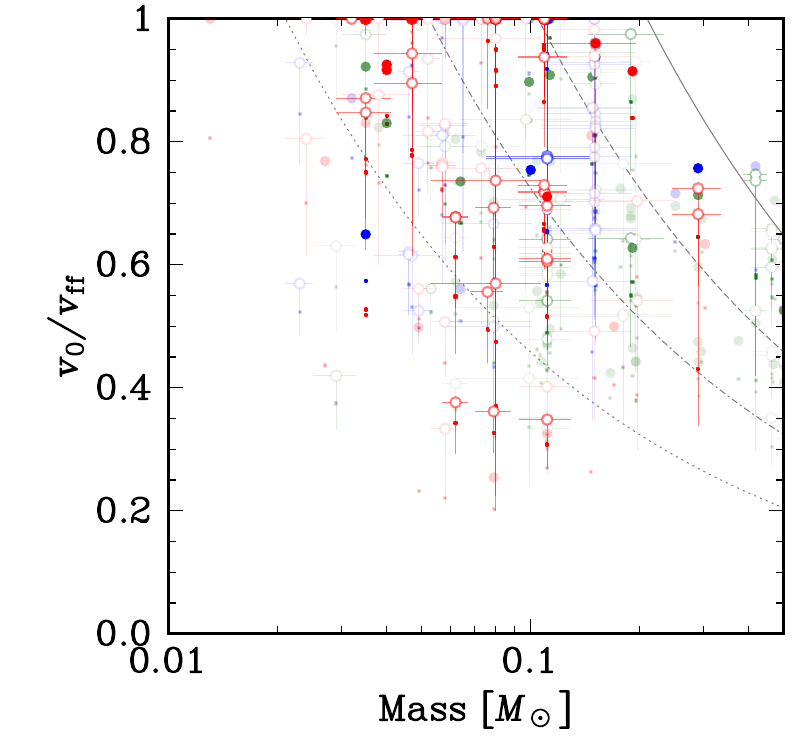}
    \includegraphics[width=\wT]{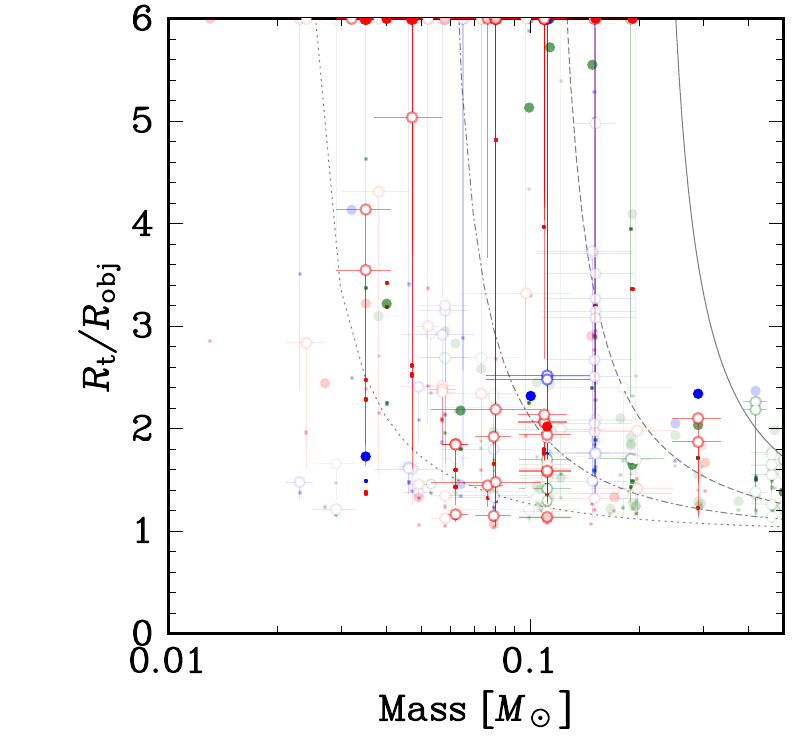}
    \includegraphics[width=\wT]{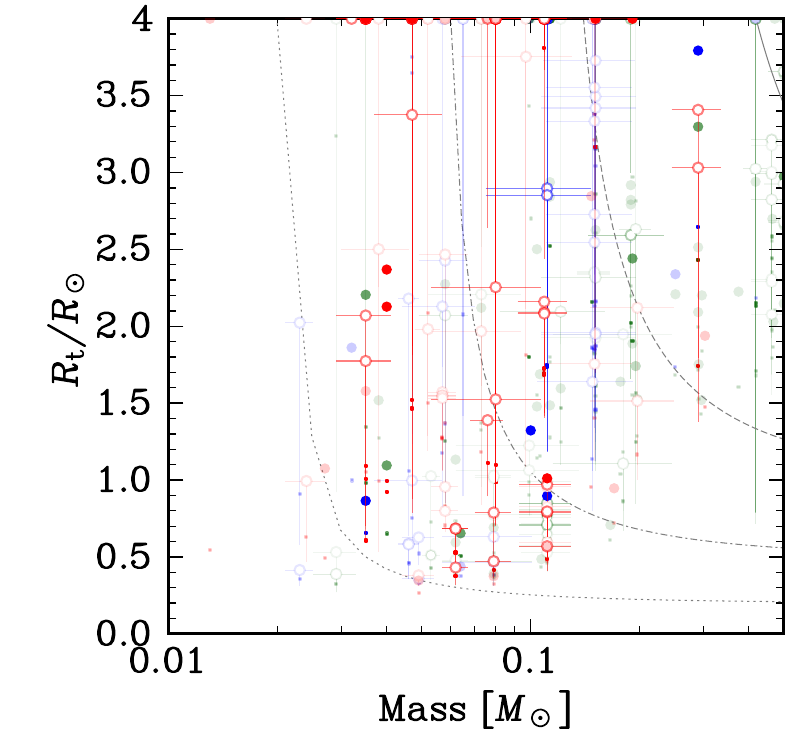}\\
    \includegraphics[width=\wT]{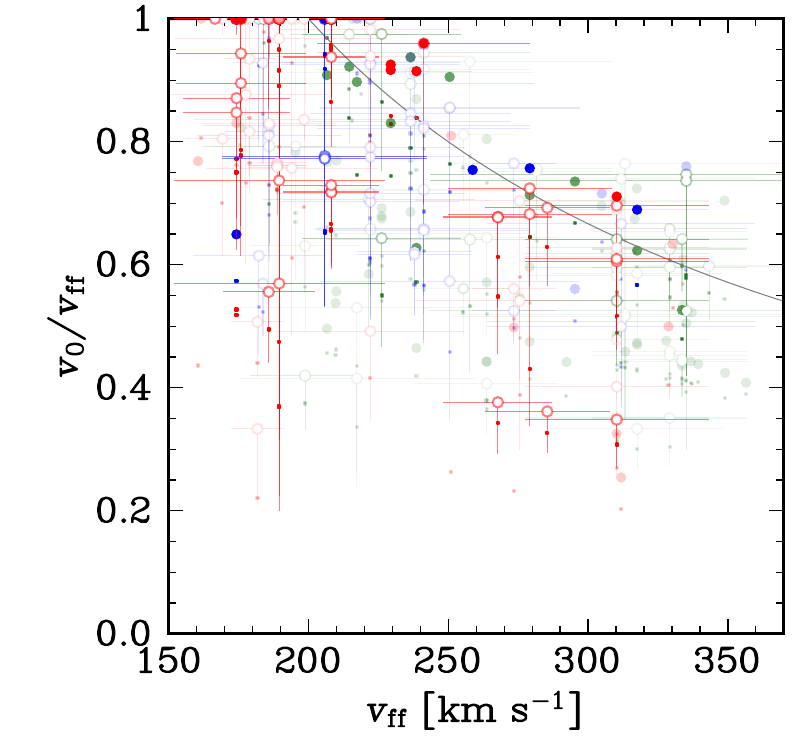}
    \includegraphics[width=\wT]{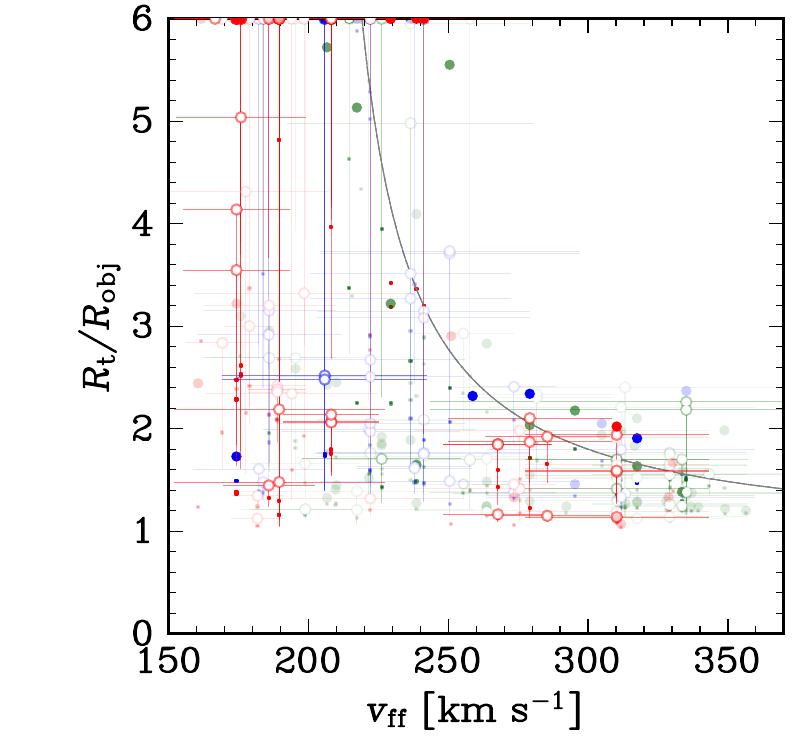}
    \includegraphics[width=\wT]{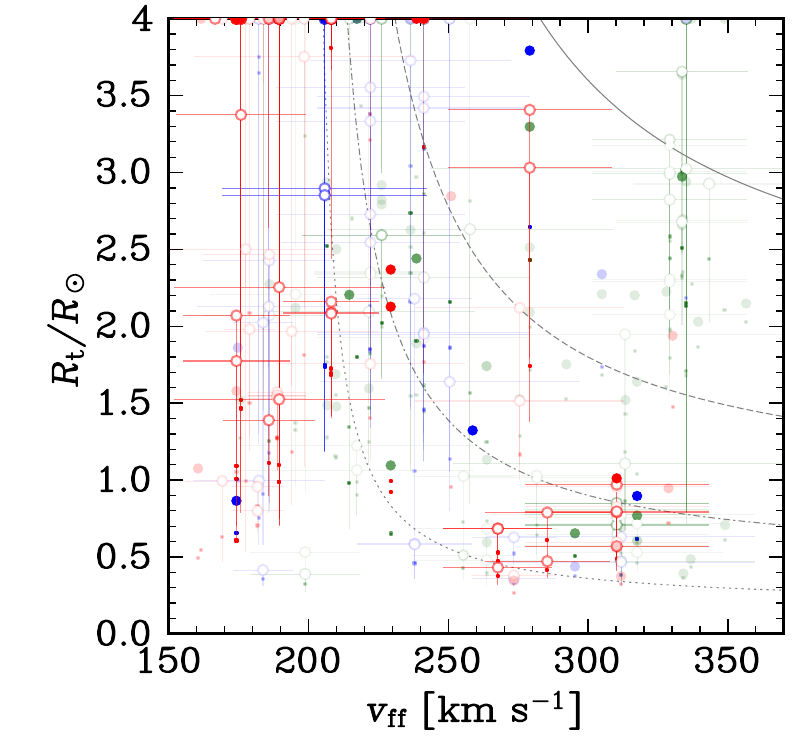}
    \caption{
    Scatter plots of the ratio of the accretion-flow velocity to the free-fall velocity (left), the truncation radius normalized by the object radius (middle), and the truncation radius in Solar-radius units using the literature object radius (right),
    shown as functions of object mass (upper) and free-fall velocity (lower).
    While the symbol rule follows that of Figure~\ref{fig:Lacc}, the circle is placed at the upper limit of the allowed range and the representative value is shown by the small bullet.
    The background lines indicate the upper limit corresponding to $v_0=200\,\kms$, set by the model parameter range; values much above this are also physically disfavored (see \S~\ref{sec:Emodel}).
    The different line styles correspond to $R/\Rsun=2$ (solid), 1 (dashed), 0.5 (dash-dotted), and 0.2 (dotted), which are used to convert the $v_0$ limit into each plotted quantity.
    Colors and symbols are described in the caption of Figure~\ref{fig:Lacc}.
    }
    \label{fig:Truncation}
\end{figure*}

The left panels of Figure~\ref{fig:Truncation} show the $v_0/\vff$ ratio as a function of mass (upper) and $\vff$ (lower). To illustrate how close the fitted solutions can remain consistent with the conventional approximation $v_0 \simeq \vff$, the circle symbols indicate the upper limits of the allowed $v_0/\vff$ ranges, while the small bullets show the representative values.
Some circles cluster near unity, suggesting that the conventional approximation $v_0 \simeq \vff$ is indeed valid in these cases. However, many circles are substantially below unity, indicating that a significant number of objects have $v_0$ much smaller than $\vff$.
This provides a natural explanation for the result in \S~\ref{sec:ClassDependency}, where some objects with relatively large masses or large $\vff$ still show hydrogen lines consistent with shock emission.
If one assumed $v_0 \simeq \vff$, shock emission would indeed become subdominant above $M\gtrsim0.2\,\Msun$, because such objects would tend to enter the strong-shock regime where hydrogen-line emission is inefficient. Our results instead suggest that $v_0$ can be substantially smaller than $\vff$, allowing even objects above this mass range to produce strong hydrogen-line emission directly in the shock-heated gas.

Although the lower-left panel of Figure~\ref{fig:Truncation} may appear to show a negative correlation between $v_0/\vff$ and $\vff$, this apparent trend should be interpreted with caution.
The blank region in the upper-right corner is largely artificial, reflecting the finite parameter range of the model grid. The black line in the lower-left panel corresponds to $v_0=200\,\kms$, which is the upper limit of the modeled $v_0$ range, showing that the model-grid upper limit causes the blank region\footnote{In fact,
$v_0/\vff$ can exceed the grid-imposed upper limit because of the uncertainty in $\vff$, propagated from the uncertainties in $(M,R)$.}.
However, this blank region is not merely unexplored but rather corresponds to the regime where shock-origin hydrogen-line emission becomes inefficient and non-shock emission is expected to dominate (see \S~\ref{sec:Emodel}).
This suggests a possible interpretation:
among high-$\vff$ objects, those with small $v_0/\vff$ can emit shock-dominated hydrogen lines, whereas those with large $v_0/\vff$ instead become non-shock-dominated and therefore are absent from Figure~\ref{fig:Truncation}. This interpretation cannot be tested directly here, because the non-shock-dominated cases are outside the scope of the present shock-emission fitting.

\subsubsection{Disk Truncation Radius}
\label{sec:Rt}
A natural explanation for small $v_0/\vff$ is that the accretion flow starts closer to the object surface than is often assumed.
The disk truncation radius $\Rt$, i.e., the starting point of the magnetospheric free-fall flow, can be derived from energy conservation as
\begin{equation}
    \frac{\Rt}{\Rs} = \left[1-\left(\frac{v_0}{\vff}\right)^2\right]^{-1}.
\end{equation}
The inferred $\Rt$ values are shown in the middle and right panels of Figure~\ref{fig:Truncation}.
Corresponding to $v_0/\vff \sim 1 $, some low-mass objects show $\Rt/\Rs \gtrsim 5$ as frequently assumed. However, some objects show substantially smaller $\Rt/\Rs$, sometime even $\lesssim 2\Rs$. As discussed in \S~\ref{sec:v0_vff}, the apparent lack of large-$\Rt$ at high mass or high $\vff$ is artificial and likely corresponds to the T-Tauri star regime where non-shock emission dominates the hydrogen line emission.

Although $\Rt\sim5\Rs$ is often used as a conventional reference value, recent observations have suggested smaller characteristic truncation radii of $\sim2.5 \Rs$ \citep{Thanathibodee+2023,gravitycollaboration_9,Armeni2024,Pittman+2025_MagSph,Venuti2026}. The representative values inferred here (small bullet rather than circles that showing upper limit) often lie in this range or below it, and are therefore broadly consistent with these recent observational constraints.

\subsubsection{Surface Dipole Magnetic Field Strength}
\label{sec:BFlux}

\begin{figure*}
    \centering
    \def\wtem{\wT}
    \includegraphics[width=\wtem]{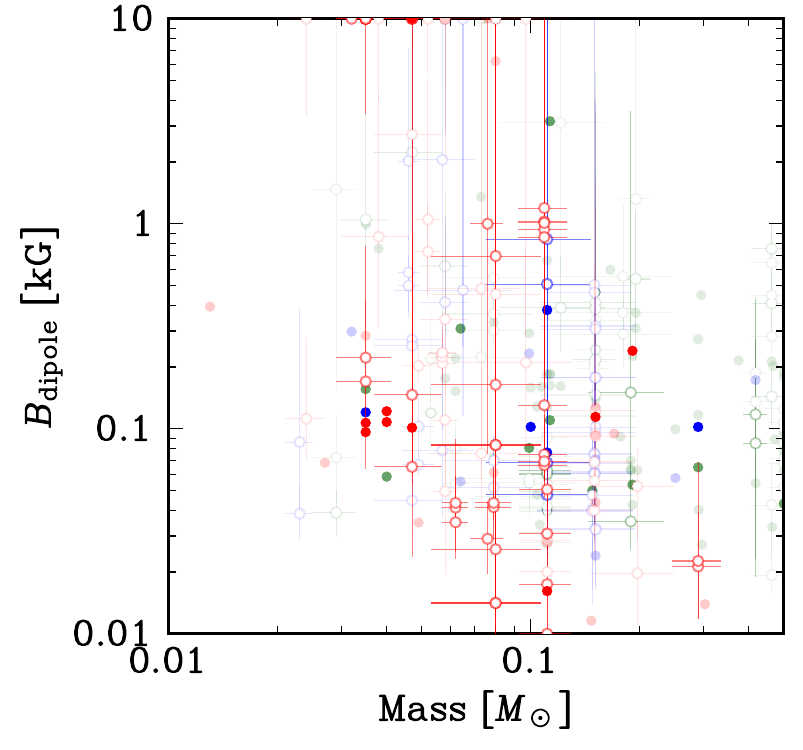}
    \includegraphics[width=\wtem]{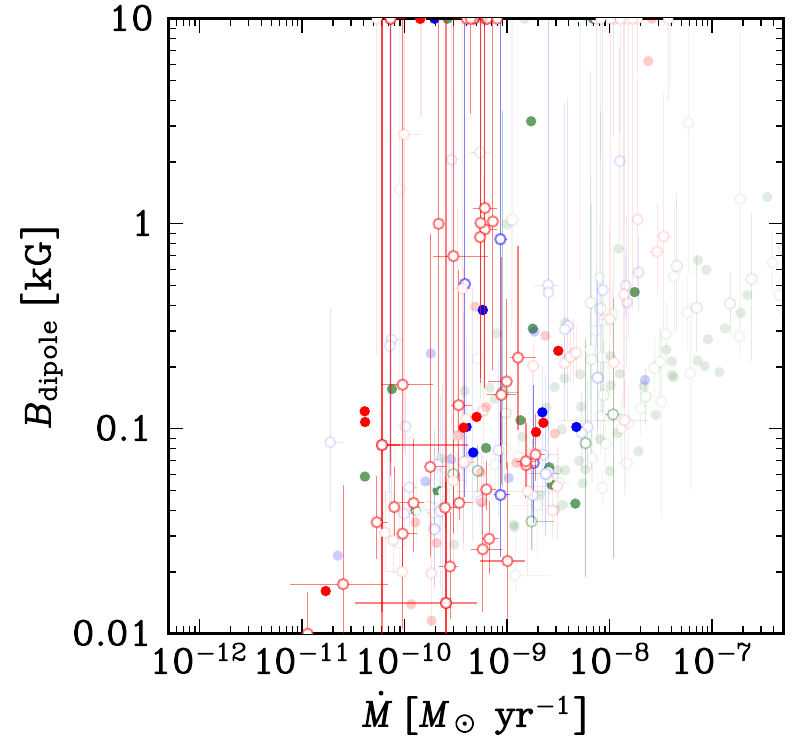}
    \includegraphics[width=\wtem]{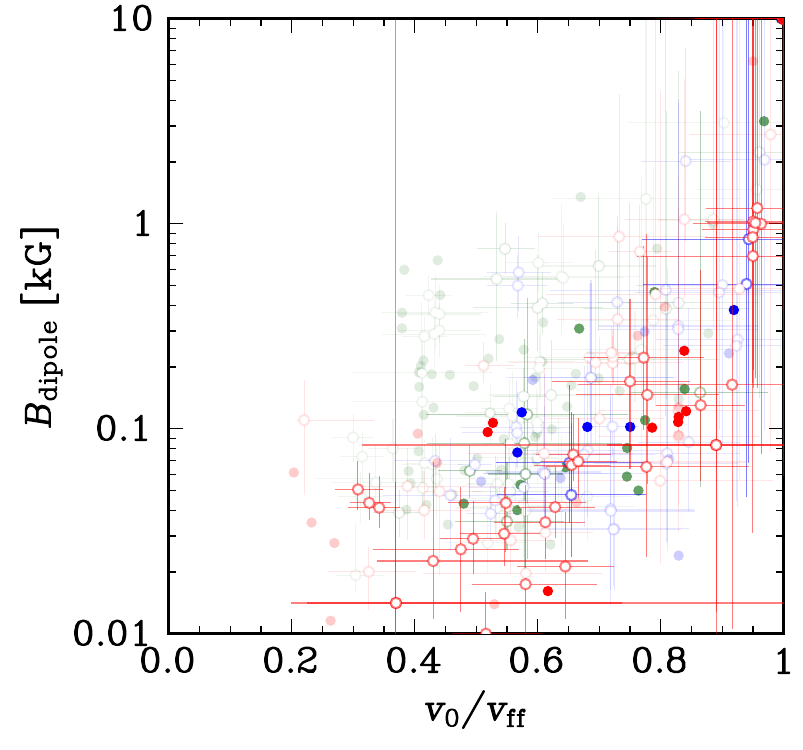}
    \caption{
    Inferred surface dipole magnetic field strength as a function of object mass (left), mass accretion rate (middle), and $v_0/\vff$ (right).
    }
    \label{fig:Bflux}
\end{figure*}

In this section, we further translate the inferred $\Rt$ into the dipole magnetic field strength $\Bs$ responsible for the magnetospheric truncation. 
Similar estimates of the magnetic field strength from the truncation radius have been made in previous studies \citep[e.g.,][]{Hasegawa2021,demars2023}.
Using the angular-momentum balance relation\footnote{The frequently used estimate obtained by balancing the magnetic pressure against the ram pressure of spherical accretion leads to the same power-law dependence \citep{Ghosh+Lamb1979,Koenigl1991}. From dimensional analysis, any derivation must lead to the same dependence \citep{MohantyShu2008,Batygin2018}.}, the truncation radius can be written as
\citep{Dangelo+Spruit2010,Takasao+2022}
\begin{equation}
\begin{split}
    \frac{\Rt}{\Rs} = &\, 2.3 \left( \frac{\Bs}{160~\mathrm{G}} \right)^{4/7}
    \left( \frac{\Rs}{2~\Rsun} \right)^{5/7}\\
    & \; \times \left( \frac{\Ms}{0.5~\Msun} \right)^{-1/7}
    \left( \frac{\Mdot}{10^{-8}~\Msun \yr^{-1}} \right)^{-2/7},
\end{split}
\end{equation}
where $\Bs$ is the dipole magnetic field strength at the object surface.
This relation can be rewritten as
\begin{equation}
\begin{split}
\label{Eq:Bflux}
    \Bs =&\, 160~\textrm{G}~  \left( \frac{\Rt/\Rs}{2.3} \right)^{7/4}
    \left( \frac{\Rs}{2~\Rsun} \right)^{-5/4}\\
    & \; \times \left( \frac{\Ms}{0.5~\Msun} \right)^{1/4}
    \left( \frac{\Mdot}{10^{-8}~\Msun \yr^{-1}} \right)^{1/2},
\end{split}
\end{equation}
or, equivalently,
\begin{equation}
\begin{split}
\label{Eq:Bflux_RtRsun}
    \Bs =&\, 160~\textrm{G}~  \left( \frac{\Rt}{4.6~\Rsun} \right)^{7/4}
    \left( \frac{\Rs}{2~\Rsun} \right)^{-3}\\
    & \; \times \left( \frac{\Ms}{0.5~\Msun} \right)^{1/4}
    \left( \frac{\Mdot}{10^{-8}~\Msun \yr^{-1}} \right)^{1/2}.
\end{split}
\end{equation}
These expressions assume that the dipole component dominates the magnetic truncation. For sources inferred to have small truncation radii, higher-order multipole components may become important, making the inferred dipole field strength an upper limit.

Figure~\ref{fig:Bflux} shows that the inferred $\Bs$ ranges from $\sim0.01$~to more than $10$~kG, while most samples cluster around $100$~G. This is broadly consistent with independent Zeeman measurements for accreting M dwarfs, which resulted in non-detections and upper limits of order $\sim10^{3}$~G \citep{Reiners2012}.
In comparison, non-accreting objects in a similar mass range have measured surface fields of order $10^{3}$~G \citep{Reiners+2009}, somewhat larger than many of our inferred values. This may suggest that low-mass objects have weaker dipole magnetic fields during the accretion phase, although the measured field components, likely the lower-order dipole field, does not necessarily reflect the total surface magnetic field strength, which can include substantial higher-order components \citep{johns-krull2007}.

In many 2D models, the truncation radius is taken to be the smaller of the magnetospheric radius and the corotation radius. In classical T~Tauri stars, the magnetospheric radius can exceed the corotation radius \citep[e.g.,][]{Bouvier+2020}. If the corotation radius were instead the quantity that set $\Rt$, our estimate of $\Bs$ would not be valid.
However, recent 3D MHD simulations suggest that the truncation radius depends only weakly on the stellar spin rate, and hence that the corotation radius may play only a limited role \citep{Takasao+2022}. This is because, near (or even inside) the magnetospheric radius, magnetic field lines rotate approximately at the local Keplerian rate rather than at the stellar spin rate, because the field-line rotation is governed by the coupled gas motion. Although the explored parameter space is still limited, these results support the above estimate of $\Bs$ as being only weakly sensitive to the stellar spin rate.
At the same time, if the gas rotation remains close to the local Keplerian speed, the free-fall assumption may break down. In that case, the smaller inferred $v_0$ would not necessarily imply a small truncation radius, and the true dipole field strength could be larger than estimated here.

\section{Discussion}
\label{sec:D}

\subsection{Origin of the Broader Component}
\label{sec:D_BC}

The broad-component (BC) excess over the shock-emission component frequently appears in our sample. In this section, we examine whether this component can be attributed to shock emission affected by unmodeled line broadening, or instead requires an additional emission component.

\subsubsection{Unmodeled Broadening as a Possible Origin of the BC}

We first examine whether line broadening not included in the model could account for the BC.
The shock-emission model includes thermal Doppler broadening, which yields a Gaussian profile, and natural and van der Waals broadening, which contribute Lorentzian wings \citep{Aoyama+2018}.
One broadening mechanism not included in the model is Stark broadeing, which may further broaden the line profile and potentially account for the BC. The FWHM of Stark broadening for \Hb can be approximated as \citep{Konjevic2012a},
\begin{align}
    w_\mathrm{Stark,\,\Hb} &= 0.94666\,\mathrm{nm} \left(\frac{n_\mathrm{e}}{10^{16}\,\cc} \right)^{1/1.49}\\
    &= 584 \,\kms
    \frac{\nu_0}{c}
    \left(\frac{n_\mathrm{e}}{10^{16}\,\cc} \right)^{1/1.49},
\end{align}
where $n_\mathrm{e}$ is the electron number density, $\nu_0$ is the line center wavelength, and $c$ is the speed of light.
Although the reference electron number density of $10^{16}\cc$ in this expression is higher than the pre-shock gas number density $n_0$, a similar power law of $\approx 2/3$ can remain valid down to $\approx 2\times 10^{12}\cc$ \citep{Palomares2012}.
Considering typical values of an ionization fraction of tens percent and a shock compression factor of several \citep{Aoyama+2018,Aoyama+2020}, the post shock $n_\mathrm{e}$ can be of the same order as $n_0$, implying Stark broadening broad enough to explain the BC, at least in some objects.

\begin{figure}
    \def\LW{\wO}
    \centering
    \includegraphics[width=\LW]{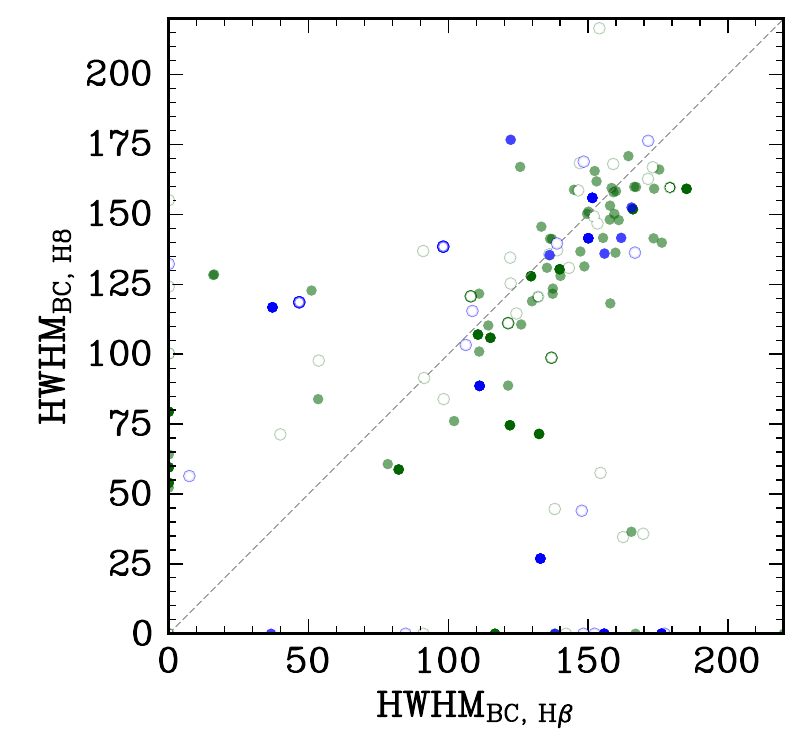}
    \includegraphics[width=\LW]{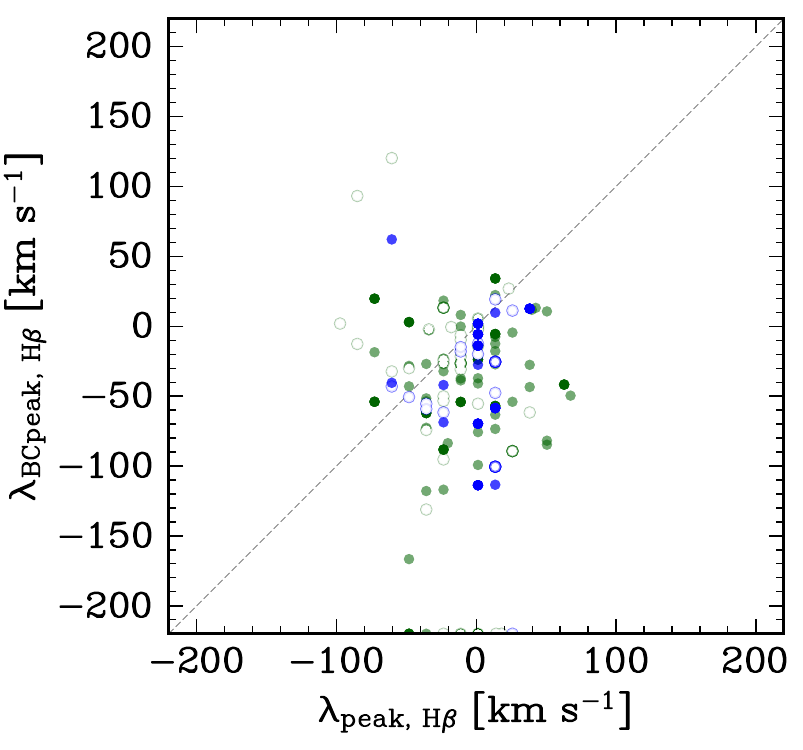}
    \caption{
    Upper: comparison between the half width at half maximum (HWHM) of the fitted BC in \Hb\ and H8.
    Lower: comparison between the observed line-peak wavelength and the fitted BC-peak wavelength.
    Colors and symbols are described in the caption of Figure~\ref{fig:Lacc}.
    }
    \label{fig:BC_Profile}
\end{figure}

However, the upper panel of Figure~\ref{fig:BC_Profile} shows that the fitted BCs in \Hb and H8 often have similar widths, although the stark broadening of H8 is expected to be broader than that of \Hb by a factor of several \cite{Stambulchik2007}.
The exceptional outliers with small HWHM are mostly associated with cases where the BC is not clearly detected.
Additionally, the BC sometimes shows a shift relative to the line peak, as shown in the lower panel of Figure~\ref{fig:BC_Profile}. Stark broadening can also induce line shifts, but these are usually redshifted \citep[e.g.,][]{Halenka2015,Oks2018}. Moreover, if the BC were produced by Stark broadening in the shock-heated gas, the same velocity shift would also affect the narrow peak formed there, so a large relative shift between the BC and the shock-emission peak would not be expected. Therefore, the BC likely arises from a region different from that producing the main peak.
Spectropolarimetric observations may help constrain this possibility: a detectable circular-polarization signal in the BC would support an origin in strongly magnetized shock-heated gas \citep{Yang2006,Johns-Krull2013}.

\begin{figure*}
    \def\LW{\wH}
    \includegraphics[width=\LW]{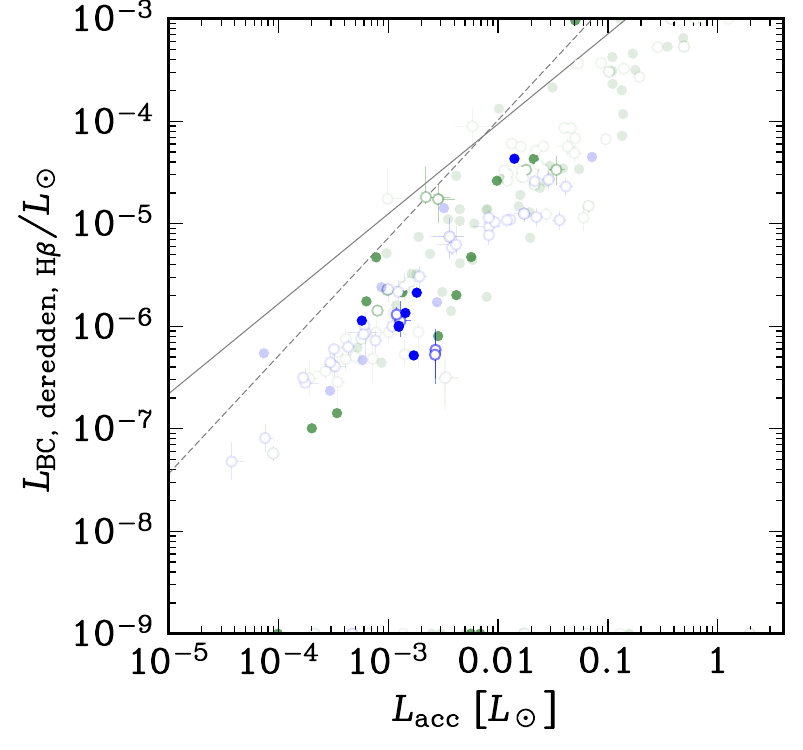}
    \includegraphics[width=\LW]{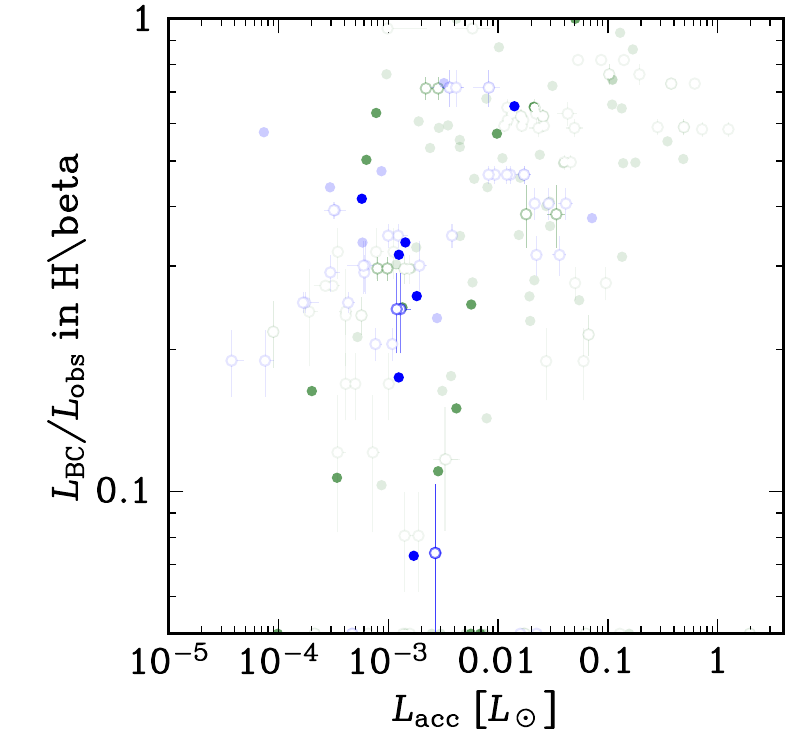}
    \\
    \includegraphics[width=\LW]{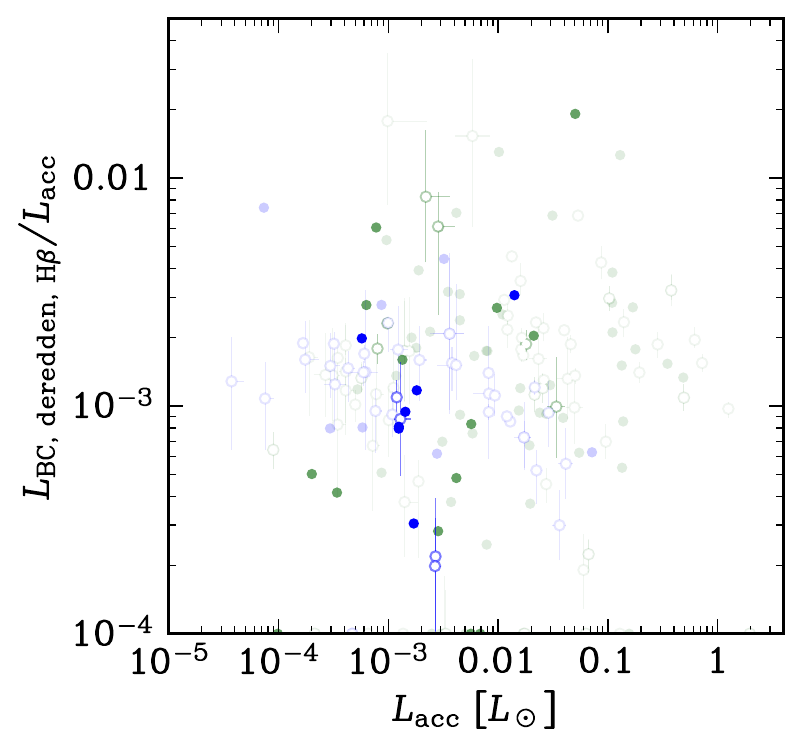}
    \includegraphics[width=\LW]{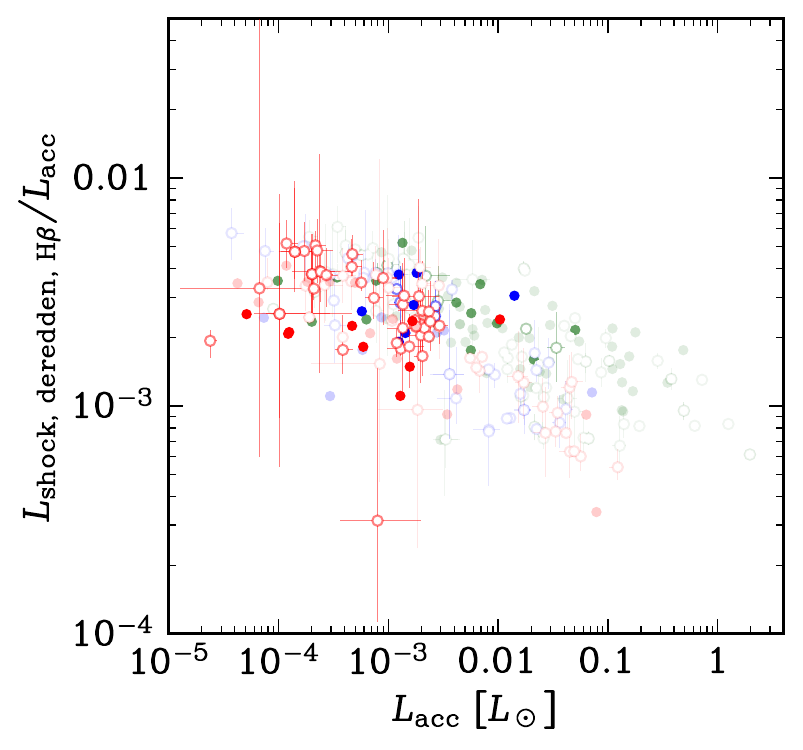}
    \caption{
    Broad-component properties of \Hb.
    Upper panels show the BC luminosity $\LBC$ in solar units (left) and $\LBC$ normalized by the total observed \Hb\ luminosity (right), both as functions of accretion luminosity.
    Lower panels show the BC (left) and shock-component (right) luminosity normalized by the accretion luminosity, as functions of accretion luminosity.
    In the upper-right panel, the empirical relation of \citet{Alcala+2017} (solid line) and the shock-model relation of \citet{Aoyama+2021} (dashed line) are also shown.
}
    \label{fig:BC_Hb}
\end{figure*}

\subsubsection{Accretion-flow Emission as a Possible Origin of the BC}

If the BC does not arise from the post-shock region, a natural alternative is emission from the magnetospheric accretion flow, as in more massive objects such as T Tauri stars (see \S~\ref{sec:I_HLMechanism}). This picture is consistent with our finding that the BC appears more frequently, and more often dominates the spectra, toward higher masses (see \S~\ref{sec:ClassDependency}).
This interpretation also provides a natural explanation for why the BC is broader than the shock-emission component. Explaining such a width by thermal Doppler broadening alone would require gas hotter than the shock-heated gas, which is unlikely. In the accretion-flow scenario, the broad width can instead be produced by the range of line-of-sight velocities in the flow, even if the gas is cooler than the post-shock region \citep[e.g.,][]{Hartmann+1994,Muzerolle+1998,Muzerolle+2001}.

To assess this accretion-flow hypothesis, the upper-left panel of Figure~\ref{fig:BC_Hb} shows the $\LBC$--$\Lacc$ relation, together with the empirical relation for TTSs of \citet{Alcala+2017} (solid line) and the shock-model relation of \citet{Aoyama+2021} (dashed line). Above $\Lacc\gtrsim10^{-2}\Lsun$, the samples follow a trend broadly similar to the \citet{Alcala+2017} relation, supporting a possible link between the BC and the accretion-flow emission traced by the empirical relation. At $\Lacc\lesssim10^{-3}\Lsun$, $\LBC$ decreases with $\Lacc$ decreasing more steeply than the \citet{Alcala+2017} relationship, suggesting that the BC emission becomes less efficient towards lower $\Lacc$.

As a possible explanation for this transition in the dominant emission mechanism, \citet{Aoyama+2021} suggested that shock emission becomes more efficient toward lower $\Lacc$ and therefore becomes dominant. The lower-right panel of Figure~\ref{fig:BC_Hb} indeed shows a negative correlation between the shock-emission efficiency, $\Lshock/\Lacc$, and $\Lacc$.
In addition, the lower-left panel suggests that the BC efficiency drops at $\Lacc \lesssim 10^{-3}\,\Lsun$.
A possible interpretation is that, at lower $\Lacc$, a larger fraction of the accretion energy is radiated efficiently in hydrogen lines from the shock-heated gas. Because such line emission may be less efficiently thermalized than X-ray/UV emission, less energy may be deposited into the surrounding accretion flow, reducing the secondary emission that would produce the BC.
However, the heating mechanism of the accretion flow itself remains unclear even in TTSs \citep{Muzerolle+2001}, and further study is needed.

\subsection[H-alpha Excess]{\Ha Excess}
\label{sec:D_Ha}

Although many objects in our sample are consistent with shock-dominated hydrogen-line emission, \Ha\ is a notable exception. Even when the higher-order hydrogen lines are well reproduced by the shock-emission model, the modeled \Ha flux is less than half the observed value in 65\,\% (52 out of 80) of \ConcShock and \ConcBC cases.
This indicates that an additional emission component contributes to \Ha, and, although it does not always dominate, often exceeds the contribution from shock emission.

\subsubsection[Shock Emission Contribution to H-alpha ]{Shock Emission Contribution to \Ha}
\label{sec:D_HaContribution}

\begin{figure*}
\def\wtem{\wH}
    \centering
    \includegraphics[width=\wtem]{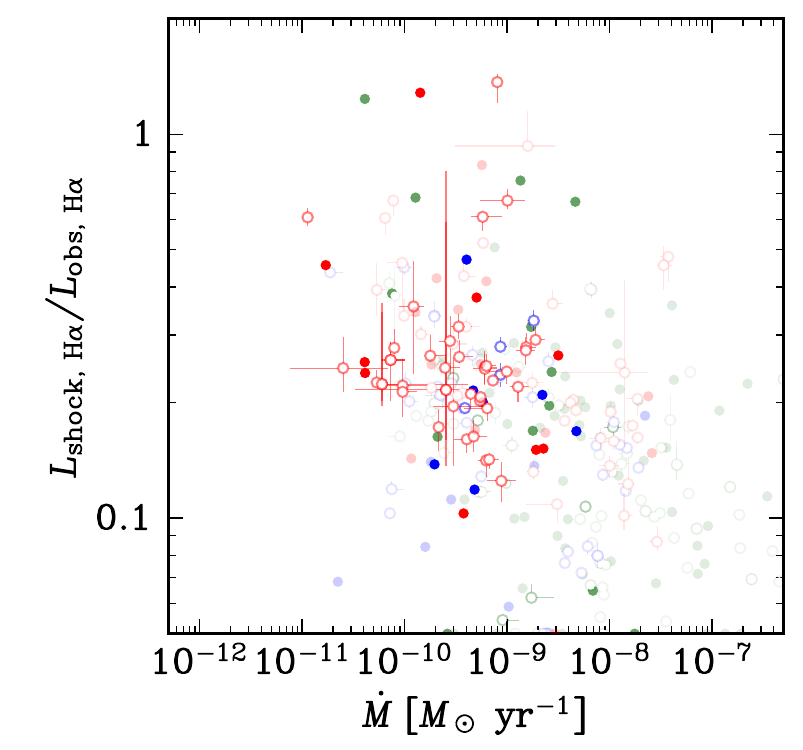}
    \includegraphics[width=\wtem]{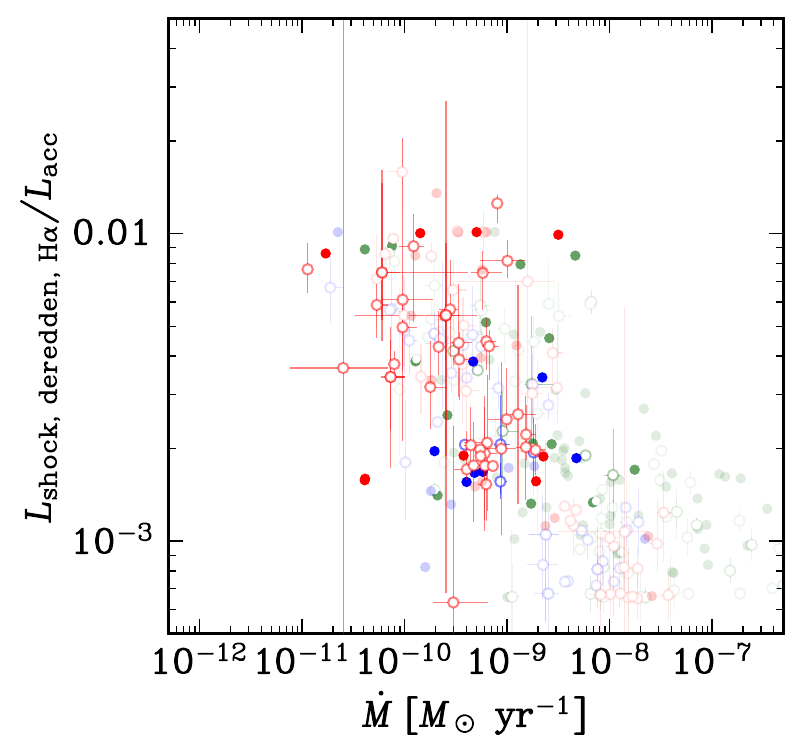}\\
    \includegraphics[width=\wtem]{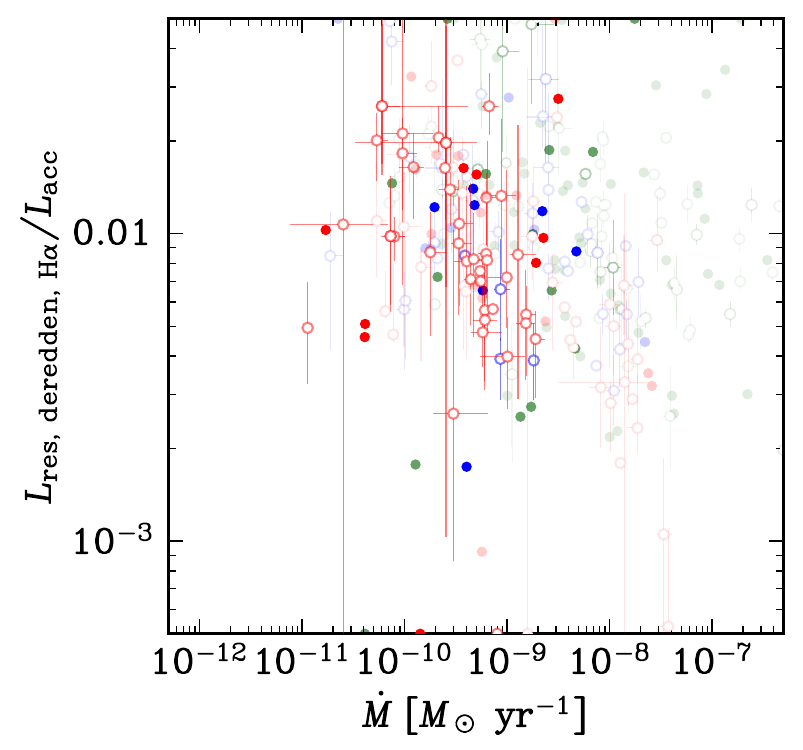}
    \includegraphics[width=\wtem]{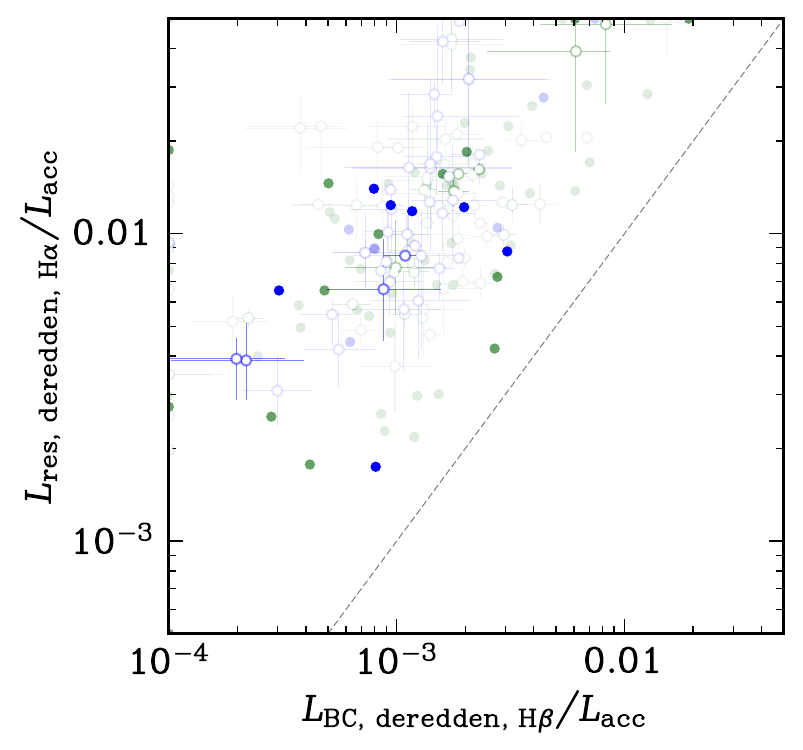}
    \caption{
    The upper-left panel shows the fraction of the shock-origin \Ha\ luminosity relative to the observed \Ha\ luminosity as a function of mass accretion rate.
    The upper-right and lower-left panels show the shock-origin and non-shock \Ha\ luminosities, respectively, normalized by the accretion luminosity, as functions of mass accretion rate.
    The lower-right panel shows the correlation between the non-shock components of \Ha\ and \Hb, both normalized by the accretion luminosity.
    }
    \label{fig:Ha_invest}
\end{figure*}

To quantify the shock-emission contribution to \Ha, we predict the shock-origin \Ha\ flux using the parameters derived from the spectral fitting of the other hydrogen lines.
The upper-left panel in Figure~\ref{fig:Ha_invest} shows that the fraction of the shock-emission component ranges from 0.01 to 1 and tends to decrease with increasing $\Mdot$, although with substantial scatter.
To investigate the origin of this trend, the upper-right and lower-left panels show the shock-origin and residual \Ha\ luminosities normalized by the accretion luminosity, respectively. These quantities represent the efficiencies of converting accretion energy into the shock-origin and non-shock \Ha\ components.
The shock-origin efficiency decreases systematically with increasing $\Mdot$, whereas the residual component shows no clear trend. The decrease in the shock-origin efficiency is naturally explained by self-absorption, or saturation, of \Ha\ in the post-shock region \citep{Aoyama+2018}.
These results suggest that the relative importance of the shock and non-shock components is largely controlled by the shock-emission efficiency.

\subsubsection[Origin of H-alpha Excess]{Origin of \Ha Excess}
\label{sec:D_HaOrigin}

Although the residual-component efficiency shows no clear correlation with any accretion properties, this efficiency is clearly correlated with the BC efficiency, as shown in the lower-right panel of Figure~\ref{fig:Ha_invest}.
This suggests that the \Ha\ excess might share the same emission mechanism as the excess-emission component that appears in other hydrogen lines. However, unlike the broad Gaussian component used to describe the BC, the \Ha\ excess is often concentrated near line center and frequently exhibits double-peaked or otherwise complex profiles. This difference may be naturally explained if the excess arises in the accretion columns \citep[e.g.,][]{Hartmann+1994,Muzerolle+1998}.
In this framework, the line profile is shaped mainly by the velocity distribution of the accelerating accretion flow: fast gas close to the stellar surface produces broad emission, whereas slower gas at larger radii near the disk contributes mainly near the line center.
Such outer-column emission may be significant only for relatively low-excitation lines such as \Ha, because the accretion columns are expected to be cooler farther from the stellar surface.

\subsubsection[H-alpha Luminosity vs Accretion Luminosity Relation]{$\LHa$--$\Lacc$ Relation}

\Ha\ is frequently used to estimate $\Lacc$ and hence $\Mdot$ because of its brightness. However, this work shows that the \Ha\ emission mechanism is often a complex mixture of shock and non-shock components, particularly in low-mass objects, suggesting that \Ha\ is not necessarily a reliable line for estimating accretion properties.

\begin{figure}
    \centering
    \includegraphics[width=\wO]{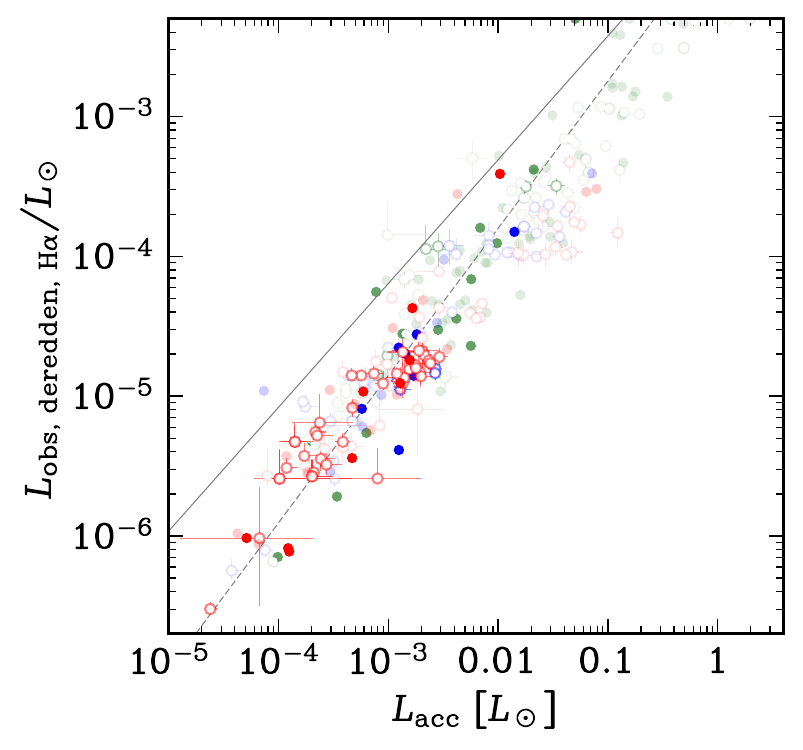}
    \includegraphics[width=\wO]{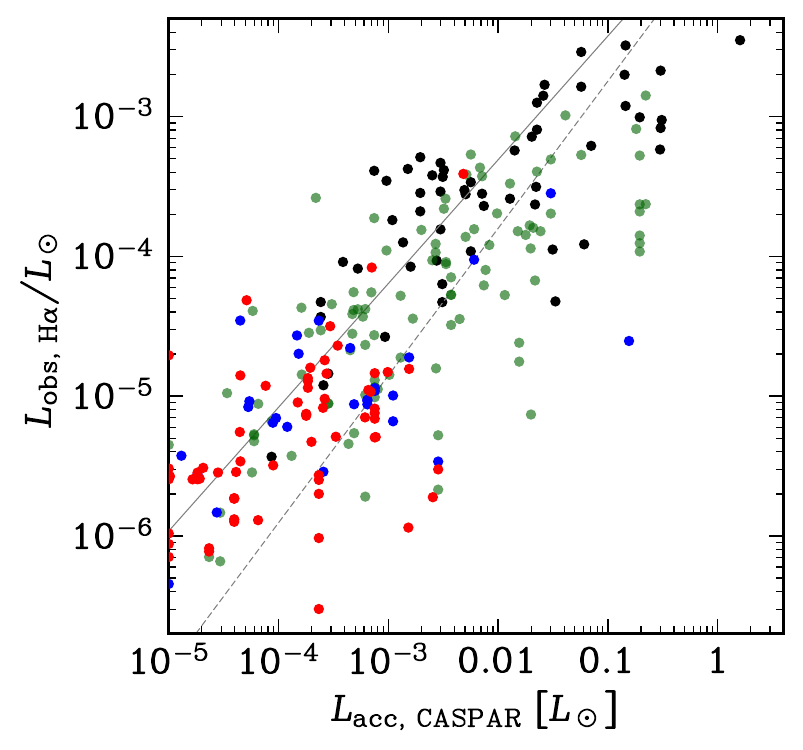}
    \caption{
    Relationship between the observed \Ha\ luminosity and $\Lacc$ derived in this work (upper) and from literature values (lower). 
    Colors and symbols are described in the caption of Figure~\ref{fig:Mdot}.
    The solid and dashed lines denote the observed-\mbox{UV}-continuum based empirical relationship of \citet{Alcala+2017} ($y=1.13\,x+1.74$) and the shock-model based relation of \citet{Aoyama+2021} ($y=0.95\,x+1.61$), respectively. In the upper panel, \Ha\ luminosities are additionally corrected using the local extinction inferred from the spectral fitting; this correction is not applied in the lower panel to keep it independent of any our fitting results.
    }
    \label{fig:Ha_Lacc}
\end{figure}

Contrary to this expectation, the upper panel of Figure~\ref{fig:Ha_Lacc} shows a clear correlation between $\LHa$ and $\Lacc$, with the sample apparently following the shock-emission prediction of \citet{Aoyama+2021} (dashed line). This agreement is non-trivial because the shock-emission component often accounts for less than 10\,\% of $\LHa$.
The agreement may be coincidental, resulting from two compensating effects: (1) the dashed-line prediction assumes optically thin shock-origin \Ha, whereas \Ha\ is optically thick in the present sample, making the shock-origin \Ha\ luminosity fainter than the prediction; and (2) the observed \Ha\ luminosity is enhanced by the non-shock component.
If it is not coincidental, the non-shock \Ha\ emission may have a similar energy-conversion efficiency to the shock emission, $\sim 0.01\,\%$ \citep{Aoyama+2018}. However, this would require the non-shock-emitting region, likely the accretion column, to reprocess a large fraction of $\Lacc$ into \Ha, which is not necessarily expected.

Comparing with the empirical $\Lacc$--$\LHa$ relationship for TTSs (\citealp{Alcala+2017}; solid line), $\Lacc$ is systematically larger by $\approx 1$ dex, because their $\Lacc$ only accounting for the continuum excess can be underestimated for low-mass objects, as discussed in \S~\ref{sec:Lacc}. 
Nevertheless, compared to the \citet{Aoyama+2021} prediction, the samples shift upward and downward at small and large $\Lacc$, respectively, indicating a gentler slope of the $\Lacc$--$\LHa$ relation that may be somewhat closer to the \citet{Alcala+2017} relationship. This trend is more apparent when the transparent points at $\Lacc \gtrsim 10^{-2}\Lsun$ are included. Although these points are less reliable individually, they may still help trace the population trend. This suggests that the empirical relationship captures the overall trend, although a correction is needed.

The lower panel of Figure~\ref{fig:Ha_Lacc} shows the $\LHa$–$\Lacc$ relationship for the same sample, but using literature values of $\Lacc$ instead of those derived in this work; namely, corrected only for interstellar $\AV$ without fitting-derived local $\AV$.
Compared with the literature values, our derived $\Lacc$ shows a tighter correlation with $\LHa$.
A similar tightning is also seen for the other hydrogen lines in Figure~\ref{fig:Lacc_comp}. For those lines, however, the tighter correlations may be partly expected, because $\Lacc$ is derived from spectral fitting that includes the same lines.
In contrast, \Ha\ is not included in the fitting, and its luminosity is often dominated by the non-shock component.
The emergence of a tighter correlation between the model-derived $\Lacc$ and the observed $\LHa$ is therefore non-trivial, and suggests that our $\Lacc$ estimates may trace the underlying accretion luminosity more reliably than the literature values.

\subsection{Implications for Accreting Planets}
\label{sec:D_Planets}
This work finds that shock-origin hydrogen-lines emission is dominant in all objects below $0.05\,\Msun$ ($\sim 50\,\MJ$).
This suggests that hydrogen lines from accreting planets are likely to be shock-dominated, although the broader component may still contribute at a subdominant level.
Consistent with this picture, previous studies have shown that hydrogen-line flux-ratio diagnostics for planetary-mass objects, such as Delorme 1 AB b \citep{Betti2022} and TWA 28 b \citep{Aoyama+2024}, are preferentially consistent with the shock-emission model, based on infrared (IR) hydrogen lines in the Paschen and Brackett series.

However, the present work is not conclusive specifically about the origin of \Ha. The shock-origin \Ha component spans 1--100\% of the observed \Ha\ even at $<0.05\,\Msun$ (see \S~\ref{sec:D_Ha} for more details), implying that the origin of \Ha can remain ambiguous. This is particularly important for accreting bona-fide planets, for which \Ha is the only confirmed hydrogen line \citep{wagner2018,haffert2019,zhou2022,Close2024, currie2022,Close2025,Li2025} and planetary accretion diagnostics therefore have to rely on \Ha\ alone.
Indeed, \citet{thanathibodee2019} modeled the observed \Ha\ emission of PDS 70 b as arising from the magnetospheric accretion flow, and the resulting $\Mdot$ inferred from $L_\Ha$ is about an order of magnitude lower than that predicted by the shock-emission model of \citet{Aoyama+2019,Aoyama+2021}. Such a flow-emission estimate could be reconciled with the shock-emission model if only $\sim$10\% of the observed \Ha\ arises from the shock, a fraction that is typical in this work.
To establish the origin of planetary \Ha\ more securely, further observations are required, including high-spectral-resolution spectroscopy to resolve the \Ha\ line profile and/or detections of additional hydrogen lines that can constrain the shock-origin contribution, as demonstrated in this work.

\subsection{Comparison with Flux-ratio-based Assessments}
\label{sec:CompFluxRatio}
Previous studies have attempted to identify hydrogen lines of shock origin using their flux ratios \citep{Betti2022,Betti2022Erratum, Aoyama+2024,Hashimoto+Aoyama2025}.
In two-dimensional flux-ratio diagrams, these studies compared the observed line ratios with predictions from the shock-emission model of \citep{Aoyama+2018} and the accretion flow model of \citep{Kwan+Fischer2011}. The emission origin was then assessed by visually examining which model distribution was more consistent with the observed ratios. Because each diagram is defined by two flux ratios sharing a common reference line, this assessment uses the integrated fluxes of only three hydrogen lines.
Among these studies, \citet{Hashimoto+Aoyama2025} used the ratios $\Hg/\Hb$ and H8$/\Hb$. All three lines, \Hb, \Hg, and H8, are included in our spectral fitting, and the samples of the two studies partially overlap. We therefore compare the two assessment below.

\begin{deluxetable}{l ccc  cc | c }
\label{tab:CompFluxRatio}
\tablecaption{Comparison of hydrogen-line-origin assessments between this study and \citet{Hashimoto+Aoyama2025}}
\tablehead{& S & B & F & N & C & Total}
\startdata
S          &  2 & 2 & 11 & 0 & 2 & 17\\
S+BC       &  2 & 1 & 5  & 0 & 0 & 8 \\
S+BC-like  &  4 & 2 & 21 & 0 & 1 & 27\\
DBC        &  3 & 1 & 6  & 1 & 0 & 11\\
N          &  0 & 0 & 6  & 3 & 0 & 9 \\ \hline
Total      & 11 & 6 & 49 & 4 & 3 & 73
\enddata
\tablecomments{
Rows: classifications of hydrogen-line origin adopted in this work (\S~\ref{sec:Class}).
Columns: classifications of hydrogen-line origin in \citet{Hashimoto+Aoyama2025}: 
accretion-shock origin (S), both shock and accretion-flow origins are plausible (B), accretion-flow origin (F), neither shock nor accretion-flow origin likely (N), and chromospheric activity (C).
}
\end{deluxetable}

The classifications adopted in this study and by \citet{Hashimoto+Aoyama2025} are listed in columns 12 and 11 of Table~\ref{tab:Summary}, respectively, and their correspondence is summarized in Table~\ref{tab:CompFluxRatio}.
The expected correspondence between the two classifications is as follows. The \ConcShock and \ConcBC cases in this study are dominated by shock-origin emission and are therefore expected to correspond primarily to the ``S'' or ``B’’ categories of \citet{Hashimoto+Aoyama2025}. Conversely, the non-shock-dominated \ConcNot and \flgDBC cases are expected to correspond primarily to their ``F'' or ``N’’ categories, and possibly to ``B.’’ Because the interpretation of the \ConcBCd cases is less clear, we exclude them from the following assessment.

The expected correspondences are indeed more common than the alternative combinations. Nevertheless, discrepancies between the two assessments occur frequently. This comparison demonstrates that flux-ratio analysis based on the integrated fluxes of three hydrogen lines provides a useful indicator of the emission origin, but does not by itself conclusively identify shock-origin hydrogen emission.

A key factor underlying the discrepancies between the two classifications is the treatment of extinction, because wavelength-dependent extinction alters the hydrogen-line flux ratios. \citet{Hashimoto+Aoyama2025} did not account for extinction, as their samples was selected to have low extinction. However, our fitting frequently derives non-zero $\AV$ even for the \citet{Hashimoto+Aoyama2025} samples, at least partly because of temporal variability (see \S~\ref{sec:extinction} and \S~\ref{sec:D_Variability}).
To constrain $\AV$ self-consistently using hydrogen-line flux ratios, at least three independent ratios from four hydrogen lines are required.
Taking flux ratios removes the emitting area as a free parameter, but leaves three parameters to constrain: $v_0$, $n_0$, and $\AV$. The two flux ratios used in the previous analysis are therefore insufficient to constrain extinction simultaneously with the shock parameters.

Another important factor is the mixture of shock and non-shock emission components. Our spectral fitting shows that the \ConcBC case is approximately one-third of the shock-dominated cases (\ConcShock\ and \ConcBC). Even among the \ConcShock cases, H8 and H9 occasionally show systematic deviations from the shock-model prediction, although the deviations remain within the observational uncertainties. One example is H8 in Par-Lup3-4 (\ObsID{203}; top row of Figure~\ref{fig:SEDs_good}). Such deviations may indicate a weak broad-component contribution even in spectra classified as \ConcShock.
Because its contribution accumulates over the line profile, a weak additional component can affect integrated flux ratios more strongly than spectral fitting, leading to a different assessment of the emission origin.

In summary, although flux ratio is an important factor in identifying the hydrogen-line origin, previously-used analysis using two-dimensional flux-ratio diagram is insufficient.
A more robust flux-ratio assessment requires at least three independent line ratios from four hydrogen lines to constrain extinction simultaneously with the shock parameters.
It is also desirable to separate the broad-component contribution before evaluating the line ratios.
We plan to develop such an updated flux-ratio framework, incorporating both extinction and multiple emission components, in a follow-up study.

\subsection{Time Variability}
\label{sec:D_Variability}

\begin{figure*}
    \centering
    \def\wtem{\wT}
    \includegraphics[width=\wtem]{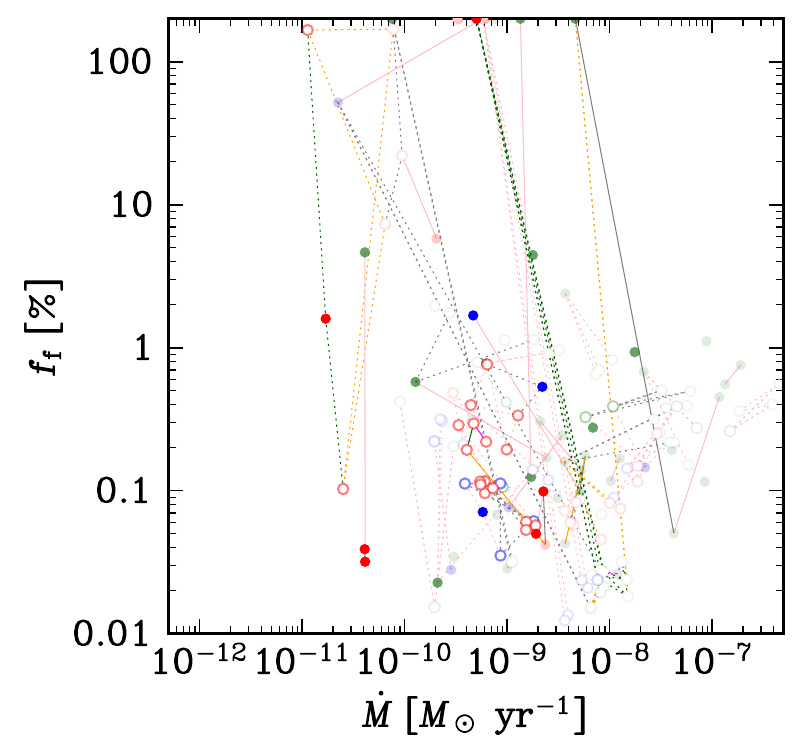}
    \includegraphics[width=\wtem]{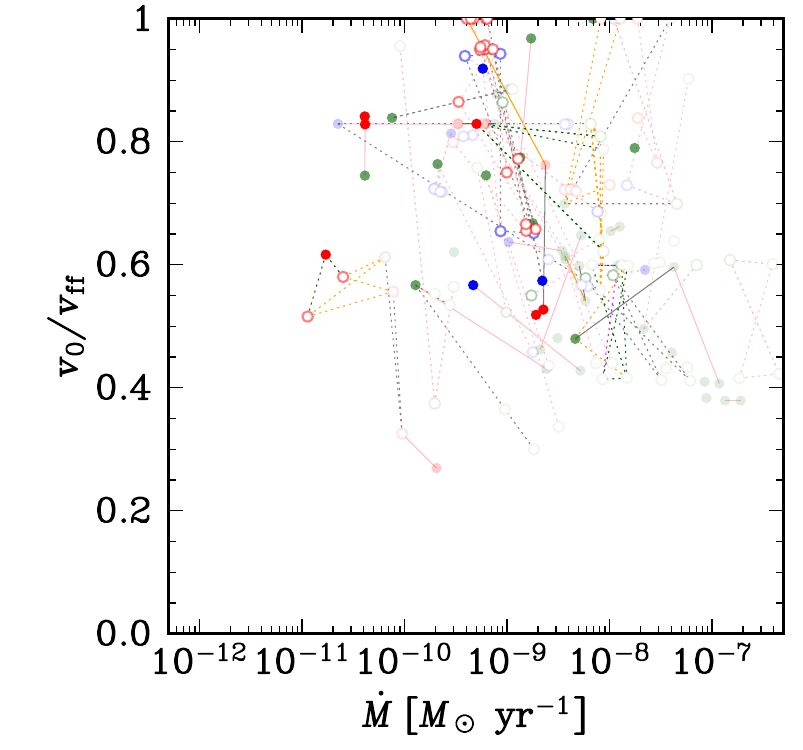}
    \includegraphics[width=\wtem]{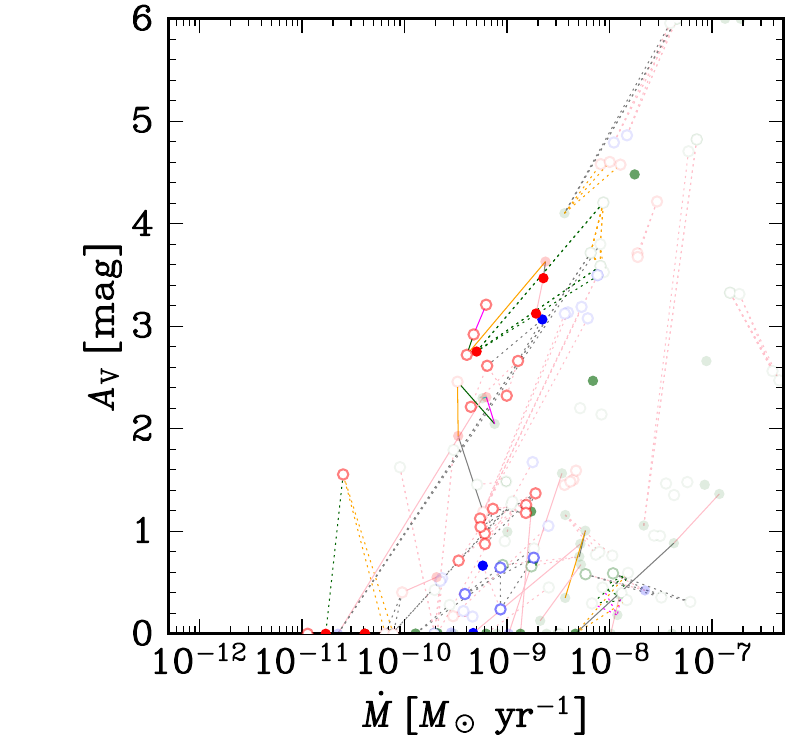}\\
    \includegraphics[width=\wtem]{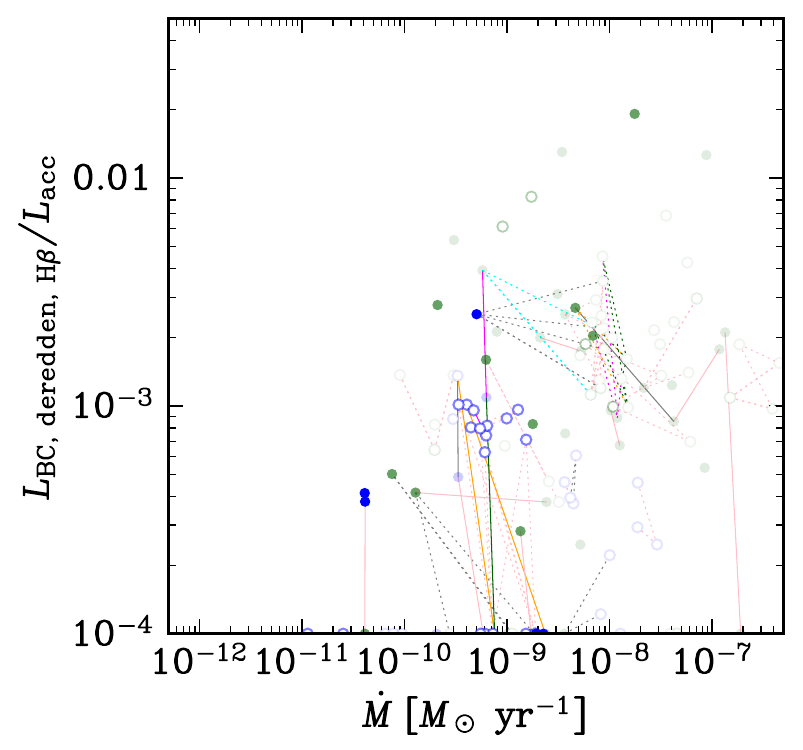}
    \includegraphics[width=\wtem]{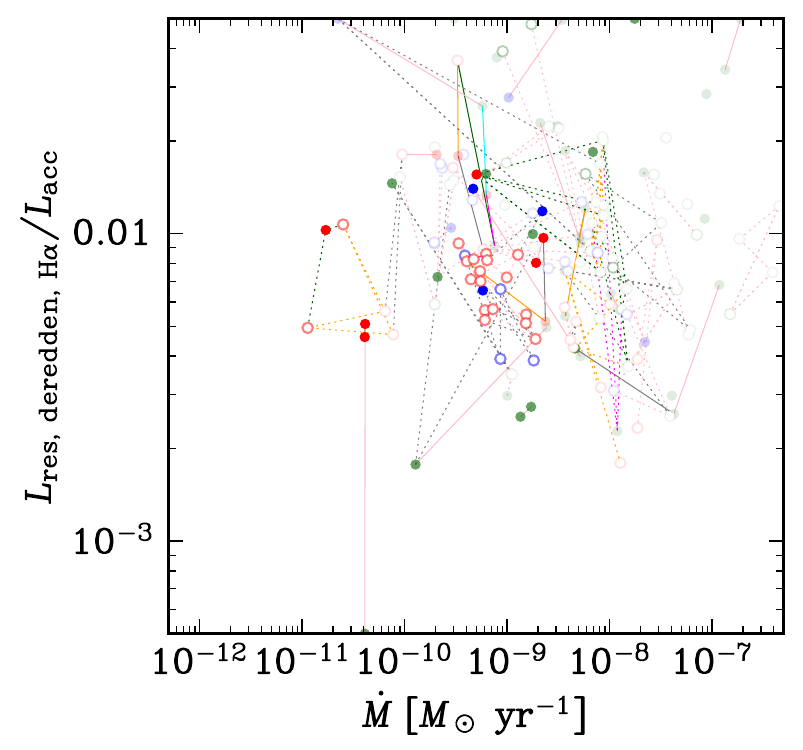}
    \includegraphics[width=\wtem]{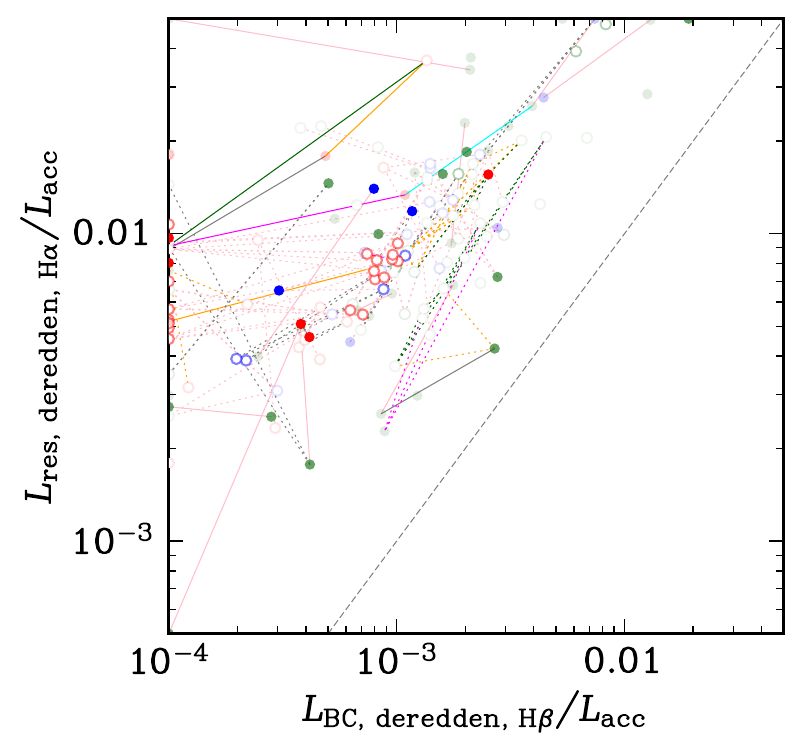}
    \caption{
    Estimates for multiply observed objects, shown as functions of the inferred mass accretion rate $\Mdot$. Thin lines connect different observing epochs for the same object: first--second (black), second--third (orange), third--fourth (green), fourth--fifth (magenta), fifth--sixth (cyan), and sixth--seventh (pink).
    A single observation can have multiple accepted estimates because of multimodal fitting; such connections are shown as dashed lines.
    The errorbars are omitted for clarity.
    }
    \label{fig:Variable}
\end{figure*}

A subset of our sample was observed multiple times. When the line fluxes and/or spectral profiles vary significantly between epochs, a likely cause is a change in the mass accretion rate, $\Mdot$. Because these repeated observations share the same basic stellar properties, they provide a useful way to reassess the dependence of the inferred quantities on $\Mdot$ with less object-to-object uncertainty than in the full sample.
Motivated by this, Figure~\ref{fig:Variable} shows the inferred parameters as a function of $\Mdot$. Different epochs of the same object are connected by colored thin lines. Dashed segments are used when a single epoch has multiple acceptable solutions.

The filling factor $\ff$ (upper left) shows both positive and negative epoch-to-epoch trends, suggesting no simple correlation with $\Mdot$. One possible feature is that $\ff$ varies more substantially at low $\Mdot$ and remains nearly constant at high $\Mdot$.
Additionally, the former large variations tend to have negative slopes, whereas the latter small variations tend to have positive slopes. The negative low-$\Mdot$ trends may reflect changes in whether a dense accretion column is located on the visible front side or hidden back side of the object.

The velocity ratio $v_0/\vff$ also shows both positive and negative trends with $\Mdot$. However, nearly vertical or horizontal connections are common, indicating that changes in $\Mdot$ and $v_0/\vff$ are not necessarily coupled.
If variations in $v_0$ reflect changes in the magnetospheric truncation radius, as discussed in \S~\ref{sec:v0_related}, this may imply that variations in the accretion rate and accretion geometry are at least partly independent.
This behavior is broadly consistent with recent 3D MHD simulations showing accretion through multiple columns with significant time variability \citep[e.g.,][]{Romanova2012,Takasao+2022,Zhu2024}. In these simulations, disk turbulence and related instabilities can alter the geometry of individual accretion columns independently of variations in the total accretion rate.

The visual extinction $\AV$ tends to increase with $\Mdot$, broadly consistent with the trend discussed in \S~\ref{sec:extinction}. Part of this correlation may be artificial, because adopting a larger $\AV$ increases the dereddened line fluxes and hence the inferred $\Mdot$. However, most connections are shallower than this artificial correlation, $\AV \propto 2.5 \log_{10}\Mdot$, suggesting that the observed positive trend is unlikely to be solely produced by this effect.
We also tested whether the inferred extinction variability affects the photospheric emission. The continuum fluxes around $1210$ and $731$\,nm, which are expected to be relatively insensitive to accretion emission, sometimes vary by up to a factor of two but show no clear correlation with the fitting-inferred variations in $\AV$. 
In most cases, the continuum fluxes vary only modestly even when the inferred $\AV$ changes by $\sim1$,mag. This lack of correlation may indicate that the inferred $\AV$ variability is partly a fitting artifact. Alternatively, if the variability is physical, the extinction may be localized around the line-emitting accretion region, possibly arising from the accretion-flow column itself. Further studies are needed to clarify the origin of the inferred variability.

To assess the non-shock component, the lower panels show the conversion efficiency from $\Lacc$ to non-shock $\Hb$ (lower left) and $\Ha$ (lower middle), and the correlation between them (lower right).
The non-shock emission efficiencies do not show a clear temporal trend with $\Mdot$. However, the non-shock \Hb\ and residual \Ha\ efficiencies correlate with each other, suggesting that they are related to a common physical origin despite their qualitatively different line profiles, consistent with the discussion in \S~\ref{sec:D_Ha}.

The absence of cleaner trends likely reflects, at least in part, the multimodality of our parameter estimates. Although Figure~\ref{fig:Variable} shows all acceptable solutions, only one mode should correspond to the actual physical state at each epoch, while the others are spurious solutions. Connections involving such artificial modes can obscure real temporal behavior. In addition, even within a single mode, the parameter uncertainties are sometimes comparable to the epoch-to-epoch variations. Thus, a better understanding of temporal accretion variability will require tighter constraints on the physical parameters, which in turn require higher-quality observations.

\subsection{Caveats}
\subsubsection{Limitation of Model Parameter Range}

We use the post-accretion-shock hydrogen-line emission model of \citet{Aoyama+2018}, but its available parameter range is not necessarily wide enough to cover the full range of very-low-mass accretors. The current pre-shock velocity grid extends up to $v_0=200\,\kms$. Above this velocity, a nearly fully ionized layer appears and becomes increasingly extended. Because such a region cools inefficiently, the cooling timescale becomes long, while the chemical timescales, especially those of hydrogen level populations and ionization, remain much shorter. This large timescale contrast makes the equations increasingly stiff and the implicit solver less stable.

A similar problem arises at high density. Larger $n_0$ leads to faster collisional transitions among hydrogen levels and to larger optical depths in hydrogen lines. As a result, line cooling becomes less efficient and the cooling timescale again becomes longer, which further increases the stiffness of the problem.
In this work, some of the line spectra were recomputed at higher spectral resolution than in \citet{Aoyama+2018}. However, for some high-$v_0$ and high-$n_0$ models, the higher-resolution calculations failed to converge for the reasons discussed above. We therefore retained the original lower-resolution spectra in this part of parameter space.
As this example also illustrates, much stronger and denser shocks remain numerically difficult to model within the current framework of \citet{Aoyama+2018}.

At the same time, stronger shocks (higher $v_0$) and denser accretion flows (higher $n_0$) are expected to correspond to regimes in which non-shock emission becomes increasingly important. Since the present model grid already reaches the regime where shock and non-shock contributions are comparable, and even cases where the shock contribution is subdominant, it is likely sufficient for the practical purpose of analysing shock-dominated hydrogen-line emission.

Nevertheless, there may still be phases in which non-shock emission remains weak even for stronger or denser accretion, so that the shock-origin component could in principle be analyzed if the non-shock contribution were separated appropriately. Therefore, extending the model to cover stronger and denser accretion shocks is one of the major directions for future work.

\subsubsection{Flux Correction}
\label{sec:D_FluxCorrection}
In the narrow-slit mode of VLT/\xshooter, slit losses can lead to an underestimate of the absolute flux \citep{Manara2021}. By comparing spectra obtained in the narrow-slit, wide-slit, and photometric modes, \citet{Alcala+2017} showed that the absolute flux in narrow-slit spectra can be lower than the true value by up to a factor of \(\sim 3\).
Our sample includes 68 observations without flux calibration for such slit loss. In addition, 44 observations are calibrated only in the UVB and VIS arms, while the NIR arm remains uncalibrated.

By comparing the fits to calibrated and uncalibrated spectra, we find that the emission-mechanism classification (\S~\ref{sec:Class}) is often unchanged by the flux calibration. This is mainly because the classification is based primarily on the spectral profiles and relative line fluxes, rather than on the absolute flux. In addition, the wavelength dependence of the slit-loss effect can be partly absorbed by the inferred extinction. We therefore include the uncalibrated spectra in our analysis.

However, the lack of absolute flux calibration can affect the inferred physical quantities. An underestimated flux scale directly leads to an underestimate of $\Lacc$ and hence $\Mdot$. On the other hand, uncalibrated spectra can also lead to an overestimate of $\AV$, as discussed above, which increases the dereddening correction. These two effects act in opposite directions when deriving intrinsic line luminosities, so the net offset in $\Lacc$ and $\Mdot$ can be smaller than the raw flux-calibration error of a factor of $\lesssim 3$ \citep{Alcala+2017}. Nevertheless, this uncertainty remains significant. We therefore flag the estimates based on uncalibrated spectra as unreliable (see \S~\ref{sec:AFit_suspicious}) and show them as transparent symbols in all plots.

In addition, the flux calibration procedures for some older data are less uniform than the more recently established procedure described by \citet{Manara2021}. For example, for the $\sigma$ Ori objects, telluric correction was not applied to the NIR-arm spectra. We do not flag such specific differences in the calibration procedure on an object-by-object basis, but they may affect the inferred quantities for some older observations.

\subsubsection{Binaries and Companions}
In our sample, nine targets\footnote{Four and two are labeled as binary and companion in CASPAR \citep{Betti+2023}, respectively. In addition, TWA27, TWA30, and TWA32 are binary though not labeled as such in the CASPAR. There might be more not-labeled binaries.} are binaries or have nearby companions. Because the \xshooter observations typically use sub-arcsecond slits, close companions can contaminate the extracted spectra. For the medium-resolution data used here ($R>4000$), the slit widths are $\le 1\farcs3$ \citep{Vernet+2011}.

For five objects, the projected separation from their companions is larger than the slit width.
SR12C, TWA~8B, TWA~30A, ISO-ChaI 143, and Sz 39 are widely separated from their companions, by
$\sim8\farcs7$ \citep{Santamaria-Miranda+2018,Santamaria-Miranda+2019},
$\sim13\farcs2$ \citep{Messina2010},
$\sim80\farcs2$ \citep{Looper2010},
$\sim18\farcs16$ \citep{Kraus+Hillenbrand2007},
and $\sim24\farcs38$ \citep{Kraus+Hillenbrand2007},
respectively, so any contamination should be very limited.

TWA~27A may be contaminated by TWA~27B, whose angular separation is $0\farcs78$ \citep{luhman2023_TWA27B}. However, at \Pb, the accretion-excess flux of TWA~27B is almost two orders of magnitude fainter than that of TWA~27A \citep{Aoyama+2024}, so the contamination is still expected to be minor. Nevertheless, the spectrum of TWA~27 is classified as \ConcBCd due to its largely consistent but statistically-significant offset from the model, inferring that this may in part reflect contamination from the companion.

TWA~32 is a nearly equal-mass binary with a reported separation of $\sim0\farcs6$ \citep{Shkolnik2011}, which is likely unresolved under the present \xshooter conditions.
Its classification as \ConcNot\ should therefore be treated with caution. However, because this single object has little impact on the overall conclusions of this paper, we retain it in the sample.

HO Cha has a close companion at a separation of only $\sim0\farcs28$ \citep{Daemgen2013} and is unresolved \citep{Manara2016}. It is classified as \ConcBCd, but the excess is fitted to another peak rather than to a broader component. Therefore, our fitting may unintendedly trance emission from one of the binary, although the estimated values should be regarded as unreliabile.

2MASS J15354856-2958551 is a binary with a separation of $\sim1$\arcsec, but it was reported to be resolved \citep{Manara2020}. our target corresponds to the eastern component.

\subsubsection[Alternative Interpretations of a Small Shock-to-Free-Fall Velocity Ratio]{Alternative Interpretations of Small $v_0/v_\mathrm{ff}$}
\label{sec:D_WeakShock}
In our estimates, $v_0/\vff$ is frequently smaller than unity. In \S~\ref{sec:v0_related}, we discussed a smaller truncation radius as the primary explanation, and the subsequent discussion of the dipole magnetic field strength is based on that interpretation. However, the essential implication of small $v_0/\vff$ is simply that the shock is weaker than expected from $\vff$. If the effective shock strength is reduced for some reason other than a smaller truncation radius, then $v_0/\vff$ need not reflect the actual flow speed ratio, and the following interpretations may become invalid.

One possible alternative is an inclined shock. For a dipolar magnetic field, the field-line geometry can be approximated as \citep{Ghosh1977,Hartmann+1994}
\begin{equation}
    r = r_\mathrm{m}\sin^2\theta,
\end{equation}
where $r_\mathrm{m}$ is the distance at which the field line crosses the disk midplane.
The cosine of the shock-inclination angle relative to the stellar surface normal can then be written as
\begin{equation}
\label{eq:ShockInclination}
    \cos\theta_\mathrm{inc.\,sh} = 2\sqrt{\frac{f_\mathrm{m}-1}{4f_\mathrm{m}-3}},
\end{equation}
where $f_\mathrm{m}\equiv r_\mathrm{m}/\Rs$.
For $f_\mathrm{m}=5$, corresponding to the classical truncation radius of $5\Rs$, one obtains $\cos\theta_\mathrm{inc.\,sh} \approx 0.97$.
Thus, the inclination effect alone is too small to explain the much lower $v_0/\vff$ values found in many of our objects.

When the magnetosonic velocity exceeds the sound speed, the effective Mach number becomes smaller, resulting in a weaker shock.
The condition that the Alfv\'en velocity perpendicular to the shock surface is comparable to $v_0$ gives a rough estimate of
\begin{equation}
\label{eq:MagnetoShock}
\begin{split}
B_\parallel \approx &\, 2 v_0 \sqrt{\pi \rho}\\
=&\, 250 \,\textrm{G} \\
&\times
\left( \frac{n_0}{10^{13}\,\cc}\right)^{1/2}
\left( \frac{v_0}{150\,\kms}\right),
\end{split}
\end{equation}
where $B_\parallel$ is the magnetic-field strength parallel to the shock surface, and the mean mass per hydrogen nuclei is $\mu_\mathrm{H}=2.27\times10^{-24}$\,g in our model.
Observationally, Zeeman-splitting measurements place upper limits of order $\sim$kG for some accreting M-type stars \citep{Reiners+2009}, while sub-kG fields are still allowed.
In our own estimates, although they are based on the assumption of pure-hydro shock, the inferred dipole field strengths can be of order a few hundred Gauss.
These values are comparable to the characteristic value in Eq.~(\ref{eq:MagnetoShock}), so the key factor is how large the shock-parallel component actually is.
Considering the shock-inclination angle in Eq.~(\ref{eq:ShockInclination}), the poloidal component of the dipole field projected parallel to the shock surface can be only a few percent of the total field strength. This suggests that magnetosonic effects may not be common, but they cannot be ruled out.

\section{Summary and Conclusions}
\label{sec:Conclusion}

We performed simultaneous spectral fitting of seven hydrogen lines (\Hb, \Hg, H6, H8, H9, \Pb, \Brg) in VLT/\xshooter archival spectra of very-low-mass accreting objects, using the hydrogen-line shock-emission model of \citet{Aoyama+2018}. Our aim was to identify the dominant hydrogen-line emission mechanism and to investigate how it changes towards the low-mass and low-accretion regime where shock emission is expected to become important. The main results of this study are as follows.

\begin{enumerate}
    \item A significant number of objects have hydrogen line dominated by the shock emission. In particular, we found only shock-dominated cases in the lowest-mass ($M<0.05\,\Msun$) and lowest-$\vff$ ($\vff<175\,\kms$) regime. This is qualitatively consistent with the theoretical prediction that hydrogen-line cooling is efficient at pre-shock velocities $v_0\lesssim200\,\kms$ (\S~\ref{sec:Class}).
    On the other hand, pure-shock-emission cases become rare above $ \sim0.2\,\Msun$ and disappear above $\sim 0.35\,\Msun$ (\S~\ref{sec:ClassDependency}).

    \item The transition from shock-dominated to non-shock-dominated hydrogen-line emission is not sharp. Even where shock emission is efficient, a broad excess component (BC) can remain non-negligible, and its relative contribution tends to increase toward higher mass and higher accretion rate. This implies that the dominant emission mechanism is governed not only by the decline of shock-emission efficiency, but also by the growth of the BC contribution (\S~\ref{sec:D_BC}). This explains naturally why previous hydrogen-line color diagnostics \citep{Hashimoto+Aoyama2025} did not reveal a clear boundary between the two regimes.

    \item A phenomenological broad-component subtraction is therefore useful for approximately isolating a shock-related narrow component from a broader excess profile. Although representing the BC by a single Gaussian is a strong assumption, it works reasonably well when the shock-emission component remains important in the observed line profile. This makes the broad-component-subtracted fit useful not only for testing the presence of shock emission but also for providing a reasonable empirical separation
    in mixed cases.

    \item The inferred pre-shock velocities are often significantly smaller than the free-fall velocity, implying smaller truncation radii (sometimes $\lesssim 2\Rs$) than are commonly assumed ($\sim 5\Rs$). Under the standard magnetospheric-accretion interpretation, these truncation radii further imply surface dipole magnetic field strengths often below one kiloGauss. This is broadly consistent with previous studies \citep{Reiners2012}, although this part of the inference remains model-dependent (\S~\ref{sec:v0_related}).
    
    \item Direct spectral fitting allows parameter estimation beyond what is possible with hydrogen-line color diagnostics alone. The inferred filling factors are typically small ($\sim0.01$--$1\%$), supporting the magnetospheric accretion scenario (\S~\ref{sec:ff}).
    In addition to the interstellar extinction, some observations infer strong local extinction, which tends to increase with accretion rate (\S~\ref{sec:extinction}) and vary with time by up to $\sim2$\,mag (\S~\ref{sec:D_Variability}).

    \item
    The inferred accretion luminosities are systematically larger than literature values based on conventional UV-continuum assumptions \citep[as reviewed in][]{Hartmann+2016}, often by an order of magnitude and in some low-mass cases by even more (\S~\ref{sec:Lacc}).
    As a result, the inferred mass-accretion rates are also systematically higher, and the $M$--$\Mdot$ relation appears flatter than previously reported in the brown-dwarf regime (\S~\ref{sec:Mdot}).

    \item The \Ha\ line is a clear exception to the other fitted hydrogen lines. Even when the other Balmer and Paschen lines are well reproduced by the shock-emission model, the predicted \Ha\ is often significantly fainter than observed. The residual \Ha\ component shows no clear dependence on the shock parameters, but correlates with the BC efficiency in \Hb, suggesting that the \Ha\ excess belongs to the same broad family of non-shock components while retaining a profile morphology specific to \Ha (\S~\ref{sec:D_Ha}).
\end{enumerate}
Overall, our results support a picture in which hydrogen-line emission in very-low-mass accretors changes gradually across parameter space, rather than switching sharply between pure shock-emission and pure non-shock-emission regimes. In this transitional regime, mixed cases appear to be common, and a phenomenological subtraction of Gaussian BC from observed spectra is useful for providing a reasonable empirical separation.

Future progress will require more realistic modeling of the non-shock component, especially accretion-column \citep[e.g.,][]{Hartmann+1994,Muzerolle+1998,Muzerolle+2001} and wind emission \citep[e.g.,][]{Kurosawa2006a,Kurosawa2011,Kurosawa+2012,Wallis+2025}, together with time-resolved spectroscopy over a wider range of masses and accretion states. In particular, a framework that treats the shock-origin narrow component and the broad excess component in a unified way will be essential for establishing hydrogen lines as quantitative probes of accretion at low mass objects.

\begin{acknowledgments}
\revise{We thank the anonymous referee for the constructive and helpful comments.}
Y.A.\ acknowledges support from the National Science Foundation of China with grants No.~W2533003 and 12233004.
G.-D.M.\ acknowledges support of the DFG priority program SPP 1992 ``Exploring the Diversity of Extrasolar Planets'' (MA~9185/1), from the Swiss National Science Foundation under grant 200021\_204847 ``PlanetsInTime'', and from the European Research Council (ERC) under the European Union's Horizon 2020 Research and Innovation Programme via the ERC Consolidator Grant ``PROTOPLANETS'' (No.~101002188; PI: M.~Benisty). Parts of this work have been carried out within the framework of the NCCR PlanetS supported by the Swiss National Science Foundation.
C.F.M.\ is funded by the European Union (ERC, WANDA, 101039452). Views and opinions expressed are however those of the author(s) only and do not necessarily reflect those of the European Union or the European Research Council Executive Agency. Neither the European Union nor the granting authority can be held responsible for them.
M.I.\ acknowledges support from the Japan Society for the Promotion of Science (JSPS) with KAKENHI No.~25K01061.
J.M.A.\ acknowledges financial support from Large Gran INAF-2024 ``Spectral Key features of Young stellar objects: Wind-Accretion LinKs Explored in the infraRed (SKYWALKER)'' CUP: C63C24001530005.

Based on observations collected at the
European Southern Observatory under ESO programmes
084.C-0269(A), 084.C-0269(B), 084.C-1095(A), 085.C-0238(A), 085.C-0764(A), 085.C-0876(A), 086.C-0173(A), 087.C-0244(A), 089.C-0143(A), 089.C-0311(A), 089.C-0538(B), 089.C-0652(A), 090.C-0050(A), 090.C-0253(A), 091.C-0195(A), 091.C-0195(B), 091.C-0195(C), 091.C-0195(D), 093.C-0097(A), 093.C-0109(A), 093.C-0506(A), 093.C-0757(A), 093.C-0769(A), 094.C-0805(A), 095.C-0134(A), 095.C-0378(A), 96.C-0455(A), 096.C-0979(A), 097.C-0349(A), 097.C-0378(A), 097.C-0592(A), 097.C-0669(A), 097.C-0681(A),
105.2061.001, 105.20NP.001, 106.20Z8.002, 106.20Z8.004, 106.20Z8.006, 106.20Z8.008, 108.22CB.001, 108.23N8.001, 109.24F7.001, 110.24BN.001, 111.255B.001, 112.25DB.001, 113.26GY.001, 287.C-5039(A),
0101.C-0527(A), 0101.C-0866(A), 0104.C-0454(A),
and data obtained from the ESO Science Archive Facility with DOI(s) under \url{https://doi.org/10.18727/archive/71} (European Southern Observatory ESO).
\end{acknowledgments}

\software{
astropy \citep{astropy:2013,astropy:2018,astropy:2022},
dustmaps \citep{dustmaps},
lmfit \citep{lmfit},
Matplotlib \citep{Hunter2007},
Numpy \citep{Harris+2020},
PyMultiNest \citep{Buchner+2014},
scipy \citep{2020SciPy-NMeth},
SUNDIALS/CVODE \citep{SUNDIALS2005,SUNDIALS2022,CVODE}
}

\section*{Data availability}
The data underlying this article are available in Zenodo at \doi{10.5281/zenodo.19398530}. A plotting script is also available at: \url{https://github.com/Yuhiko/PlottingScript_Aoyama2026}.

Additional figures are available in the online material at the same repository.
The CASPAR dataset we use is available in Zenodo at \doi{10.5281/zenodo.10150171}.

\appendix

\section{Target List}
\label{sec:A_Targets}
Table~\ref{tab:Targets} lists the physical properties of all 164 targets.

\begin{deluxetable}{rlccccc}[hb]
%\tablewidth{0pt} 
%\tabletypesize{\footnotesize}
%\tablenum{1}
\tablecaption{Physical properties of targets.} \label{tab:Targets}
\tablehead{ 
\colhead{Obj.~ID} &
\colhead{2MASS Name} 
& \colhead{$M/\Msun$} & \colhead{$R/\Rsun$} &\colhead{$\vff/[\kms]$} 
& \colhead{$d$/[pc]} & \colhead{$A_\mathrm{V0}$/[mag]}
% {Obj. ID} &
% {2MASS Name} 
% & {$M/\Msun$} & {$R/\Rsun$} &{$\vff/[\kms]$} 
% & {$d$/[pc]} & {$A_\mathrm{V0}$/[mag]}
}
%\decimalcolnumbers
\digitalasset
\startdata
  1 & SR12C & $0.013\pm0.006$ & $0.19\pm0.09$ & $160\pm50$ & $125\pm0$ & $0.12\pm0.02$ \\
  2 & J04141458+2827580 & $0.29\pm0.02$ & $1.49\pm0.08$ & $270\pm10$ & $129.9\pm0.5$ & $0.075\pm0.008$ \\
  3 & J04141760+2806096 & $0.084\pm0.028$ & $0.89\pm0.19$ & $190\pm40$ & $134.6\pm0.8$ & $0.11\pm0.01$ 
\enddata
%\tablerefs{
\tablecomments{
\revise{
Columns~2--4 are from CASPAR \citep{CASPAR}. 
Column~5 is derived using GAIA parallax \citep{gaia2020a}. When GAIA parallax data are unavailable, the CASPAR value is adopted and its uncertainty is set to zero to avoid inflating the error bars due to the distance uncertainty.}
Column~6 is derived following \citet{Edenhofer+2024}.
Table~\ref{tab:Targets} is published in its entirety in the machine-readable format. A portion is shown here for guidance regarding its form and content.%
}
\end{deluxetable}

In the publicly available flux-calibrated spectra of \citet{Venuti+2019}, we found that the VIS-arm flux of \ObsID{125} (TWA30, 2011 July 15)
is fainter by a factor of ten than expected from both the uncalibrated ESO archive spectrum and the flux levels in the UVB and NIR arms. This is likely because the VIS-arm spectrum is reported per $\AA$, whereas the other spectra are reported per nm. We therefore multiply the VIS-arm flux by a factor of ten.

\section{Line Flux Measurement}
\label{sec:A_FluxMeasure}

We measure the observed line flux by directly integrating the flux density over wavelength bins:
\begin{equation}
    F=\sum_i F_{\lambda,i}\,\Delta\lambda_i,
\end{equation}
where $F_{\lambda,i}$ and $\Delta\lambda_i$ denote the observed flux density per unit wavelength and the wavelength width of the $i$th spectral bin, respectively.

To define the integration range, we first determine the wavelength interval for the reference line \Hb, $\lambda_{\Hb,\,\min}$--$\lambda_{\Hb,\,\max}$, by requiring
\begin{equation}
    F_{\lambda}(\lambda_{\Hb,\,\min/\max}) = 1\sigma_{F,\lambda}.
\end{equation}
We then convert these boundaries into Doppler velocities relative to the line center,
\begin{equation}
    v_{\mathrm{D},\min/\max}
    = c\left(\frac{\lambda_{\Hb,\,\min/\max}}{\lambda_{\Hb,\,0}}-1\right),
\end{equation}
where $c$ is the speed of light. The same $v_{\mathrm{D},\min/\max}$ range is then applied to all the other lines, and the flux is integrated over the corresponding wavelength interval for each line.

The uncertainty on the measured line-integrated flux is computed as
\begin{equation}
    \sigma_F=\sqrt{\sum_i (\Delta\lambda_i\,\sigma_{F,\lambda,i})^2},
\end{equation}
over the same integration range.
For non-detected lines, we place upper limits on their fluxes. To remain consistent with the 3-$\sigma$ detection criterion adopted in the spectral fitting, the upper limit is taken to be three times the above flux uncertainty.
Some of our target was not corrected for the slit losses in the narrow-slit mode of VLT/\xshooter. The effect of this is discussed in \S~\ref{sec:D_FluxCorrection}.

Table~\ref{tab:Luminosity} lists the line luminosities normalized by the solar luminosity $\Lsun$.
Both the observed total luminosity and the broad-component luminosity are corrected for the interstellar extinction following \citet{Edenhofer+2024}, but not for the local extinction derived from our fitting (\S~\ref{sec:extinction}).
Additional data, including luminosities corrected for the local extinction, are available as online material.
\begin{deluxetable}{rrll cccc c ccc}[hb]
\tabletypesize{\footnotesize}
\tablecaption{Measured luminosity of hydrogen lines} \label{tab:Luminosity}
\digitalasset
\tablehead{
\colhead{Obj. ID} & \colhead{Obs. ID} & \colhead{Obj. Name/} & 
\colhead{Cal.} &
\multicolumn{8}{c}{ Luminosity/[$10^{-7}L_\odot$] } 
\\
&& \colhead{Method/Source} &
& \colhead{\Ha} & \colhead{\Hb} 
& \colhead{\Hg} 
& \colhead{\Hd} %& \colhead{H7} 
& \colhead{H8} 
&\colhead{H9} &\colhead{\Pb} & \colhead{\Brg}
}
%\decimalcolnumbers
\startdata 
  1 &  1 & SR12C & F & $7.08\pm0.05$ & $0.74\pm0.02$ & $0.41\pm0.02$ & $0.31\pm0.02$ & $0.23\pm0.02$ & $0.2\pm0.03$ & $0.0\pm2.0$ & $0.01\pm0.9$ \\
&&BC& F & $6.0\pm0.2$ &---&---&---&---& $0.034\pm0.004$ &---&---\\
  2 &  2 & J04141458+2827580 & T & $3140\pm20$ & $617\pm5$ & $374\pm4$ & $299\pm3$ & $200\pm4$ & $142\pm6$ & $91\pm22$ & $47\pm16$ \\
&&BC& T & $2100\pm100$ & $88\pm17$ & $50\pm7$ & $51\pm11$ & $31\pm5$ & $28\pm5$ &---&---\\
  3 &  3 & J04141760+2806096 & T & $1118\pm2$ & $95.7\pm0.3$ & $34.6\pm0.4$ & $21.6\pm0.3$ & $13.3\pm0.4$ & $11.3\pm0.4$ & $611\pm2$ & $279\pm3$ \\
&&BC& T & $1093\pm9$ & $104\pm4$ & $33\pm2$ & $20\pm1$ & $12.0\pm0.5$ & $10.3\pm0.3$ & $480\pm30$ & $164\pm9$
\enddata 
\tablecomments{
Table~\ref{tab:Luminosity} is published in its entirety in machine-readable format. A portion is shown here for guidance regarding its form and contents.
\revise{``Obj. ID'' and ``Obs. ID'' indicate the object and observation IDs used throughout the paper.
The third column gives the object name for the first row, which lists the total observed luminosity, and ``BC'' for the second row, which lists the broad-component luminosity.
``Cal.'' indicates whether the slit-loss correction was applied (T: yes; F: no).}
}
\end{deluxetable}

\section{Detailed Procedure and Full Results of Fitting}
\label{sec:AFit}

\subsection{Details of the Fitting Procedure}
\label{sec:AFit_FitDetail}

We fit the shock-model spectra to the observations within a Bayesian framework.
Assuming Gaussian uncertainties, the likelihood is
\begin{equation}
\mathcal{L}(\theta) \propto \exp\!\left[-\frac{1}{2}\chi^2(\theta)\right],
\end{equation}
where the $\chi^2$ is defined as
\begin{equation}
\label{eq:chi2}
\chi^2(\theta)=\sum_i
\left[
\frac{\Fobs(\lami)-\Fsyn(\lami;\theta)}{\FN(\lami)}
\right]^2.
\end{equation}
Here $\Fobs$ and $\FN$ are the observed spectral flux density and its $1\sigma$ uncertainty, respectively, and $\Fsyn$ is defined in Eq~(\ref{eq:fsyn}).

For posterior exploration, we adopt a nested-sampling algorithm and use the MultiNest library \citep{Feroz+Hobson2008,Feroz+2009,Feroz+2019} through the Python interface PyMultiNest \citep{Buchner+2014}.
Nested sampling is designed for Bayesian evidence estimation, where the likelihood is weighted by the prior volume. As a result, highly localized but high-likelihood solutions may not always be well represented by the posterior samples. In this work, our priority is to identify parameter sets that best reproduce the observed spectra, even if the acceptable region in parameter space is narrow. Therefore, we use nested sampling to explore the posterior distribution and to identify possible multimodal structure in parameter space, and then apply an additional post-processing procedure when selecting the best-$\redchi$ solutions (\S~\ref{sec:FitGoodness}) and when estimating representative parameter values (\S~\ref{sec:SED_Estimate}).

\paragraph{Fitting window.}
To define the spectral region used in the fit, we first select wavelength points around \Hb\ satisfying $\Fobs > \FC + 3\FCN$, where $\FC$ and $\FCN$ denote the continuum level and its uncertainty. We then apply the same Doppler-velocity range to the other fitted lines. This choice allows us to treat the line set uniformly, including weak lines whose peaks fall below $3\FCN$.

\paragraph{Decorrelation of Fitting Parameters}
Our fitting has six free parameters: the accretion flow velocity ($v_0$) and hydrogen-nuclei number density ($n_0$) immediately upstream of the shock front, the emitting area ($\Semit$), the wavelength shift ($f_\lambda$), the visual extinction magnitude ($\AV$), and extinction-curve parameter ($\RV$). However, some of these parameters are not independent and are expected to be strongly correlated. We therefore replace $\Semit$ for decorrelation.

Since the line flux is largely proportional to the accretion luminosity (see Eq.~[\ref{eq:Lacc1}]), the observed flux can be expressed as
\begin{equation}
    F_\mathrm{line} = f_\mathrm{line} \frac{1}{2} \mu_\mathrm{H} v_0^3 n_0 \Semit  10^{-f_\mathrm{ext} \AV/2.5},
\end{equation}
where $f_\mathrm{line}$ is the fraction of the accretion luminosity emitted in the line and $f_\mathrm{ext}$ is the ratio of the extinction at the line wavelength to the visual extinction. We therefore use a normalized emitting area,
\begin{equation}
    \Semit' \equiv  \Semit v_0^3 n_0  10^{-\AV/2.5}.
\end{equation}
This reparameterization significantly reduces the correlations among the fitting parameters and improves the fitting performance.

\subsection{Broad-component Subtracted Fit}
\label{sec:AFit_BCS}

For the broad-component-subtracted (BCS) fit, we first estimate the shock-related narrow component by fitting the shock-emission model only to the line-core region, defined by $F > 0.5\,F_{\mathrm{peak}}$, where $F_{\mathrm{peak}}$ is the peak flux of each line. The fitting procedure is otherwise the same as in \S~\ref{sec:SpectralFit_method}, and we adopt the best-fit solution with the smallest $\redchi$.

We then model the residual spectrum obtained after subtracting this peak-fit shock model, assuming that the broad component can be approximated by a single Gaussian profile. The target flux for the broad-component fit is defined as
\begin{equation}
    F_{\mathrm{t,\,BC}} = F_{\mathrm{obs}} - F_{\mathrm{PF}},
\end{equation}
where $F_{\mathrm{PF}}$ is the best-fit peak-fit shock model, and $F_{\mathrm{obs}}$ is the observed flux.
This procedure is not intended to provide a physical decomposition of the shock component; rather, it is intended to effectively mask the regions where the narrow shock component is dominant.
Since the peak fit is performed without explicitly accounting for a potential BC, $F_{\mathrm{t,\,BC}}$ often becomes nearly zero around the peak of $F_{\mathrm{obs}}$, where the S/N ratio is high and the fit is most strongly constrained. Therefore,
we intentionally inflate the per-pixel error in regions where the narrow component is dominant, and adopt
\begin{equation}
    \delta F_{\mathrm{t,\,BC}} = \sqrt{ \delta F_{\mathrm{obs}}^2 + F_{\mathrm{PF}}^2 },
\end{equation}
where $\delta F_{\mathrm{obs}}$ is the $1\sigma$ uncertainty of the observed flux. For the numerical Gaussian fitting, we use the least-squares optimizer \texttt{astropy.modeling.fitting.LevMarLSQFitter}.

For the broad-component fit, we define the wavelength boundaries of the fitting region using the condition that the continuum subtracted flux $F_{\mathrm{obs}}=0$, rather than the $3\sigma$ criterion used in the fiducial spectral fitting, in order to better capture faint broad emission. In addition, we reject broad-component fits with a peak signal-to-noise ratio ${\rm S/N}<1$ and set the broad-component flux to zero in such cases, thereby avoiding unphysical fits to very low-amplitude but extremely broad profiles.

Finally, we refit the shock-emission model to the spectra after subtracting the fitted broad component. For this broad-component-subtracted (BCS) spectrum, the target flux is
\begin{equation}
    F_{\mathrm{t,\,BCS}} = F_{\mathrm{obs}} - F_{\mathrm{BCF}},
\end{equation}
and the corresponding uncertainty is taken as
\begin{equation}
    \delta F_{\mathrm{t,\,BCS}} = \sqrt{\delta F_{\mathrm{obs}}^2 + \delta F_{\mathrm{BCF}}^2},
\end{equation}
where $F_{\mathrm{BCF}}$ and $\delta F_{\mathrm{BCF}}$ are the fitted broad-component flux and its $1\sigma$ uncertainty, respectively.

\subsection{Representative Parameter Values and Acceptable Ranges}
\label{sec:AFit_procedure}
We do not directly adopt the parameter summaries returned by PyMultiNest. Nested sampling is designed to maximize the Bayesian evidence, i.e., likelihood weighted by prior volume, and can therefore disfavor highly localized but high-likelihood solutions. In this work, however, our priority is to identify parameter sets that best reproduce the observed spectra, even if the acceptable region in parameter space is narrow.

We therefore proceed in a mode-based manner. First, we identify likelihood modes from the MultiNest samples. In addition to the final reported modes, we also consider candidate modes that may have been discarded during the nested-sampling procedure, because such modes can still contain high-likelihood solutions with small posterior volume. To judge whether a parameter set belongs to a given mode, we use the Mahalanobis distance defined from the sample covariance of each mode.

For each mode, we then refine the best-fit point by a local least-squares minimization and adopt that point as the representative parameter set. Modes whose best-fit likelihood does not satisfy our acceptance criteria are rejected. The acceptance criteria are the same as those used for the emission-mechanism classification in \S~\ref{sec:Class}: $\Pvalue \geq 0.05$ for the fiducial fit and $\redchi \leq$ 5 for the BCS fit.

To estimate the acceptable range of each parameter, we use the one-parameter 1-$\sigma$ criterion
\begin{equation}
    \chi^2 < \chi^2_{\mathrm{best,\,mode}} + \Delta\chi^2_{1\sigma,\,1},
\end{equation}
where $\chi^2_{\mathrm{best,\,mode}}$ is the minimum $\chi^2$ within the mode and $\Delta\chi^2_{1\sigma,\,1}\simeq 1.0$ is the 1-$\sigma$ increment for one degree of freedom. We determine the range of each parameter by testing whether a fixed value of the target parameter can be accommodated by any acceptable parameter set while allowing the other parameters to vary freely.
The $\chi^2$ exploration is performed using SciPy \citep{2020SciPy-NMeth}.

In deriving additional physical properties from the above estimated model parameters, we treat error propagation as follows.

For quantities derived purely from the fitted model parameters, we assess the acceptable range of the derived variable in the same manner as for the directly fitted parameters, as described above.

When combining a parameter range derived from the model fitting with literature parameters given with standard deviations, we first evaluate the derived quantity at the edge values of the fitted range and then expand the resulting interval by linearly propagating the literature uncertainties.

All model parameters, derived properties, and plotting scripts are available as online material (see the DATA AVAILABILITY section).

Tables~\ref{tab:FidFit} and \ref{tab:BCSFEstimate} list all fitting results for the fiducial (\S~\ref{sec:SpectralFit_method}) and BCS (\S~\ref{sec:M_BCSFit}) fits, respectively. Similarly, Table~\ref{tab:combined} lists all combined estimates, excluding observations that are not well fitted by the model.

\begin{deluxetable}{rrllccccccll}[ht]
\tablecaption{Estimated Properties by Spectral Fitting} \label{tab:FidFit}
\digitalasset
\tablehead{ \colhead{Obs. ID} & \colhead{Obj. ID}& \colhead{2MASS Name} & \colhead{Cal.}
&\colhead{$v_0/[\kms]$}  &\colhead{$n_0 / [\cc]$}     &\colhead{$S_\mathrm{emit}/[0.01\RS^2]$} &\colhead{$(f_\lambda-1)/[10^{-6}]$} &\colhead{$A'_\mathrm{V}$} &\colhead{$R_\mathrm{V}$} 
%&\colhead{$\dot{M}\, / [10^{-9}\MSyr]$}
&\colhead{$\chi^2_\mathrm{red,best}$} &\colhead{p-value}
}
%\decimalcolnumbers
\startdata
  1 &  1 & SR12C & F & $130^{132}_{123}$ & $14.5^{14.6}_{14.4}$ & $0.061^{0.061}_{0.025}$ & $-120^{-120}_{-130}$ & $2.8^{3.4}_{1.8}$ & $3.5^{4.8}_{2.6}$ & $0.31$ & $0.99$ \\
  7 &  7 & J04262939+2624137 & F & $199$ & $14.4^{14.6}_{14.3}$ & $0.28^{0.36}_{0.22}$ & $-63^{-59}_{-68}$ & $1.4^{1.5}_{1.2}$ & $5.5^{5.5}_{5.3}$ & $0.85$ & $0.86$ \\
&&Mode002& F & $199$ & $14.4^{14.6}_{14.3}$ & $0.28^{0.36}_{0.28}$ & $-63^{-59}_{-68}$ & $1.4^{1.5}_{1.3}$ & $5.5^{5.5}_{5.3}$ & $0.85$ & $0.86$ \\
&&Mode003& F & $136$ & $14.6$ & $0.33$ & $-72$ & $1.4$ & $5.2$ & $1.1$ & $0.1$ \\
  8 &  7 & J04262939+2624137 & F & $136^{137}_{135}$ & $14.8^{14.9}_{14.7}$ & $0.26^{0.32}_{0.22}$ & $-75^{-70}_{-81}$ & $1.5^{1.6}_{1.3}$ & $5.5^{5.5}_{4.9}$ & $0.8$ & $0.94$ \\
&&Mode002& F & $199$ & $14.7^{14.8}_{14.5}$ & $0.18^{0.23}_{0.14}$ & $-64^{-59}_{-70}$ & $1.4^{1.5}_{1.3}$ & $5.5^{5.5}_{5.2}$ & $0.8$ & $0.93$ \\
&&Mode003& F & $136$ & $14.7$ & $0.27$ & $-74$ & $1.4$ & $5.3$ & $0.81$ & $0.92$ 
\enddata
\tablecomments{Table~\ref{tab:FidFit} is published in its entirety in the machine-readable format. A portion is shown here for guidance regarding its form and contents. 
}
\end{deluxetable}

\begin{deluxetable}{rrllccccccll}[ht]
\tablewidth{0pt} 
\tablecaption{Estimated Properties by Broad-component Subtracted Spectral Fitting} \label{tab:BCSFEstimate}
\tablehead{ \colhead{Obs. ID} & \colhead{Obj.~ID}& \colhead{2MASS Name}  & \colhead{Cal.}
&\colhead{$v_0/[\kms]$}  &\colhead{$n_0 / [\cc]$}     &\colhead{$S_\mathrm{emit}/[0.01\RS^2]$} &\colhead{$(f_\lambda-1)/[10^{-6}]$} &\colhead{$A'_\mathrm{V}$} &\colhead{$R_\mathrm{V}$} 
%&\colhead{$\dot{M}\, / [10^{-9}\MSyr]$}
&\colhead{$\chi^2_\mathrm{red,best}$} &\colhead{p-value}
}
\digitalasset
%\decimalcolnumbers
\startdata
  1 &  1 & SR12C & F & $130^{132}_{123}$ & $14.5^{14.6}_{14.4}$ & $0.061^{0.11}_{0.061}$ & $-120^{-120}_{-130}$ & $2.8^{3.4}_{1.8}$ & $3.5^{4.8}_{2.6}$ & $0.29$ & $0.99$ \\
  3 &  3 & J04141760+2806096 & T & $189^{191}_{188}$ & $14.9$ & $7.3^{7.7}_{7.3}$ & $-64^{-58}_{-69}$ & $11$ & $3.3^{3.4}_{3.2}$ & $5.1$ & $<0.05$ \\
  5 &  5 & J04233919+2456141 & F & $138$ & $14.9$ & $9.1^{9.4}_{9.1}$ & $-130$ & $3.0^{3.1}_{3.0}$ & $2.9^{3.0}_{2.7}$ & $3.3$ & $<0.05$ \\
  6 &  6 & J04245708+2711565 & F & $139^{139}_{138}$ & $14.7$ & $0.74^{0.78}_{0.7}$ & $100$ & $0.0$ & $5.0^{5.5}_{2.8}$ & $1.3$ & $<0.05$ \\
  7 &  7 & J04262939+2624137 & F & $135^{136}_{134}$ & $14.4^{14.5}_{14.3}$ & $0.56^{0.66}_{0.46}$ & $-63^{-59}_{-68}$ & $1.6^{1.7}_{1.5}$ & $5.5^{5.5}_{4.5}$ & $0.77$ & $0.96$ \\
&&Mode002& F & $189^{193}_{185}$ & $13.8^{13.9}_{13.7}$ & $0.83^{0.99}_{0.83}$ & $-52^{-48}_{-56}$ & $1.5^{1.6}_{1.4}$ & $5.5^{5.5}_{5.1}$ & $0.84$ & $0.88$ \\
\enddata
\tablecomments{Table~\ref{tab:BCSFEstimate} is published in its entirety in the machine-readable format. A portion is shown here for guidance regarding its form and contents. 
}
\end{deluxetable}

\begin{deluxetable}{rrllcccccc}[ht]
\tablewidth{0pt} 
\tablecaption{Inferred properties of accreting objects} 
\label{tab:combined}
\tablehead{
\colhead{Obs. ID} & \colhead{Obj. ID} &
\colhead{Name} &\colhead{Cal.}
&\colhead{$\Mdot/[10^{-9}\,\Msun\,\mathrm{yr}^{-1}]$}  &\colhead{$\ff\,\%$}
&\colhead{$v_0/\vff\,\%$}  &\colhead{$\Rt / [\Rs]$} &\colhead{$\Rt / [\Rsun]$}    
&\colhead{$B$/[Gauss]} 
}
%\decimalcolnumbers
\digitalasset
\startdata
\multicolumn{9}{c}{ \it \bf Fiducial}\\
%%%%%%%%%%%%%%%%%%%%%%%%%%%%%%%%%%%%%%%%%%%%%%%%%%%%%%%%%%%%%%%%%%%%%%%%%%%%%%%%%%%%%%%%%%%%%%%%%%%%%%%%%%%%%%%%
  1 &  1 & SR12C & F & $0.488^{0.488}_{0.198}$ & $0.13^{0.26}_{0.003}$ & $80^{100}_{51}$ & $2.8^{\infty}_{1.3}$ & $0.54^{\infty}_{0.13}$ & $390^{\infty}_{26}$ \\
  7 &  7 & J04262939+2624137 & F & $3.63$ & $0.072^{0.082}_{0.063}$ & $72^{76}_{68}$ & $2.0^{2.3}_{1.8}$ & $1.2^{1.5}_{1.0}$ & $200^{280}_{150}$ \\
%%%%%%%%%%%%%%%%%%%%%%%%%%%%%%%%%%%%%%%%%%%%%%%%%%%%%%%%%%%%%%%%%%%%%%%%%%%%%%%%%%%%%%%%%%%%%%%%%%%%%%%%%%%%%%%%
\tableline
\multicolumn{9}{c}{ \it \bf S+BC}\\
%%%%%%%%%%%%%%%%%%%%%%%%%%%%%%%%%%%%%%%%%%%%%%%%%%%%%%%%%%%%%%%%%%%%%%%%%%%%%%%%%%%%%%%%%%%%%%%%%%%%%%%%%%%%%%%%
 22 & 11 & J04390163+2336029 & F & $0.579$ & $0.071^{0.092}_{0.049}$ & $91^{100}_{75}$ & $6.4^{\infty}_{2.3}$ & $6.4^{\infty}_{1.9}$ & $370^{\infty}_{50}$ \\
 23 & 11 & J04390163+2336029 & F & $0.862^{1.07}_{0.749}$ & $0.035^{0.056}_{0.021}$ & $65^{77}_{53}$ & $1.7^{2.5}_{1.3}$ & $1.7^{2.8}_{1.1}$ & $47^{120}_{23}$ \\
&&Mode002& F & $0.387$ & $0.11^{0.14}_{0.078}$ & $93^{100}_{77}$ & $8.5^{\infty}_{2.4}$ & $8.5^{\infty}_{2.1}$ & $500^{\infty}_{46}$ \\
%%%%%%%%%%%%%%%%%%%%%%%%%%%%%%%%%%%%%%%%%%%%%%%%%%%%%%%%%%%%%%%%%%%%%%%%%%%%%%%%%%%%%%%%%%%%%%%%%%%%%%%%%%%%%%%%
\tableline
\multicolumn{9}{c}{ \it \bf S+BC-like}\\
%%%%%%%%%%%%%%%%%%%%%%%%%%%%%%%%%%%%%%%%%%%%%%%%%%%%%%%%%%%%%%%%%%%%%%%%%%%%%%%%%%%%%%%%%%%%%%%%%%%%%%%%%%%%%%%%
  5 &  5 & J04233919+2456141 & F & $221^{228}_{221}$ & $0.55^{0.61}_{0.5}$ & $44^{45}_{42}$ & $1.2$ & $1.4^{1.5}_{1.3}$ & $440^{490}_{410}$ \\
  6 &  6 & J04245708+2711565 & F & $10.0^{10.3}_{9.69}$ & $0.02^{0.022}_{0.017}$ & $39^{40}_{37}$ & $1.1^{1.2}_{1.1}$ & $2.0^{2.1}_{1.9}$ & $62^{68}_{56}$
\enddata
\tablecomments{Table~\ref{tab:combined} is published in its entirety in the machine-readable format. A portion is shown here for guidance regarding its form and contents.}
\end{deluxetable}

\subsection{Additional Flag for Less-confident Estimate}
\label{sec:AFit_suspicious}

Even for parameter sets that pass the acceptance criteria in Appendix~\ref{sec:AFit}, we assign additional flags to some less-confident estimates, in order to retain potentially useful information while distinguishing such cases from more reliable estimates.

\begin{itemize}
    \item Estimates with $v_0 > \vff$ are unlikely. We therefore flag cases in which the lower bound of the estimated $v_0$ range is still above $\vff$ as less confident.
    
    \item Estimates with $\ff > 1.0$ are also unlikely. However, the shock front can in principle be displaced above the object surface. In addition, if the accretion flow extends beyond the magnetospheric regime---for example, accretion from the nebula at a very early stage---the shock may form at a larger radius. Nevertheless, such a distant shock surface corresponds to a shallower gravitational potential, a slower accretion flow, and hence a weaker shock. Therefore, excessively large $\ff$ remains unlikely. As a somewhat arbitrary criterion, we flag cases with $\ff > 1.5$ as less confident.

    \item Estimates with $n_0 > 10^{14}\cc$ rely on extrapolation beyond the original grid, which extends only up to $n_0 = 10^{14}\cc$ \citep{Aoyama+2018}. We nevertheless perform this extrapolation because a significant number of estimates lie close to the $n_0 = 10^{14}\cc$ boundary, and we wish to assess the possible parameter range above this limit. We therefore flag cases in which the lower bound of the estimated $n_0$ range is already above this limit as less confident.
\end{itemize}

When an observation has more than one estimate without a less-confident flag, the less-confident modes are omitted from Tables~\ref{tab:combined} and from the analysis in \S~\ref{sec:SED_Estimate}. When an observation has only less-confident mode(s), we retain them.
These observations are flagged as ``\flgQ'' in Table~\ref{tab:Summary} and are shown with transparent symbols in the figures where the inferred parameters are plotted.
Additionally, the estimates derived from the data without flux calibration are also shown as less reliable in the plots, though they are not flagged in the emission-mechanism classification (see also \S~\ref{sec:D_FluxCorrection}).

Tables~\ref{tab:FidFit} and \ref{tab:BCSFEstimate} list all modes, including those with less-confident flags. The mode IDs are identical across Tables~\ref{tab:Luminosity}, \ref{tab:FidFit}, \ref{tab:BCSFEstimate}, and \ref{tab:combined}.

\bibliographystyle{yahapj}
\bibliography{reference}

\end{document}